\documentclass[10pt,letterpaper]{article}
\usepackage[top=0.85in,left=2.75in,footskip=0.75in]{geometry}

\usepackage{amsmath,amssymb}

\usepackage{changepage}

\usepackage[utf8x]{inputenc}

\usepackage{textcomp,marvosym}

\usepackage{cite}

\usepackage{nameref,hyperref}

\usepackage[right]{lineno}

\usepackage{microtype}
\DisableLigatures[f]{encoding = *, family = * }

\usepackage[table]{xcolor}

\usepackage{array}

\newcolumntype{+}{!{\vrule width 2pt}}

\newlength\savedwidth

\newcommand\thickhline{\noalign{\global\savedwidth\arrayrulewidth\global\arrayrulewidth 2pt}%
\hline
\noalign{\global\arrayrulewidth\savedwidth}}

\raggedright
\usepackage[aboveskip=1pt,labelfont=bf,labelsep=period,justification=raggedright,singlelinecheck=off]{caption}

\makeatletter
\renewcommand{\@biblabel}[1]{\quad#1.}
\makeatother

\usepackage{lastpage,fancyhdr,graphicx}
\usepackage{epstopdf}
\fancyheadoffset[L]{2.25in}
\fancyfootoffset[L]{2.25in}
\DeclareMathOperator*{\argmax}{arg\,max}
\DeclareMathOperator*{\argmin}{arg\,min}
\usepackage{algorithm} 
\usepackage{algpseudocode}
\usepackage[graphicx]{realboxes} 
\usepackage{caption}
\usepackage{subcaption}
\usepackage{booktabs}
\usepackage{amsfonts}
\usepackage{pdflscape}
\usepackage{tabularx}
\usepackage[figuresright]{rotating}

\graphicspath{{./img/}}
\usepackage{tikz,xcolor,hyperref}

\definecolor{lime}{HTML}{A6CE39}
\DeclareRobustCommand{\orcidicon}{%
	\begin{tikzpicture}
	\draw[lime, fill=lime] (0,0) 
	circle [radius=0.16] 
	node[white] {{\fontfamily{qag}\selectfont \tiny ID}};
	\draw[white, fill=white] (-0.0625,0.095) 
	circle [radius=0.007];
	\end{tikzpicture}
	\hspace{-2mm}
}

\foreach \x in {A, ..., Z}{%
	\expandafter\xdef\csname orcid\x\endcsname{\noexpand\href{https://orcid.org/\csname orcidauthor\x\endcsname}{\noexpand\orcidicon}}
}
\newcommand{\orcid}[1]{\href{https://orcid.org/#1}{\textcolor[HTML]{A6CE39}{\aiOrcid}}}

\begin{document}
\vspace*{0.2in}

\begin{flushleft}
{\Large
\textbf\newline{Empowering differential networks using Bayesian analysis} 
}
\newline
\\

\newcommand{\orcidauthorA}{0000-0003-4235-6147}
\newcommand{\orcidauthorB}{0000-0003-4793-5674}
\newcommand{\orcidauthorC}{0000-0002-5881-9241}

Jarod Smith \textsuperscript{\orcidA{} 1\Yinyang},
Andri{\"e}tte Bekker\textsuperscript{\orcidB{}1\Yinyang},
Mohammad Arashi\textsuperscript{\orcidC{}2,1\Yinyang*}
\\
\bigskip
\textbf{1} Department of Statistics, University of Pretoria,
Pretoria, 0002, South Africa
\\
\textbf{2} Department of Statistics, Faculty of Mathematical Sciences, Ferdowsi University of Mashhad, Mashhad, Iran
\\
\bigskip

%
%
\Yinyang These authors contributed equally to this work.





* arashi@um.ac.ir

\end{flushleft}
\section*{Abstract}
Differential networks (DN) are important tools for modeling the changes in conditional dependencies between multiple samples. A Bayesian approach for estimating DNs, from the classical viewpoint, is introduced with a computationally efficient threshold selection for graphical model determination. The algorithm separately estimates the precision matrices of the DN using the Bayesian adaptive graphical lasso procedure. Synthetic experiments illustrate that the Bayesian DN performs exceptionally well in numerical accuracy and graphical structure determination in comparison to state of the art methods. The proposed method is applied to South African COVID-19 data to investigate the change in DN structure between various phases of the pandemic.



\section*{Introduction}\label{sec_1_introduction}

Probabilistic networks are becoming ever-present in a multitude of scientific disciplines. These networks aim to illustrate the relationships, if any, between the components of complex systems \cite{Ali_shojaie}. If the data is assumed to be Gaussian distributed with mean $\boldsymbol{\mu}$ and covariance matrix $\mathbf{\Sigma}$; the precision matrix $\mathbf{\Theta}$ directly determines the conditional dependence relations and structure of the undirected graphical model  \cite{koller2009probabilistic}.\par

Numerous statistical methods exist for Gaussian covariance and precision matrix estimation, as well as graphical model determination. In particular, from a frequentist approach, \cite{meinshausen2006high} introduce a computationally efficient neighborhood selection procedure.  The lasso is used for covariance estimation which enjoys consistency for sparse
high-dimensional graphs. The approach is quite effective, in that the sparse precision matrix is estimated by fitting the lasso to each variable using the remaining as predictors. Finally, the estimated precision matrix entry $(\theta_{ij})$ is non-zero if the estimated coefficient of $i$ on $j$ or vice versa is non-zero. Importantly, their algorithm can consistently estimate the set of non-zero entries in $\mathbf{\Theta}$, \cite{lauritzen1996graphical}. For a penalised likelihood methodology for sparse precision matrix estimation see \cite{yuan_and_lin_glasso, friedman2008sparse}. More so,  \cite{banerjee_glasso} estimate the undirected graphical model using both a block coordinate descent algorithm, as well as Nesterov’s first order method \cite{nesterov2005smooth}. Additionally, \cite{cai2011constrained} propose a $\ell_1$ constraint estimation technique for both sparse and non-sparse high dimensional matrices with applicability on a wide range of sparsity patterns and class of matrices; precision estimation in Gaussian graphical models for example. For a joint graphical model estimation approach see \cite{guo2011joint,danaher2014joint}. \par

Fully Bayesian treatments of Gaussian graphical models are, also, well rooted in literature. In particular, \cite{wang_2012_bayes} introduce the Bayesian adaptive graphical lasso (BAGLASSO) which utilises a generalised Pareto distribution in the hierarchical formulation of the Bayesian graphical lasso. \cite{banerjee_2015_bayes} provide a method for graphical model determination by invoking positive prior mass on the event that there is no conditional dependencies between variables. In terms of joint graphical model inference from a Bayesian perspective see \cite{peterson2015bayesian}. Lastly,  \cite{williams2018bayesian} propose using Kullback-Leibler divergence and cross-validation for graphical model structure estimation. \par

\subsection*{Background}\label{subsec:background}{

DN analysis is a statistical methodology that involves functions of at least two graphical models. Numerous measures exist for comparing and evaluating the differences between these graphical structures \cite{Ali_shojaie}. Let $\mathcal{G}=(\mathcal{V}, \mathcal{E})$ define a graphical model with nodes $\mathcal{V}=\{1,2,...,p\}$ and a set of edges $\mathcal{E} \subseteq \mathcal{V} \times \mathcal{V}$. The graph visually depicts the conditional dependence structure between the nodes of the system. For this work, the focus will be on the difference of two Gaussian graphical models, $\mathcal{G}_1$ and $\mathcal{G}_2$ that share the same set of nodes $\mathcal{V}$. In particular, the edge sets given here are equivalent to the adjacency matrices obtained from the Gaussian graphical model estimation. More specifically, assume that the observations, $\mathbf{x}_1, \mathbf{x}_2, ... ,\mathbf{x}_{n_1}$ and $\mathbf{y}_1, \mathbf{y}_2, ... ,\mathbf{y}_{n_2}$ are generated from a $p$ variate Gaussian distribution, $N_p(\boldsymbol{\mu}_1,\mathbf{\Sigma}_1)$ and $N_p(\boldsymbol{\mu}_2,\mathbf{\Sigma}_2)$, respectively. The interest here is estimating the DN ($\mathbf{\Delta}=\mathbf{\Sigma}_2^{-1}-\mathbf{\Sigma}_1^{-1}$), that is the difference between two precision matrices. DN analysis is becoming increasingly popular and important, for example  in biological systems where protein interaction networks can be utilised as informative biosignatures for prevalent diseases \cite{chuang2007network,taylor2009dynamic}. The fundamental idea here is that, if two molecules interact with one another then a statistical dependency between them should be observed. Additionally, another application of DNs is multivariate statistical quadratic discriminant analysis \cite{li2015sparse,jiang2018direct}, under the Gaussian distribution assumption.

Recently, a plethora of statistical techniques have emerged for estimating DNs. These techniques can largely be classified into two main categories. The first estimating the individual precision matrices, $\mathbf{\Theta}_1$ and $\mathbf{\Theta}_2$ separately; where the estimated DN is the difference between the estimated precision matrices. For example, the methods and references for Gaussian graphical model estimation outlined in the introduction can be used to directly to estimate $\mathbf{\Delta}$. The second methodology estimates both the precision matrices simultaneously. The approach here, typically penalises a joint loss function for both precision matrices. \cite{chiquet2011inferring} provide a methodology for inference and estimation of functions of Gaussian graphical models. In particular, the Intertwined Graphical LASSO (IGL) approach biases the estimation of the precision matrices towards a common value. More so, their Graphical Cooperative-LASSO (GCL) utilises a group-penalty for solutions that favour a common sparsity pattern. \cite{guo2011joint} and \cite{zhu2018multiple} estimate separate graphical models using a joint penalised loss function. \cite{zhao2014direct} propose a method for estimating $\mathbf{\Delta}$ directly which relaxes the need for both individual precision matrices to be sparse nor be estimated directly. Similarly, \cite{yuan2017differential} and \cite{jiang2018direct} utilise an alternating direction method of multipliers (ADMM) algorithm for estimating $\mathbf{\Delta}$ from their joint $\ell_1$ penalised convex loss function. More recently, \cite{tang2020fast} introduce a computationally efficient iterative shrinkage-thresholding algorithm for minimising the $\ell_1$ loss function defined in  \cite{jiang2018direct}, namely

 \begin{equation}\label{Jiang_2018_loss}
      L_1(\mathbf{\Delta}) = \frac{1}{2}\mathrm{trace}(\mathbf{\Delta}^{\top} \mathbf{S_1} \mathbf{\Delta} \mathbf{S_2})-\mathrm{trace}(\mathbf{\Delta}(\mathbf{S_1}-\mathbf{S_2})),
 \end{equation}
 
\noindent is convex and $\mathbf{S}_1$ and $\mathbf{S}_2$ are the sample covariance matrices. The DN estimate is obtained by minimising the penalised loss Eq \eqref{Jiang_2018_loss}. An analogous symmetric convex loss function and estimator is proposed by \cite{yuan2017differential}.
 






 The shrinkage-thresholding algorithm proposed by \cite{tang2020fast}, based on the fast-iterative shrinkage-thresholding algorithm in \cite{beck2009fast}, aims to minimise  Eq \eqref{Jiang_2018_loss} without computing matrix inverses in the estimation process. The objective function is given by
 
 \begin{equation*}
        \argmin_{\mathbf{\Delta} \in \mathbb{R}^{p \times p}}  
       L_1(\mathbf{\Delta}) + \lambda \Arrowvert \mathbf{\Delta} \Arrowvert_1,
 \end{equation*}


\noindent where $\Arrowvert \mathbf{\Delta} \Arrowvert_1=\sum_{i<j}\sum_{i=1}^{p}|\hat{\Theta}_{2:ij}-\hat{\Theta}_{1:ij}|$. The optimisation objective converges to the solution sequentially using a quadratic approximation and a gradient descent algorithm. The efficiency of the procedure is attested to this approach, resulting in superior computational complexity in contrast to the ADMM approaches by  \cite{yuan2017differential} and \cite{jiang2018direct}.
}

\subsection*{Contributions}{
In this paper, the objective is to develop a framework for Bayesian DN estimation, which remains unexplored. In particular, the BAGLASSO  is adapted for the Bayesian DN estimation; noting that frequent references thereto are included throughout the development of the novel DN architecture.  The block Gibbs sampler is used for estimating each component of the DN. An adjusted edge inclusion threshold, based on a conjugate Wishart prior, for graphical structure learning is also exhibited. Comparisons in synthetic data studies illustrate that the Bayesian DN is proficient in both graphical structure learning and matrix estimation, when compared to current state of the art methods. \par

Finally, it is worth noting the main contributions of this paper. First, the  novel Bayesian approach to estimating DNs, using the Bayesian adaptive graphical lasso (BAGLASSO), is introduced. A threshold selection strategy for graphical structure determination, based on a conjugate Wishart prior is explored. An application to South African COVID-19 data is investigated to examine the change in DN structure of key daily metrics between various phases of the pandemic. Lastly, an R package has been developed for the BAGLASSO block Gibbs sampler. The Markov Chain Monte Carlo (MCMC) sampler simulates precision matrices from the posterior distribution of the BAGLASSO. The R package is available on The Comprehensive R Archive Network (CRAN)  \href{https://cran.r-project.org/web/packages/abglasso/index.html}{abglasso}.\par
}


\section*{The Bayesian DN}\label{sec:adBN}

A fully Bayesian treatment of DNs remains unexplored and the novel methodology here aims to develop a simple yet highly accurate Bayesian DN estimation procedure. The novel contribution utilises the BAGLASSO as a launching point to separately estimate the components of the DN. Moreover, the framework has been develop for low to moderate, $p \in \{10-100\}$, dimensions where $n \geq p$. \par
 
\subsection*{The Bayesian graphical lasso prior}\label{subsec:bayes_graphical_lasso}

Recall that the graphical lasso objective is maximising the penalized log-likelihood

\begin{equation} \label{graphical_lasso}
   \argmax_{\mathbf{\Theta}\in \mathbb{M}^{+}} 
    \log(\mathrm{det}\mathbf{\Theta})-\mathrm{trace}(\frac{\mathbf{S}}{n}\mathbf{\Theta}) - \rho\Arrowvert \mathbf{\Theta} \Arrowvert_1
\end{equation}

\noindent where $M^+$ is the space of positive definite matrices and $\mathbf{S}$ is the sample covariance matrix. More over, $\rho \geq 0$ is the shrinkage parameter and $\mathbf{\Theta}=(\theta_{ij})$ is the precision matrix. The Bayesian connection to the graphical lasso problem is the maximum a posteriori (MAP) estimate, assuming a random sample from $N_p(\boldsymbol{\mu},\mathbf{\Theta}^{-1})$, of the following


\begin{equation} \label{bayes_glasso_prior}
    p\left(\mathbf{\Theta}\:|\lambda\right)=C^{-1}\prod_{i<j}\bigg\{\mathrm{DE}(\theta_{ij}\:|\:\lambda)\bigg\}
    \prod_{i=1}^{p}\bigg\{\mathrm{EXP}(\theta_{ii}\:|\:\lambda)\bigg\}\:\:\:\:(\mathbf{\Theta}\in \mathbb{M}^{+}).
\end{equation}



\noindent The prior distribution is given by the product of a double exponential (DE) with form $p(y)=\lambda/2\exp(-\lambda|y|)$ for the off diagonal elements and an exponential (EXP) with form $p(y)=\lambda \exp(-\lambda y)1_{y>0}$, otherwise. The value of $\mathbf{\Theta}$ which maximizes the posterior density is the graphical lasso estimate when $\rho=\lambda/n$. 


\subsection*{Hierarchical representation}\label{subsubsec:hierarchical_rep}
\cite{wang_2012_bayes} propose a hierarchical representation of the graphical lasso prior Eq \eqref{bayes_glasso_prior}, using the Bayesian lasso formulation in \cite{park2008bayesian}. The Gibbs sampler in \cite{park2008bayesian} utilises the structure of the double exponential distribution as a scale mixture of Gaussians (assuming independence of the conditional double exponential priors) (\cite{andrews1974scale}; \cite{west1987scale}) to simulate from the desired posterior distribution. The positive definite constraint on the precision matrix in the graphical lasso Eq \eqref{bayes_glasso_prior} implies that the Gaussian components for $\theta_{ij}$ in the scale mixture formulation are no longer independent given the scale parameters. To address this issue, the hierarchical representation of the graphical lasso prior is given by

\begin{equation}\label{wang_hier_glasso_prior_1}
    p(\mathbf{\theta}\:|\:\mathbf{\tau},\lambda)=C_{\mathbf{\tau}}^{-1}
    \prod_{i<j}\bigg\{\frac{1}{\sqrt{2\pi\tau_{ij}}}\exp(-\frac{\theta_{ij}^2}{2\tau_{ij}})\bigg\}
    \prod_{i=1}^{p}\bigg\{\frac{\lambda}{2}\exp(-\frac{\lambda}{2}\theta_{ii})\bigg\}
    \:\:\:\:(\mathbf{\Theta}\in \mathbb{M}^{+}),
\end{equation}


\noindent where $\boldsymbol{\theta} = \{\theta_{ij}\}_{i\leq j}$ is a vector of the upper triangular matrix entries of $\mathbf{\Theta}$ and $\boldsymbol{\tau}=\{\tau_{ij}\}_{i<j}$ the scale parameters. The normalising constant has no closed-form solution. Obtaining the marginal distribution Eq \eqref{bayes_glasso_prior}, \cite{wang_2012_bayes} propose an exponential mixing density with rate parameter $\lambda^2/2$. Simple substitution yields that the mixing density circumvents the intractable normalising constant. Finally, the hierarchical representation in Eq \eqref{wang_hier_glasso_prior_1} is used in the development of the data-augmented block Gibbs sampler, available in the supplementary material, with a target distribution given by 





\begin{equation} \label{target_dist}
    \begin{aligned}
        p(\mathbf{\Theta},\mathbf{\tau}\:|\:\mathbf{Y},\lambda)
        \propto
        \mathrm{det}\mathbf{\Theta}^{\frac{n}{2}}
        \exp\{-\mathrm{trace}(\frac{1}{2}\mathbf{S}\mathbf{\Theta})\}
        \prod_{i<j}\bigg\{&\tau_{ij}^{-\frac{1}{2}}\exp(-\frac{\theta_{ij}^2}{2\tau_{ij}})\exp(-\frac{\lambda^2}{2}\tau_{ij})\bigg\}\\
        & \times 
        \prod_{i=1}^{p}\bigg\{\exp(-\frac{\lambda}{2}\theta_{ii})\bigg\} \:\:\:\:(\mathbf{\Theta}\in \mathbb{M}^{+}).
    \end{aligned}
\end{equation}

\subsection*{BAGLASSO}\label{subsubsec:bayes_adap_graph_lasso}


It is well known that the double exponential prior in Eq \eqref{bayes_glasso_prior} may over-shrink (under-shrink) large (small) coefficients in $\mathbf{\Theta}$. The limitations within a regression context have been studied in (\cite{li2010bayesian}; \cite{griffin2010inference}; \cite{carvalho2010horseshoe}) with alternative proposals. The BAGLASSO, Bayesian analog to the adaptive graphical lasso, exploit the framework and flexibility of the hierarchical representation in Eq \eqref{wang_hier_glasso_prior_1} to address the aforementioned limitation

\begin{equation}
    \log(\mathrm{det} \mathbf{\Theta})-\mathrm{trace}(\frac{\mathbf{S}}{n}\mathbf{\Theta}) -\lambda \sum_{1\leq i\leq p} \sum_{1\leq j\leq p}\xi_{ij}|\theta_{ij}|,
\end{equation}

\noindent where $\xi_{ij}=1/|\Tilde{\theta}_{ij}|$ are the adaptive weights and the weight matrix ($\Tilde{\theta}_{ij}$) is the sample precision matrix. \par

The Bayesian graphical lasso Eq \eqref{bayes_glasso_prior} enables the selection of an appropriate hyperprior on the shrinkage parameter $\lambda$, recall that $\rho = \lambda/n$ in the Bayesian formulation of Eq \eqref{graphical_lasso}. Adhering to the positive definite constraint on $\mathbf{\Theta}$, the normalising constant in Eq \eqref{bayes_glasso_prior} for a single prior $\lambda$ for all elements in $\mathbf{\Theta}$ can be obtained by applying the substitution $\mathbf{\Tilde{\Theta}}=\lambda\mathbf{\Theta}$. Thereafter, a diffuse gamma prior $\lambda \sim \mathrm{GA}(r,s)$ and corresponding conditional posterior  $\lambda \sim \mathrm{GA}(r + p(p+1),s+\Arrowvert \mathbf{\Theta} \Arrowvert_1 / 2)$ can be obtained and sampled from. When allowing for individual $\lambda_{ij}$'s for different $\theta_{ij}$'s, the normalising constant $C$ will inevitably depend on $\lambda_{ij}$ and a hierarchical formulation can be used to construct a set of prior distributions for various $\lambda_{ij}$. In particular, assuming a random sample from $N_p(\boldsymbol{\mu},\mathbf{\Theta}^{-1})$, 


\begin{equation*}
    p(\mathbf{\Theta} \:|\: \{\lambda_{ij}\}_{i\leq j}) = 
    C^{-1}_{\{\lambda_{ij}\}_{i\leq j}}
    \prod_{i<j}\bigg\{\mathrm{DE}(\theta_{ij}\:|\:\lambda_{ij})\bigg\}
    \prod_{i=1}^{p}\bigg\{\mathrm{EXP}(\theta_{ii}\:|\:\frac{\lambda_{ii}}{2})\bigg\}\:\:\:\:(\mathbf{\Theta}\in \mathbb{M}^{+}),
\end{equation*}

\begin{equation} \label{unique_lambda_hier_glasso}
    p(\{\lambda_{ij}\}_{i<j}\:|\:\{\lambda_{ii}\}_{i=1}^p) \propto
    C_{\{\lambda_{ij}\}_{i\leq j}}
    \prod_{i<j}\mathrm{GA}(r,s).
\end{equation}

\noindent The normalising constant $C_{\{\lambda_{ij}\}_{i\leq j}}$ is intractable and the set $\{\lambda_{ii}\}_{i=1}^p$ are hyperparameters for the diagonal elements of $\mathbf{\Theta}$. As it happens, the computation of $\lambda_{ij}$ is simplified by circumventing the intractable normalising constant. \par

The BAGLASSO selects the amount of shrinkage $\lambda_{ij}$ proportionally to the current value of $\theta_{ij}$. To see this, \cite{wang_2012_bayes} demonstrate that the conditional posterior, $\lambda_{ij}\:|\:\mathbf{\Theta} \sim \mathrm{GA}(r+1,|\theta_{ij}|+s)$, has an expected value that is inversely related to magnitude of $\theta_{ij}$. The data augmented block Gibbs sampler for the hierarchical representation in Eq \eqref{unique_lambda_hier_glasso} is the fundamental building block upon which the novel Bayesian DN is devised.


\subsection*{Technicalities on conditional dependencies}\label{subsubsec:technicalties}
A conjugate Wishart prior may be used to infer on events such as $\theta_{ij}=0$ when a purely Bayesian posterior inference regarding network structure is desired. An alternative thresholding strategy is presented which is an adaption of the recommendation by \cite{carvalho2010horseshoe}. In particular the conjugate Wishart $\mathrm{W}(3,\epsilon\mathbf{I}_p)$ prior is used. The corresponding posterior is   $\mathrm{W}(3+n,(\mathbf{S}+\epsilon \mathbf{I}_p)^{-1})$, where $\epsilon = 0.001$ and $\mathbf{I}_p$ a $p$ dimensional identity matrix. The posterior samples are used to compute the posterior distribution of the $p \times p$ partial correlation matrix. The recommended strategy here suggests $\{\theta_{ij}\neq 0\}$ if

\begin{equation} \label{threshold_rule_smith}
    |E_h(\rho_{ij}\:|\:\mathbf{Y})|>\eta.
\end{equation}\par

\noindent The Bayesian posterior thresholding recommendation by \cite{wang_2012_bayes} claim that $\{\theta_{ij}\neq 0\}$ if and only if

\begin{equation} \label{threshold_rule_wang}
   \frac{\Tilde{\rho}_{ij}}{E_g(\rho_{ij}\:|\:\mathbf{Y})} >\eta.
\end{equation}

\noindent Noting that $\Tilde{\rho}_{ij}$ is the posterior sample mean estimate of the partial correlation under graphical lasso priors Eq \eqref{bayes_glasso_prior}; $g$ is the standard conjugate Wishart $\mathrm{W}(3,\mathbf{I}_p)$ and $h$ the standard conjugate Wishart $\mathrm{W}(3,\epsilon\mathbf{I}_p)$. Moreover, $\eta  \in \{0-1\}$ with the lower and upper bounds resulting in a completely dense or sparse estimate, respectively.\par

\begin{figure}[hb]
\begin{adjustwidth}{-2cm}{0in}
  \centering
  \begin{tabular}[c]{ccc}
  \centering
    \begin{subfigure}[c]{0.35\textwidth}
      \includegraphics[width=\textwidth]{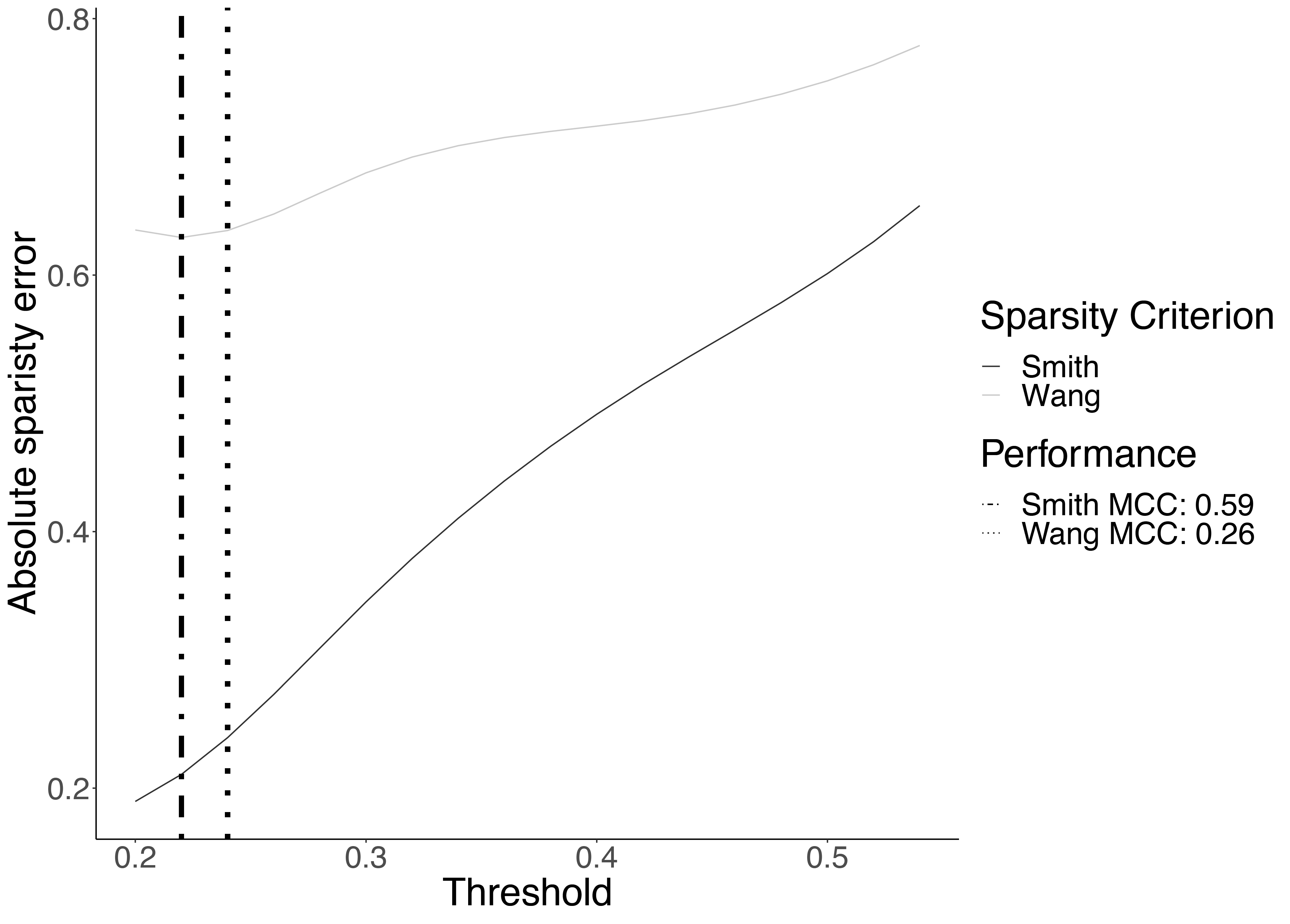}
         \caption{Model 1: AR(1).}
         \label{fig:thresh_study_p10_mod_1}
    \end{subfigure}&
    \begin{subfigure}[c]{0.35\textwidth}
      \includegraphics[width=\textwidth]{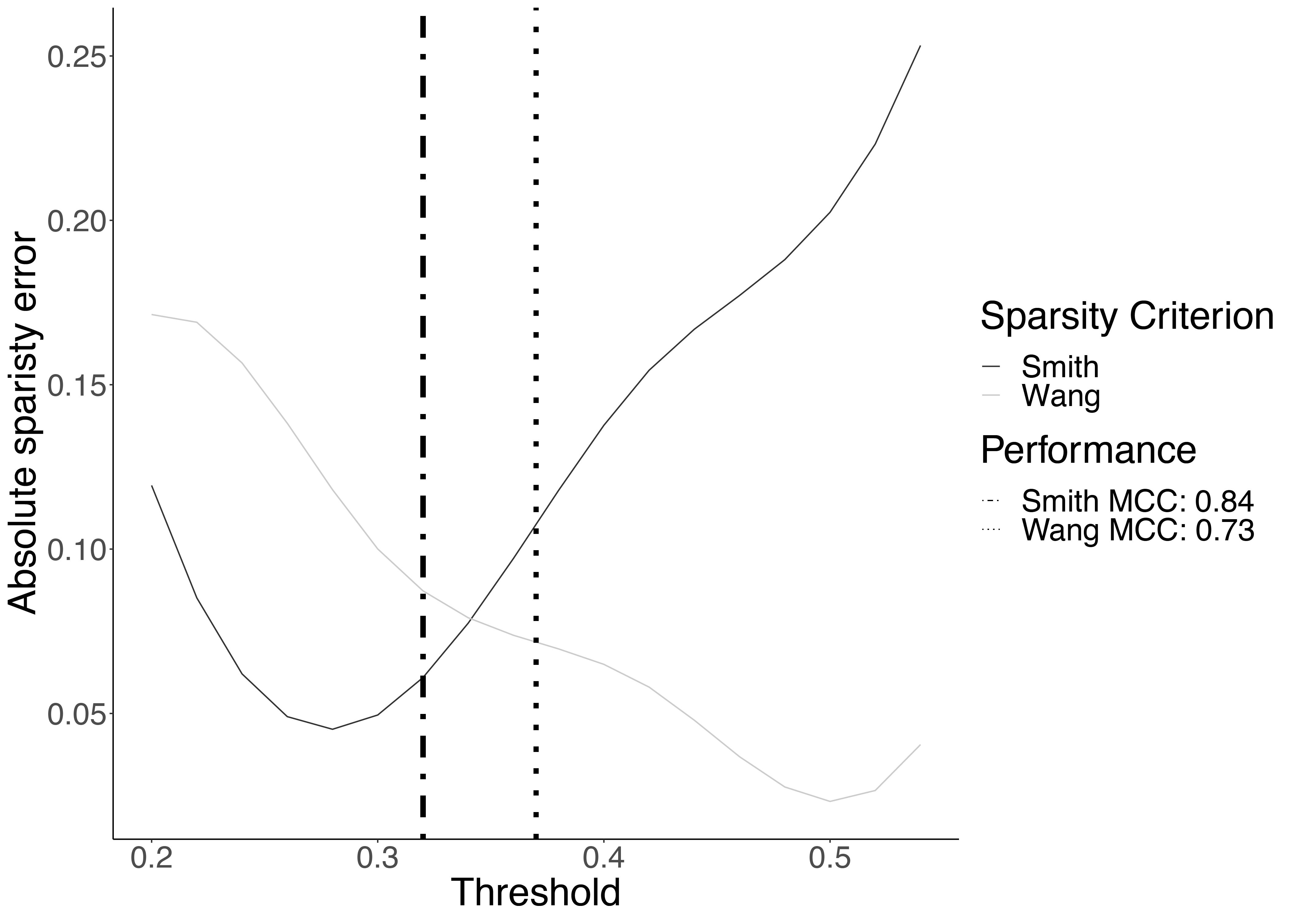}
         \caption{Model 2: AR(2).}
         \label{fig:thresh_study_p10_mod_2}
    \end{subfigure}&
    \begin{subfigure}[c]{0.35\textwidth}
      \includegraphics[width=\textwidth]{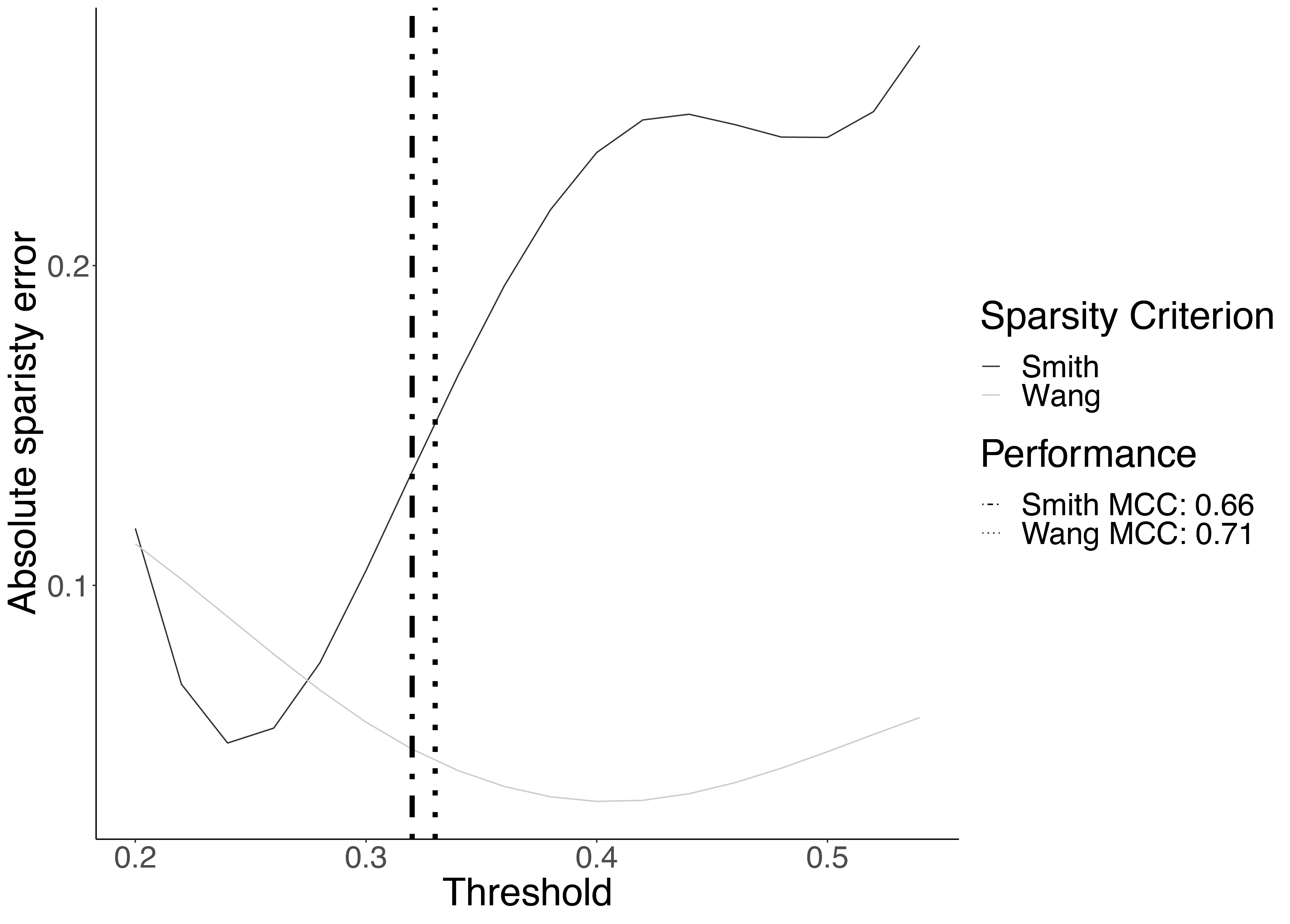}
         \caption{Model 3: at most $80\%$ sparse.}
         \label{fig:thresh_study_p10_mod_3}
    \end{subfigure}\\
    \begin{subfigure}[c]{0.35\textwidth}
      \includegraphics[width=\textwidth]{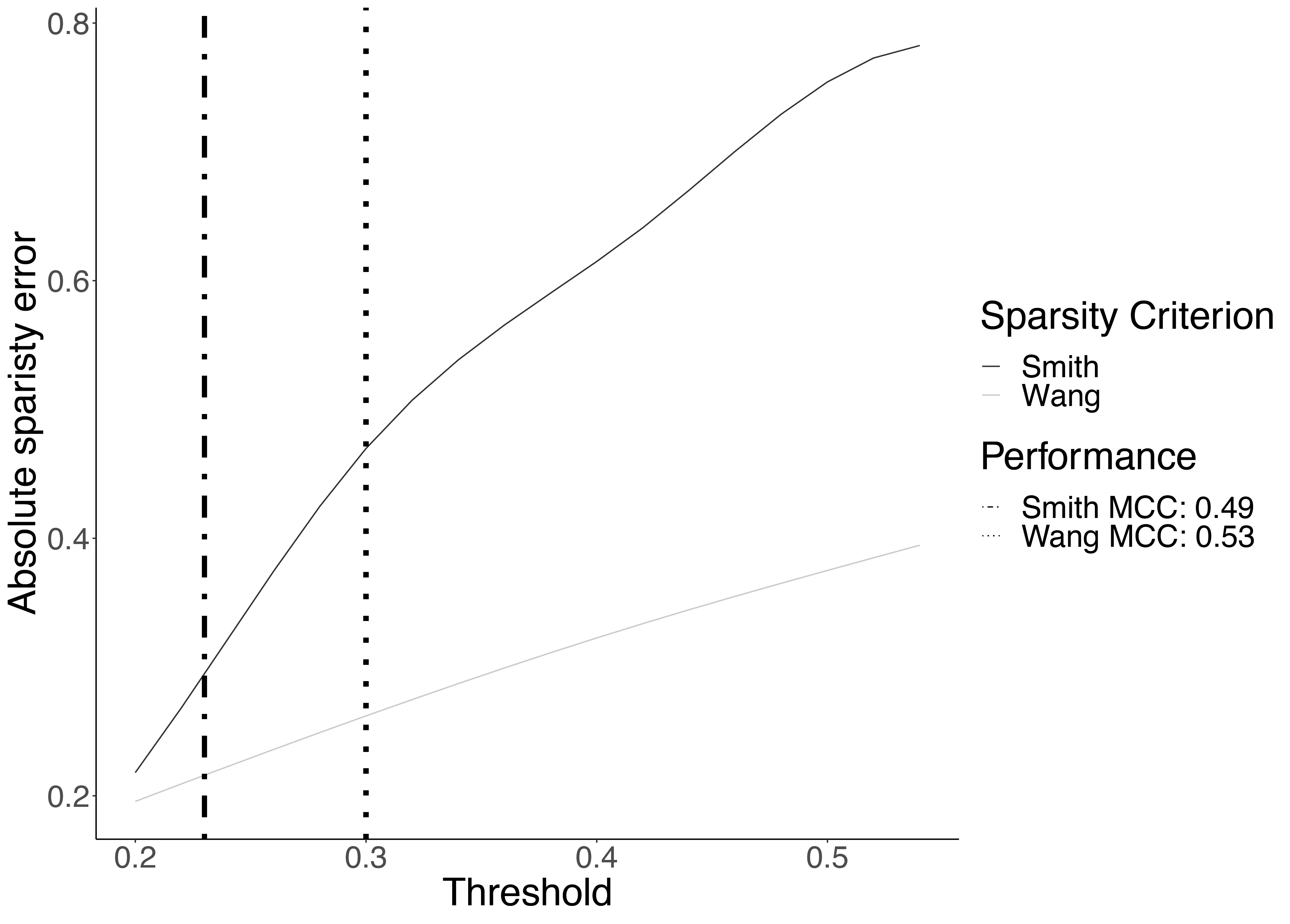}
        \caption{Model 4: at most $40\%$ sparse.}
         \label{fig:thresh_study_p10_mod_4}
    \end{subfigure}&
    \begin{subfigure}[c]{0.35\textwidth}
      \includegraphics[width=\textwidth]{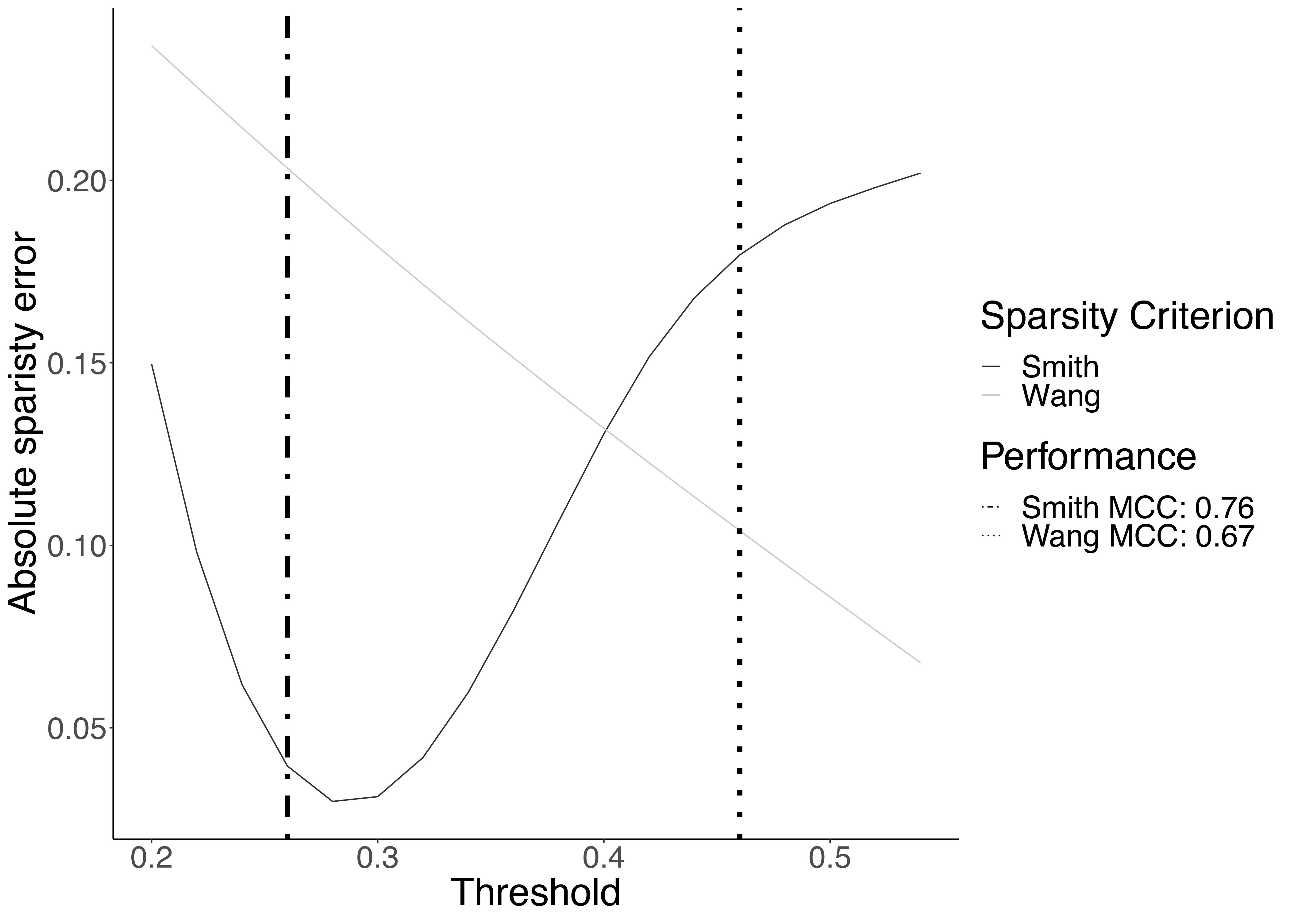}
        \caption{Model 5: scale-free.}
         \label{fig:thresh_study_p10_mod_5}
    \end{subfigure}&
    \begin{subfigure}[c]{0.35\textwidth}
      \includegraphics[width=\textwidth]{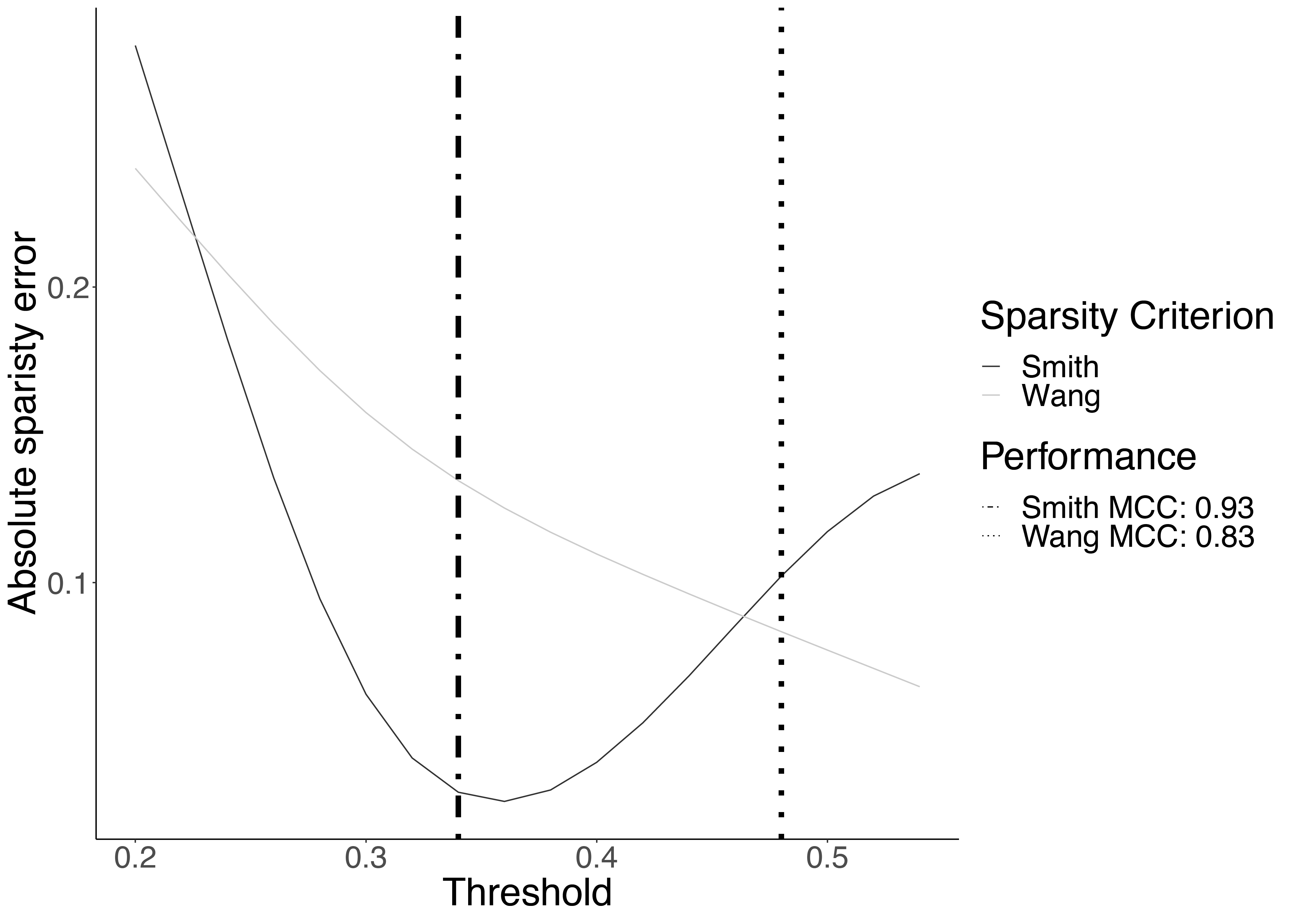}
        \caption{Model 6: band.}
         \label{fig:thresh_study_p10_mod_6}
    \end{subfigure}\\
    \begin{subfigure}[c]{0.35\textwidth}
      \includegraphics[width=\textwidth]{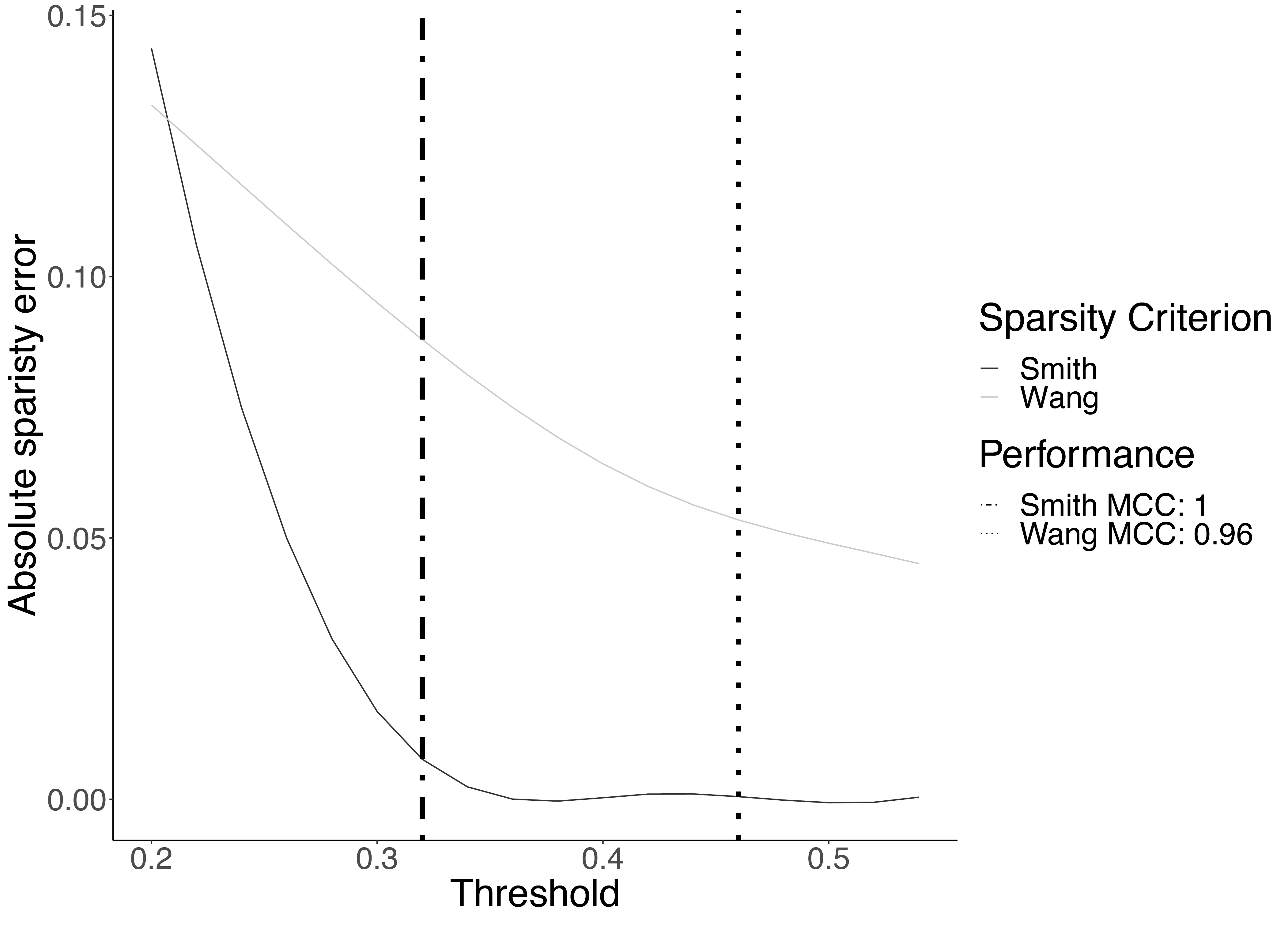}
         \caption{Model 7: cluster.}
         \label{fig:thresh_study_p10_mod_7}
    \end{subfigure}&
    \begin{subfigure}[c]{0.35\textwidth}
      \includegraphics[width=\textwidth]{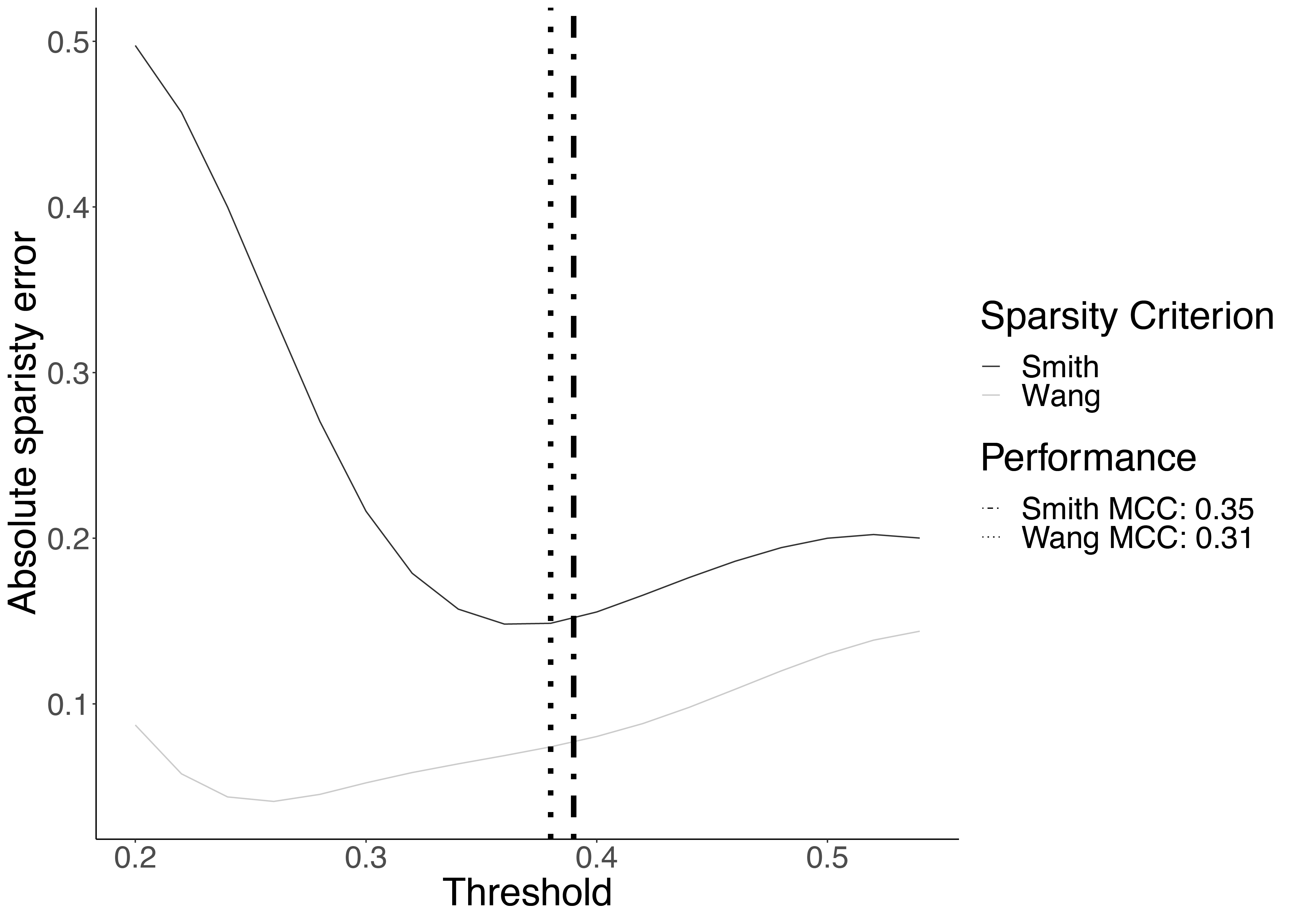}
         \caption{Model 8: star.}
         \label{fig:thresh_study_p10_mod_8}
    \end{subfigure}
    &
    \begin{subfigure}[c]{0.35\textwidth}
     \includegraphics[width=\textwidth]{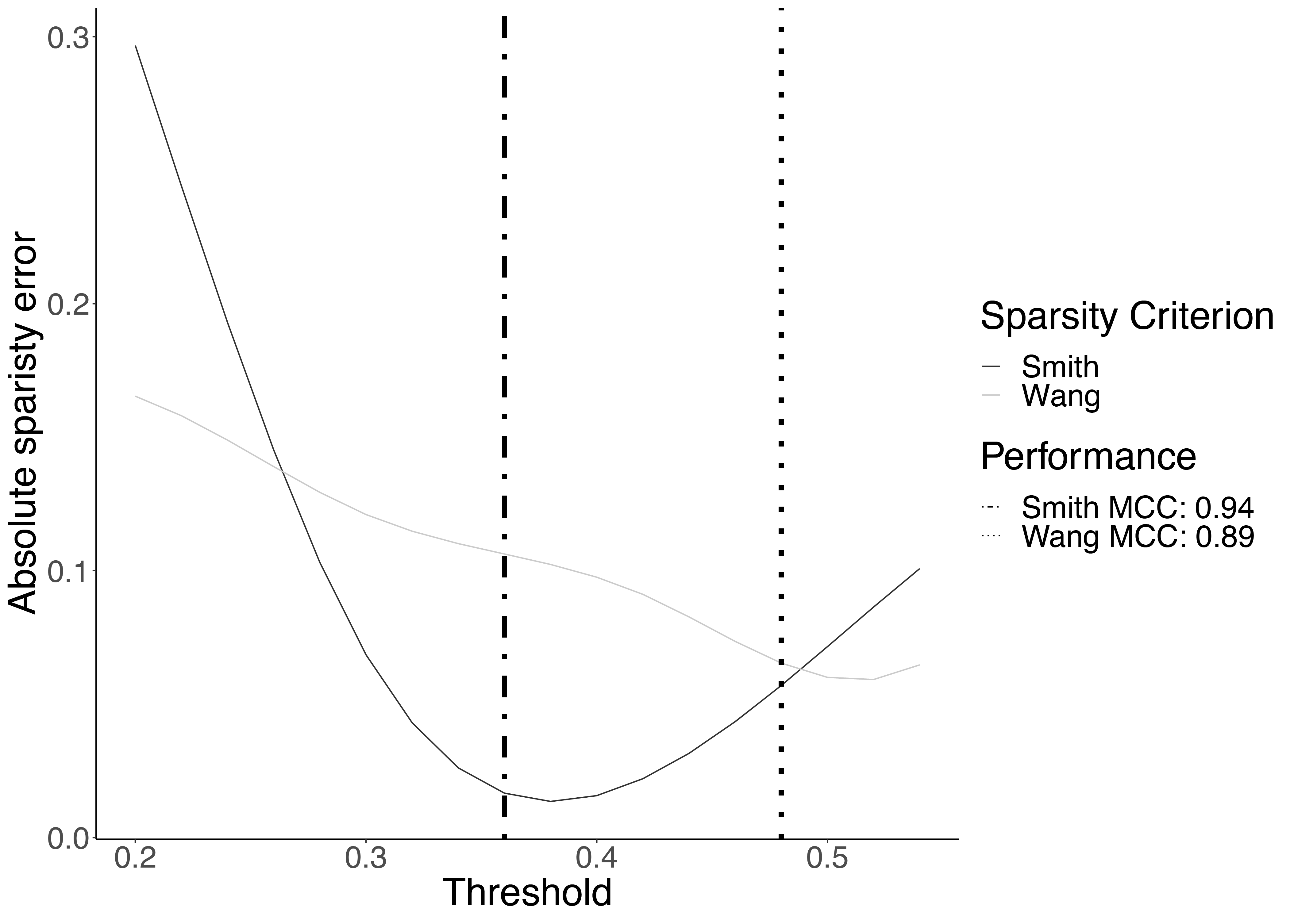}
         \caption{Model 9: circle.}
         \label{fig:thresh_study_p10_mod_9}
    \end{subfigure}
   \\
  \end{tabular} 
    \caption{
    \bf The median of the absolute sparsity error and best performing MCC for various graph structures under varying thresholds for each Bayesian sparsity criterion in Eq \eqref{threshold_rule_wang} (dotted) and Eq \eqref{threshold_rule_smith} (dot-dash) for dimension $p=10$. The best performing threshold is indicated by a vertical line with the accompanying MCC value displayed in the legend.}
    \label{fig:smooth_wang_thresh_comparison_p10}
\end{adjustwidth}       
\end{figure}

The original recommendation for $\eta$ in Eq \eqref{threshold_rule_wang} is $0.5$. The forthcoming synthetic data analysis section describes the simulation procedure, as well as, illustrates the performance of the Bayesian DN with regards to different graph structures, namely an AR($1$), AR($2$), sparse random, scale-free, band, cluster, star and circle. The goal here, however, is to suggest a suitable sparsity threshold region under the varying graph structures. The Bayesian DN is applied across all graph structures with thresholds, $\eta$,  in the range of $0.2$ and $0.6$ in increments of $0.02$. The absolute sparsity error is computed for each graph structure and threshold candidate for both Bayesian sparsity criterion in Eq \eqref{threshold_rule_wang} and Eq \eqref{threshold_rule_smith}. The results are based on the median of $40$ replications and the Matthews Correlation Coefficient (MCC), see \cite{matthews1975comparison}, is used to determine the best performing threshold. Figure \ref{fig:smooth_wang_thresh_comparison_p10} (\ref{fig:thresh_study_p10_mod_1}) - (\ref{fig:thresh_study_p10_mod_9}) display the optimal threshold, based on the top performing MCC, for each graph structure and Bayesian sparsity criterion for $p=10$. The optimal threshold plots for $p=30$ and $p=100$ are available in the supplementary material. The optimal threshold based on Eq \eqref{threshold_rule_smith} , $\eta^*$, for the Bayesian DN is, in most cases, in the neighborhood of the minimum absolute sparsity error and in the region of $\eta^*  \in \{0.2-0.4\}$. Both Bayesian sparsity criterion candidates perform comparably well noting, however, that Eq \eqref{threshold_rule_smith} requires less computation.\par

\section*{Synthetic data analysis}\label{sec:syn_data_analysis}
The synthetic experiment is designed to test the parameter estimation and graphical structure determination of the DN estimation for both the novel Bayesian approach (referred to as `B-net') and the iterative shrinkage-thresholding estimator (referred to as `D-net') from \cite{tang2020fast}. The iterative shrinkage-thresholding estimator uses the lasso penalty and Bayesian Information Criterion (BIC) for model estimation and selection, respectively. For all simulations, the assumption is that the observations, $\mathbf{x}_1, \mathbf{x}_2, ... ,\mathbf{x}_{n_1}$ and $\mathbf{y}_1, \mathbf{y}_2, ... ,\mathbf{y}_{n_2}$ are generated from a Gaussian $N_p(0,\mathbf{\Sigma}_{1})$ and $N_p(0,\mathbf{\Sigma}_{2})$ respectively. The true DN is

\begin{equation*}
    \mathbf{\Delta}=\mathbf{\Sigma}_{2}^{-1} - \mathbf{\Sigma}_{1}^{-1},
\end{equation*}

\noindent where the true precision matrices are $\mathbf{\Theta}_1=\mathbf{\Sigma}_{1}^{-1}$ and $\mathbf{\Theta}_2=\mathbf{\Sigma}_{2}^{-1}$. The Bayesian DN applies the BAGLASSO Eq \eqref{unique_lambda_hier_glasso} to each sample, i.e. separately estimates the precision matrices. Furthermore, for excellent performance set $r = 10^{−2}$ and $s = 10^{−6}$, see supplementary material for more details, for the hyperparameters of the prior distributions of $\lambda_{ij}$ for $i < j$ and $\lambda_{ii} = 1$ for $i = 1,\ldots,p$. The iterative shrinkage-thresholding approach jointly estimates the precision matrices for Eq \eqref{Jiang_2018_loss}. The following $9$ graphical structure variations are considered - where the structure of each is applied to each component in the DN's composition to achieve the desired structure in the DN itself - in the simulation:

\begin{itemize}
  \item \emph{structure 1}: An AR($1$) model.
    \begin{itemize}
     \item \emph{Component 1}: $\theta_{ij}=0.7^{|i-j|}$.
     \item \emph{Component 2}: $\theta_{ij}=0.75^{|i-j|}$.
   \end{itemize}
  \item \emph{structure 2}: An AR($2$) model.
   \begin{itemize}
     \item \emph{Component 1}: $\theta_{ii}=0.1, \theta_{i,i-1}= \theta_{i-1,i}= 0.05$ and $\theta_{i,i-2}= \theta_{i-2,i} = 0.025$.
     \item \emph{Component 2}: $\theta_{ii}=1, \theta_{i,i-1}= \theta_{i-1,i}= 0.5$ and $\theta_{i,i-2}= \theta_{i-2,i} = 0.25$.
   \end{itemize}
  \item \emph{structure 3}: A sparse random model where both components have approximately up to $80\%$ off-diagonal elements set to zero.
  \item \emph{structure 4}: A moderately sparse random model where both components have approximately up to $40\%$ off-diagonal elements set to zero.
  \item \emph{structure 5}: A scale-free model where the second component is a scalar multiple of the first.
  \item \emph{structure 6}: A band or diagonal model.
   \begin{itemize}
     \item \emph{Component 1}: $\theta_{ii}=1,\theta_{ij}=0.2$ for $1 \leq i \neq j \leq p/2$, $\theta_{ij}=0.5$ for $p/2+1 \leq i \neq j \leq p$ and $\theta_{ij}=0$ otherwise. 
     \item \emph{Component 2}: $\theta_{ii}=1,\theta_{ij}=0.7$ for $1 \leq i \neq j \leq p/2$, $\theta_{ij}=0.9$ for $p/2+1 \leq i \neq j \leq p$ and $\theta_{ij}=0$ otherwise.
   \end{itemize}
  \item \emph{structure 7}: A cluster model containing two disjoint groups.
   \begin{itemize}
     \item \emph{Component 1}: $\theta_{ii}=1,\theta_{ij}=0.5$ for $1 \leq i \neq j \leq p/2$, $\theta_{ij}=0.5$ for $p/2+1 \leq i \neq j \leq p$ and $\theta_{ij}=0$ otherwise. 
     \item \emph{Component 2}: $\theta_{ii}=1,\theta_{ij}=0.9$ for $1 \leq i \neq j \leq p/2$, $\theta_{ij}=0.9$ for $p/2+1 \leq i \neq j \leq p$ and $\theta_{ij}=0$ otherwise.
   \end{itemize}
  \item \emph{structure 8}: A star model with every node connected to the first node.
   \begin{itemize}
     \item \emph{Component 1}: $\theta_{ii}=1, \theta_{1,i}= \theta_{i,1}= 0.1$ and $\theta_{i,j}= 0.$ otherwise.
     \item \emph{Component 2}: $\theta_{ii}=1, \theta_{1,i}= \theta_{i,1}= 2.1$ and $\theta_{i,j}= 0.$ otherwise.
   \end{itemize}
  \item \emph{structure 9}: A circular model.
   \begin{itemize}
     \item \emph{Component 1}: $\theta_{ii}=2, \theta_{i,i-1}= \theta_{i-1,i}= 1$ and $\theta_{1,p}= \theta_{p,1} = 0.45$.
     \item \emph{Component 2}: $\theta_{ii}=4, \theta_{i,i-1}= \theta_{i-1,i}= 2$ and $\theta_{1,p}= \theta_{p,1} = 0.95$.
   \end{itemize}
\end{itemize}
The sample sizes and dimensions for each model are $n_1=n_2\in\{50,100,200\}$ and $p_1=p_2 \in\{10,30,100\}$, respectively. The Bayesian estimates are based on $10000$ Monte Carlo iterations after $5000$ burn-in iterations. 
To assess the performance of DN matrix estimation, six loss functions are considered and defined in Table \ref{tab:syn_loss_functions}, where $p$ denotes the dimension and $\gamma_i$ the $i^{th}$ eigenvalue, respectively. Notice that some loss functions utilise the true DN matrix and its estimates, while others utilise the eigenvalues and their respective estimates. Table \ref{tab:syn_loss_results} reports the median of L1, L2, EL1, EL2, MAXEL1 and MINEL1 for $p = 10,\:30,\:100$ in structures $1-9$ based on $40$ replications. For each scenario, the best performing measure is boldfaced.
\begin{table}[h]
\begin{adjustwidth}{-2cm}{0in}
\caption{
\bf Loss functions used in the synthetic data analysis to assess the numerical accuracy of the B-net and D-net estimates.}
\centering
\begin{tabular}{l|l|l}
\hline                              
\bf Measure & \bf Loss function & \bf Abbreviation  \\
\thickhline
Matrix $L_1$-norm & $\Arrowvert \hat{\mathbf{\Delta}} -  \mathbf{\Delta}\Arrowvert_1 = max_{1 \leq j \leq p}\sum_{i=1}^p|\hat{\Delta}_{ij}-\Delta_{ij}|$              & L1              \\

Frobenius loss &  $\Arrowvert \hat{\mathbf{\Delta}} -  \mathbf{\Delta}\Arrowvert_F$, where $\Arrowvert A\Arrowvert_F^2=trace(AA^{\top})$ &  L2             \\

$L_1$ eigenvalue loss                 & $\sum_{i=1}^p|\hat{\gamma}_i-\gamma_i|/p$              &   EL1             \\

$L_2$ eigenvalue loss                  & $\sum_{i=1}^p(\hat{\gamma}_i-\gamma_i)^2/p$              & EL2              \\

$L_1$ loss on the largest eigenvalue  & $|\hat{\gamma}_{max}-\gamma_{max}|$              & MAXEL1              \\

$L_1$ loss on the smallest eigenvalue & $|\hat{\gamma}_{min}-\gamma_{min}|$              &  MINEL1          \\    
\hline
\end{tabular}

\label{tab:syn_loss_functions}
\end{adjustwidth}
\end{table}


 The eigenvalue based loss functions are designed to investigate the extremes of the eigenvalue spectrum. In particular, the MAXEL1 loss function highlights which estimator is favourable in a principal component setting, \cite{banerjee2016regularized}. A couple of observations are worth noting from Table \ref{tab:syn_loss_results} and Table \ref{tab:syn_sdlos_results}. First, the D-net estimator performs better with the AR($1$) structure. Second, the B-net estimator performs exceptionally well in remaining structures. Third, the standard errors for both DN estimation techniques remain relatively consistent throughout the dimension spectrum considered, noting that the D-net estimator yields, in general, better results. This may be due to the fact that the best performing tuning parameter in the solution path leads to highly sparse estimates. The B-net estimation procedure inherits the utilisation of multiple penalty parameters in the precision matrix estimation, leading to robust estimation of the precision matrices. \par
 
To assess the performance on graphical structure determination, the specificity, sensitivity, false negative rate, f1 score and the MCCs are computed and defined in Table \ref{tab:performance_measures}. Noting that, TP, TN, FP and FN denote the number of true positives, true negatives, false positives and false negatives, respectively. Values of specificity, sensitivity, f1-score and MCC closer to one, imply better classification performance. The closer the values of false negative rate are to zero the better. The sparsity for the B-net estimator is determined by the thresholding rule in Eq \eqref{threshold_rule_smith} and Figures (\ref{fig:thresh_study_p10_mod_1}) - (\ref{fig:thresh_study_p10_mod_9}). Similarly, the best performing tuning parameter in the solution path of the D-net algorithm determines the sparsity of the estimator.  The median performance scores, based on $40$ repetitions, for each graphical structure is presented in Table \ref{tab:syn_perf_results}. The main diagonals of the adjacency matrices were not included in the scoring. \par

\begin{sidewaystable}[!ht]
\caption{
\bf Summary of L1, L2, EL1, EL2, MAXEL1 and MINEL1 for an AR($1$), AR($2$), sparse random, scale-free, band, cluster, star and circle graphical model. The median loss values reported here are based on 40 replications for both the B-net and D-net estimators. The best performing values are boldfaced.}

\scriptsize
\begin{tabular}{lllllllllllllllllllll} 
\hline
\multicolumn{1}{c}{} & \multicolumn{2}{c}{\bf AR(1)}                      & \multicolumn{2}{c}{\bf AR(2)}                      & \multicolumn{2}{c}{\bf S80}                        & \multicolumn{2}{c}{\bf S40}                        & \multicolumn{2}{c}{\bf SF}    & \multicolumn{2}{c}{\bf Band}                                               & \multicolumn{2}{c}{\bf Cluster}                      & \multicolumn{2}{c}{\bf Star}                       & \multicolumn{2}{c}{\bf Circle}                      \\ 
\cmidrule(lr){2-3}\cmidrule(lr){4-5}\cmidrule(lr){6-7}\cmidrule(lr){8-9}\cmidrule(lr){10-11}\cmidrule(lr){12-13}\cmidrule(lr){14-15}\cmidrule(lr){16-17}\cmidrule(lr){18-19}
\multicolumn{1}{c}{} & \multicolumn{1}{c}{B-net} & \multicolumn{1}{c}{D-net} & \multicolumn{1}{c}{B-net} & \multicolumn{1}{c}{D-net} & \multicolumn{1}{c}{B-net} & \multicolumn{1}{c}{D-net} & \multicolumn{1}{c}{B-net} & \multicolumn{1}{c}{D-net} & \multicolumn{1}{c}{B-net} & \multicolumn{1}{c}{D-net} & \multicolumn{1}{c}{B-net} & \multicolumn{1}{c}{D-net} & \multicolumn{1}{c}{B-net} & \multicolumn{1}{c}{D-net} & \multicolumn{1}{c}{B-net} & \multicolumn{1}{c}{D-net} & \multicolumn{1}{c}{B-net} & \multicolumn{1}{c}{D-net}  \\ 
\hline
\multicolumn{21}{c}{p=10} 
\\
\\
L1      & 1.04 & \textbf{0.64}    & \textbf{1.13} & 1.49    & \textbf{2.26} & 3.41   & \textbf{6.03} & 7.56   & \textbf{2.24} & 2.79   & \textbf{1} & 1.21     & \textbf{0.85} & 1.68   & 18.47 & \textbf{18}    & 2.73 & \textbf{2} \\ 
L2      & 0.91 & \textbf{0.64}    & \textbf{1.41} & 2.09    & \textbf{2.21} & 3.8    & \textbf{6.09} & 7.45   & \textbf{2.14} & 2.22   & \textbf{1.39}& 1.91   & \textbf{1.05} & 2.48   & 8.73 & \textbf{8.49}   & \textbf{3.87} & 4.3 \\ 
EL1     & 0.14 & \textbf{0.1}     & \textbf{0.24} & 0.53    & \textbf{0.23} & 0.61   & \textbf{0.63} & 1.12   & \textbf{0.11} & 0.48   & \textbf{0.22} & 0.43  & \textbf{0.25} & 0.58   & \textbf{1.19} & 1.2    & \textbf{0.42} & 1.23 \\ 
EL2     & 0.03 & \textbf{0.03}    & \textbf{0.09} & 0.35    & \textbf{0.1} & 0.47    & \textbf{0.47} & 1.73   & \textbf{0.02} & 0.49   & \textbf{0.07} & 0.21  & \textbf{0.1} & 0.52    & \textbf{6.02} & 7.2    & \textbf{0.27} & 1.83 \\ 
MAXEL1  & \textbf{0.17} & 0.51    & \textbf{0.49} & 1.07    & \textbf{0.2} & 0.72    & \textbf{0.7} & 1.57    & \textbf{0.15} & 1.37   & \textbf{0.44} & 0.55  & \textbf{0.64} & 1.42   & \textbf{5.45} & 6      & \textbf{0.58} & 1.91 \\ 
MINEL1  & 0.28 & \textbf{0.06}    & \textbf{0.09} & 0.45    & \textbf{0.16} & 0.67   & \textbf{0.9} & 1.09    & \textbf{0.15} & 1.37   & \textbf{0.44} & 0.49  & 0.36 & \textbf{0.22}   & \textbf{5.52} & 6      & \textbf{0.6} & 1.91 \\ 
\\
\multicolumn{21}{c}{p=30}  
\\
\\
L1     & 1.86 & \textbf{1.27}    & \textbf{1.04} & 1.52    & \textbf{10.48} & 13.66  & \textbf{25.16} & 30.8   & 5.98 & \textbf{5.21}   & \textbf{0.97} & 1       & 5.89 & \textbf{5.65}   & 58.07 & \textbf{58}      & 2.72 & \textbf{2.06} \\ 
L2     & 1.98 & \textbf{1.35}    & \textbf{1.93} & 3.82    & \textbf{13.11} & 17.03  & \textbf{26.02} & 32     & 3.77 & \textbf{2.87}   & \textbf{2.21} & 3.67    & 8.39 & \textbf{8.2}    & 15.26 & \textbf{15.23}   & \textbf{4.96} & 7.65 \\ 
EL1    & 0.16 & \textbf{0.12}    & \textbf{0.17} & 0.58    & \textbf{0.77} & 2.4     & \textbf{1.75} & 4.67    & \textbf{0.22} & 0.31   & \textbf{0.22} & 0.59    & \textbf{0.72} & 0.74   & \textbf{0.71} & 0.72     & \textbf{0.15} & 1.25 \\ 
EL2    & \textbf{0.04} & 0.06    & \textbf{0.04} & 0.43    & \textbf{0.75} & 7.36    & \textbf{3.93} & 28.49   & \textbf{0.1} & 0.27    & \textbf{0.07} & 0.42    & \textbf{2.1} & 2.21    & \textbf{7.29} & 7.73     & \textbf{0.04} & 1.92 \\ 
MAXEL1 & \textbf{0.56} & 1.07    & \textbf{0.22} & 1.2     & \textbf{1.04} & 3.41    & \textbf{2.84} & 8.34    & \textbf{1.05} & 1.62   & \textbf{0.49} & 0.81    & \textbf{5.41} & 5.55   & \textbf{10.45} & 10.77   & \textbf{0.64} & 1.93 \\ 
MINEL1 & 0.38 & \textbf{0.14}    & \textbf{0.16} & 0.53    & \textbf{1.14} & 3.4     & \textbf{3.03} & 8.23    & \textbf{0.85} & 1.62   & \textbf{0.49} & 0.81    & \textbf{0.19} & 0.35   & \textbf{10.45} & 10.77   & \textbf{0.64} & 1.93 \\ 
\\
\multicolumn{21}{c}{p=100}                                                      \\&&&&&&&&&&&&&&&&&&&&\\
L1     & 2.11 & \textbf{1.33}    & \textbf{1.02} & 1.35    & \textbf{44.18} & 46.52   & \textbf{91.89} & 96.26    & 7.04 & \textbf{7.03}   & \textbf{0.97} & 1     & 19.88 & \textbf{19.61}  & \textbf{198.16} & 198.32   & 2.81 & \textbf{2.03} \\ 
L2     & 3.36 & \textbf{2.62}    & \textbf{3.1} & 7.07     & \textbf{58.3} & 62.13    & \textbf{103.75} & 108.41  & 5.56 & \textbf{3.88}   & \textbf{3.87} & 7.04  & 28.2 & \textbf{28}      & \textbf{28.17} & 28.19     & 9.17 & \textbf{14.09} \\ 
EL1    & 0.14 & \textbf{0.13}    & \textbf{0.15} & 0.63    & \textbf{3.37} & 5.17     & \textbf{5.04} & 9.08      & \textbf{0.11} & 0.23   & \textbf{0.27} & 0.63  & \textbf{0.78} & {0.78}    & \textbf{0.4} & 0.4         & \textbf{0.18} & 1.27 \\ 
EL2    & \textbf{0.03} & 0.07    & \textbf{0.03} & 0.5     & \textbf{14.36} & 36.49   & \textbf{33.86} & 114.22   & \textbf{0.03} & 0.15   & \textbf{0.1} & 0.49   & \textbf{7.69} & 7.83    & \textbf{7.78} & 7.83       & \textbf{0.06} & 1.98 \\ 
MAXEL1 & \textbf{0.46} & 1.29    & \textbf{0.07} & 1.35    & \textbf{6.22} & 9.97     & \textbf{10.41} & 18.07    & \textbf{0.1} & 1.77    & \textbf{0.53} & 1     & \textbf{19.4} & 19.58   & \textbf{19.72} & 19.79     & \textbf{0.83} & 1.98 \\ 
MINEL1 & 0.48 & \textbf{0.17}    & \textbf{0.31} & 0.67    & \textbf{6.18} & 10.04    & \textbf{9.86} & 17.81     & \textbf{0.1} & 1.77    & \textbf{0.53} & 1     & \textbf{0.2} & 0.38     & \textbf{19.72} & 19.79     & \textbf{0.83} & 1.98 \\

\hline
\end{tabular}

\label{tab:syn_loss_results}
\end{sidewaystable}
\clearpage
\begin{sidewaystable}[!ht]
\caption{
\bf Summary of L1, L2, EL1, EL2, MAXEL1 and MINEL1 for an AR($1$), AR($2$), sparse random, scale-free, band, cluster, star and circle graphical model. The median standard errors reported here are based on 40 replications for both the B-net and D-net estimators. The best performing values are boldfaced.}
\centering
\noindent\setlength\tabcolsep{4pt}%
\scriptsize
\begin{tabular}{lllllllllllllllllllll} 

\hline
\multicolumn{1}{c}{} & \multicolumn{2}{c}{\bf AR(1)}                      & \multicolumn{2}{c}{\bf AR(2)}                      & \multicolumn{2}{c}{\bf S80}                        & \multicolumn{2}{c}{\bf S40}                        & \multicolumn{2}{c}{\bf SF}    & \multicolumn{2}{c}{\bf Band}                                               & \multicolumn{2}{c}{\bf Cluster}                      & \multicolumn{2}{c}{\bf Star}                       & \multicolumn{2}{c}{\bf Circle}                      \\ 
\cmidrule(lr){2-3}\cmidrule(lr){4-5}\cmidrule(lr){6-7}\cmidrule(lr){8-9}\cmidrule(lr){10-11}\cmidrule(lr){12-13}\cmidrule(lr){14-15}\cmidrule(lr){16-17}\cmidrule(lr){18-19}
\multicolumn{1}{c}{} & \multicolumn{1}{c}{B-net} & \multicolumn{1}{c}{D-net} & \multicolumn{1}{c}{B-net} & \multicolumn{1}{c}{D-net} & \multicolumn{1}{c}{B-net} & \multicolumn{1}{c}{D-net} & \multicolumn{1}{c}{B-net} & \multicolumn{1}{c}{D-net} & \multicolumn{1}{c}{B-net} & \multicolumn{1}{c}{D-net} & \multicolumn{1}{c}{B-net} & \multicolumn{1}{c}{D-net} & \multicolumn{1}{c}{B-net} & \multicolumn{1}{c}{D-net} & \multicolumn{1}{c}{B-net} & \multicolumn{1}{c}{D-net} & \multicolumn{1}{c}{B-net} & \multicolumn{1}{c}{D-net}  \\ 
\hline
\multicolumn{21}{c}{p=10} 
\\
\\
L1      & 0.19 & \textbf{0.05}     & 0.14 & \textbf{0.09}    & \textbf{0.72} & 1.16    & \textbf{1.37} & 2.42    & 0.16 & \textbf{0.01}   & 0.29 & \textbf{0.19}   & 0.33 & \textbf{0.12}   & 0.79 & \textbf{0.12}   & 0.69 & \textbf{0.12} \\ 
L2      & 0.15 & \textbf{0.03}     & 0.15 & \textbf{0.04}    & \textbf{0.78} & 1.02    & \textbf{1.07} & 2.11    & 0.09 & \textbf{0.01}   & 0.37 & \textbf{0.15}   & 0.37 & \textbf{0.05}   & 0.42 & \textbf{0.06}   & 0.89 & \textbf{0.01} \\ 
EL1     & 0.02 & \textbf{0.01}     & 0.07 & \textbf{0.04}    & \textbf{0.12} & 0.24    & \textbf{0.19} & 0.77    & 0.02 & \textbf{0.01}   & \textbf{0.09} & 0.1    & 0.08 & \textbf{0.03}   & 0.06 & \textbf{0.02}   & 0.14 & \textbf{0.03} \\ 
EL2     & \textbf{0.01} & \textbf{0.01}     & \textbf{0.05} & \textbf{0.05}    & \textbf{0.11} & 0.41    & \textbf{0.39} & 3.51    & {0.01} & \textbf{0.01}   & \textbf{0.06} & 0.1    & 0.09 & \textbf{0.04}   & 0.8 & \textbf{0.29}    & 0.17 & \textbf{0.1} \\ 
MAXEL1  & 0.11 & \textbf{0.01}      & 0.25 & \textbf{0.08}    & \textbf{0.27} & 0.5     & \textbf{0.57} & 1.28    & 0.08 & \textbf{0.01}   & \textbf{0.17} & 0.18   & 0.3 & \textbf{0.08}    & 0.4 & \textbf{0.12}    & 0.48 &\textbf {0.1} \\ 
MINEL1  & 0.13 & \textbf{0.04}     & 0.1 & \textbf{0.08}     & \textbf{0.26} & 0.57    & \textbf{0.64} & 1.46    & 0.08 & \textbf{0.01}   & \textbf{0.17} & 0.22   & 0.12 & \textbf{0.09}   & 0.38 & \textbf{0.12}   & 0.5 & \textbf{0.1} \\ 
\\
\multicolumn{21}{c}{p=30}  
\\
\\
L1     & 0.22 & \textbf{0.01}     & 0.14 & \textbf{0.11}    & 1.45 & \textbf{1.25}    & \textbf{1.63} & 2.28   & 0.11 & \textbf{0.01}   & 0.16 & \textbf{0.04}   & 0.08 & \textbf{0.04}  & 0.22 & \textbf{0.03}   & 0.5 & \textbf{0.05} \\ 
L2     & 0.2 & \textbf{0.01}         & 0.14 & \textbf{0.01}    & \textbf{1.03} & 1.23    & \textbf{1.07} & 1.23   & 0.05 & \textbf{0.01}   & 0.27 & \textbf{0.2}    & 0.04 & \textbf{0.01}     & 0.06 & \textbf{0.01}   & 1.07 & \textbf{0.01} \\ 
EL1    & 0.02 & \textbf{0.01}        & 0.03 & \textbf{0.02}    & \textbf{0.2} & 0.25     & \textbf{0.18} & 0.3    &\textbf {0.01} & \textbf{0.01}   & \textbf{0.04} & \textbf{0.04}   & \textbf{0.01} & \textbf{0.01}        & \textbf{0.01} & \textbf{0.01}      & 0.05 & \textbf{0.01} \\ 
EL2    & \textbf{0.01} & \textbf{0.01}        & \textbf{0.01} & 0.03    & \textbf{0.39} & 1.58    & \textbf{0.86} & 4.6    & \textbf{0.01} & \textbf{0.01}   & \textbf{0.03} & 0.07   & 0.03 & {0.02}  & 0.15 & \textbf{0.05}   & 0.05 & \textbf{0.03} \\ 
MAXEL1 & 0.15 & \textbf{0.02}     & 0.12 & \textbf{0.06}    & \textbf{0.61} & 1.31    & \textbf{0.76} & 1.58   & 0.06 & \textbf{0.01}          & \textbf{0.1} & 0.11    & 0.04 & \textbf{0.03}  & 0.11 & \textbf{0.04}   & 0.28 & \textbf{0.04} \\ 
MINEL1 & 0.16 & \textbf{0.02}     & 0.07 & \textbf{0.05}    & \textbf{0.61} & 1.37    & \textbf{0.83} & 1.6    & 0.03 & \textbf{0.01}          & \textbf{0.11} & 0.12   & 0.05 & \textbf{0.03}  & 0.11 & \textbf{0.04}   & 0.28 & \textbf{0.04} \\ 
\\
\multicolumn{21}{c}{p=100}                                                      \\&&&&&&&&&&&&&&&&&&&&\\
L1     & 0.24 & \textbf{0.01}    & 0.12 & \textbf{0.02}     & \textbf{2.26} & 2.85    & \textbf{2.49} & 4.02   & 0.1 & \textbf{0.01}    & \textbf{0.01} & \textbf{0.01}      & 0.03 & \textbf{0.01}     & \textbf{0.07} & 0.36   & 0.19 & \textbf{0.01} \\ 
L2     & 0.2 & \textbf{0.01}        & 0.05 & \textbf{0.01}        & 1.17 & \textbf{1.1}     & 1.85 & \textbf{1.71}   & \textbf{0.01} & \textbf{0.01}   & 0.02 & \textbf{0.01}   & 0.02 & \textbf{0.01}     & \textbf{0.01} & 0.05   & 1.31 & \textbf{0.01} \\ 
EL1    & \textbf{0.01} & \textbf{0.01}       & \textbf{0.01} & \textbf{0.01}        & 0.17 & \textbf{0.11}    & 0.22 & \textbf{0.16}   & \textbf{0.01} & \textbf{0.01}      & \textbf{0.01} & \textbf{0.01}      & \textbf{0.01} & \textbf{0.01}        & \textbf{0.01} & \textbf{0.01}         & 0.04 & \textbf{0.01} \\ 
EL2    & \textbf{0.01} & \textbf{0.01}       & \textbf{0.01} & \textbf{0.01}           & \textbf{1.45} & 1.56    & \textbf{2.98} & 3.84   & \textbf{0.01} & \textbf{0.01}      & \textbf{0.01} & \textbf{0.01}      & \textbf{0.01} & \textbf{0.01}     &\textbf {0.06} & 0.08   & \textbf{0.01} & \textbf{0.01} \\ 
MAXEL1 & 0.19 & \textbf{0.01}    & 0.07 & \textbf{0.01}     & \textbf{0.75} & 0.89    & \textbf{1.08} & 1.32   & 0.03 & \textbf{0.01}   & \textbf{0.03} & \textbf{0.03}   & \textbf{0.01} & \textbf{0.01}  & \textbf{0.07} & 0.1    & 0.14 & \textbf{0.01} \\ 
MINEL1 & 0.12 & \textbf{0.01}    & 0.05 & \textbf{0.01}     & \textbf{0.6} & 0.89    & \textbf{0.89} & 1.52    & 0.03 & \textbf{0.01}   & \textbf{0.03} & \textbf{0.03}   & \textbf{0.01} & \textbf{0.01}     & \textbf{0.07} & 0.1    & 0.14 & \textbf{0.01} \\ 

\hline
\end{tabular}

\label{tab:syn_sdlos_results}
\end{sidewaystable}
\clearpage

\begin{sidewaystable}[!ht]
\centering
\caption{
\bf Summary of SE, SP, F1, MC and for an AR($1$), AR($2$), sparse random, scale-free, band, cluster, star and circle graphical model. The median performance scores reported here are based on 40 replications for both the B-net and D-net estimators. The best performing values are boldfaced and scores that were not attainable due to single class classification are encoded as NA.}
\noindent\setlength\tabcolsep{4pt}%
\scriptsize
\begin{tabular}{lllllllllllllllllllll} 

\hline
\multicolumn{1}{c}{} & \multicolumn{2}{c}{\bf AR(1)} & \multicolumn{2}{c}{\bf AR(2)}                      & \multicolumn{2}{c}{\bf S80}                        & \multicolumn{2}{c}{\bf S40}                        & \multicolumn{2}{c}{\bf SF}    & \multicolumn{2}{c}{\bf Band}                                               & \multicolumn{2}{c}{\bf Cluster}                      & \multicolumn{2}{c}{\bf Star}                       & \multicolumn{2}{c}{\bf Circle}                      \\ 
\cmidrule(lr){2-3}\cmidrule(lr){4-5}\cmidrule(lr){6-7}\cmidrule(lr){8-9}\cmidrule(lr){10-11}\cmidrule(lr){12-13}\cmidrule(lr){14-15}\cmidrule(lr){16-17}\cmidrule(lr){18-19}
\multicolumn{1}{c}{} & \multicolumn{1}{c}{B-net} & \multicolumn{1}{c}{D-net} & \multicolumn{1}{c}{B-net} & \multicolumn{1}{c}{D-net} & \multicolumn{1}{c}{B-net} & \multicolumn{1}{c}{D-net} & \multicolumn{1}{c}{B-net} & \multicolumn{1}{c}{D-net} & \multicolumn{1}{c}{B-net} & \multicolumn{1}{c}{D-net} & \multicolumn{1}{c}{B-net} & \multicolumn{1}{c}{D-net} & \multicolumn{1}{c}{B-net} & \multicolumn{1}{c}{D-net} & \multicolumn{1}{c}{B-net} & \multicolumn{1}{c}{D-net} & \multicolumn{1}{c}{B-net} & \multicolumn{1}{c}{D-net}  \\ 
\hline
\multicolumn{21}{c}{p=10} 
\\
\\
SE &  \textbf{0.72} & 0.11    &  \textbf{0.82} & 0.21    &  \textbf{0.6} & 0.53    &  \textbf{0.65} & 0.36   &  \textbf{0.89} & NA    &  \textbf{0.89} & 0.44   &  \textbf{1} & 0.25   &  \textbf{0.22} & 0.11    &  \textbf{0.9} & 0 \\ 
SP &  \textbf{1} &  \textbf{1}          &  \textbf{0.97} & 0.7     &  \textbf{0.89} & 0.82   & 0.87 & {0.94}   &  \textbf{0.9} & NA     &  \textbf{0.95} & 0.93   &  \textbf{1} & 0.87   &  \textbf{0.8} & 1        & 0.95 &  \textbf{0.98} \\ 
PR &  \textbf{0.87} & 0.5     &  \textbf{0.3} & 0.12     &  \textbf{0.22} &  \textbf{0.22}   & {0.72} & 0.56   &  \textbf{0.18} & NA    &  \textbf{0.16} & 0.11   &  \textbf{0.4} & 0.16   &  \textbf{0.06} & 0.02  &  \textbf{0.2} & 0 \\ 
MC &  \textbf{0.45} & 0.11    &  \textbf{0.76} & -0.07   &  \textbf{0.54} & 0.39   &  \textbf{0.43} & 0.24   &  \textbf{0.71} & NA    &  \textbf{0.83} & 0.41   &  \textbf{1} & 0.14    &  \textbf{0.11} & 0.3    &  \textbf{0.88} & -0.07 \\ 
F1 &  \textbf{0.84} & 0.2     &  \textbf{0.84} & 0.24    &  \textbf{0.67} & 0.56   &  \textbf{0.77} & 0.51   &  \textbf{0.76} & NA    &  \textbf{0.86} & 0.4    &  \textbf{1} & 0.34    &  \textbf{0.25} & 0.2    &  \textbf{0.9} & 0 \\ 
FNR &  \textbf{0.28} & 0.89  &  \textbf{0.18} & 0.79    &  \textbf{0.4} & 0.47    &  \textbf{0.35} & 0.64   &  \textbf{0.11} & NA    &  \textbf{0.11} & 0.56   &  \textbf{0} & 0.75    &  \textbf{0.78} & 0.89   &  \textbf{0.1} & 1 \\ 
\\
\multicolumn{21}{c}{p=30}  
\\
\\
SE &  \textbf{0.31} & 0.01   &  \textbf{0.75} & 0.04   & { \textbf0.23} & 0.08   &  \textbf{0.37} & 0.03   &  \textbf{0.79} & NA    &  \textbf{0.9} & 0.24    &  \textbf{0.89} & 0.02    &  \textbf{0.03} & 0.02    &  \textbf{0.97} & 0 \\ 
SP & 0.97 &  \textbf{1}      &  \textbf{0.99} & 0.93   & 0.98 &  \textbf{0.99}   & 0.82 &  \textbf{1}      &  \textbf{0.92} & NA    &  \textbf{1} &  \textbf{1}         & {0.99} & 0.98    &  \textbf{1} & \textbf{1}          &  \textbf{1} & 0.99 \\ 
MC &  \textbf{0.17} & 0.02   &  \textbf{0.8} & -0.04   &  \textbf{0.35} & 0.16   &  \textbf{0.16} & 0.05   &  \textbf{0.52} & NA    &  \textbf{0.92} & 0.46   &  \textbf{0.89} & -0.02   & 0.06 &  \textbf{0.08}    &  \textbf{0.95} & -0.03 \\ 
F1 &  \textbf{0.48} & 0.01   &  \textbf{0.82} & 0.04   &  \textbf{0.36} & 0.14   &  \textbf{0.53} & 0.06   &  \textbf{0.53} & NA    &  \textbf{0.93} & 0.38   &  \textbf{0.93} & 0.04    &  \textbf{0.06} & 0.03    &  \textbf{0.95} & 0 \\ 
FNR &  \textbf{0.69} & 0.99  &  \textbf{0.25} & 0.96   &  \textbf{0.77} & 0.92   &  \textbf{0.63} & 0.97   &  \textbf{0.21} & NA    &  \textbf{0.1} & 0.76    &  \textbf{0.11} & 0.98    &  \textbf{0.97} & 0.98    &  \textbf{0.03} & 1 \\ 
\\
\multicolumn{21}{c}{p=100}                                                      \\&&&&&&&&&&&&&&&&&&&&\\
SE &  \textbf{0.21} & 0     &  \textbf{0.73} & 0   &  \textbf{0.04} & 0.01   &  \textbf{0.15} & 0      &  \textbf{0.49} & NA    &  \textbf{0.95} & 0.05   &  \textbf{0.79} & 0   & 0.02 &  \textbf{0.12}   &  \textbf{0.99} & 0 \\ 
SP &  \textbf{0.99} & 1     &  \textbf{1} &  \textbf{1}      &  \textbf{1} &  \textbf{1}         & 0.92 &  \textbf{1}      &  \textbf{0.98} & NA    &  \textbf{1} &  \textbf{1}         &  \textbf{1} & \textbf{1}      &  \textbf{1} &  \textbf{1}         &  \textbf{1} &  \textbf{1} \\ 
MC &  \textbf{0.33} & 0.01  &  \textbf{0.82} & 0   &  \textbf{0.14} & 0.05   &  \textbf{0.07} & 0.01   &  \textbf{0.39} & NA    &  \textbf{0.95} & 0.22   &  \textbf{0.81} & 0    & 0.08 &  \textbf{0.35}  &  \textbf{0.98} & -0.01 \\ 
F1 &  \textbf{0.34} & 0     &  \textbf{0.82} & 0   &  \textbf{0.08} & 0.01   &  \textbf{0.25} & 0      &  \textbf{0.4} & NA     &  \textbf{0.95} & 0.1    &  \textbf{0.88} & 0   & 0.04 &  \textbf{0.22}   &  \textbf{0.98} & 0 \\ 
FNR &  \textbf{0.79} & 1    &  \textbf{0.27} & 1   &  \textbf{0.96} & 0.99   &  \textbf{0.85} & 1      &  \textbf{0.51} & NA    &  \textbf{0.05} & 0.95   &  \textbf{0.21} & 1   & 0.98 &  \textbf{0.88}   &  \textbf{0.01} & 1 \\

\hline
\end{tabular}
\label{tab:syn_perf_results}
\end{sidewaystable}
\clearpage

\begin{sidewaysfigure}[ht]
  \centering
  \begin{tabular}[c]{ccc}
  \centering
    \begin{subfigure}[c]{0.3\textwidth}
      \includegraphics[width=\textwidth]{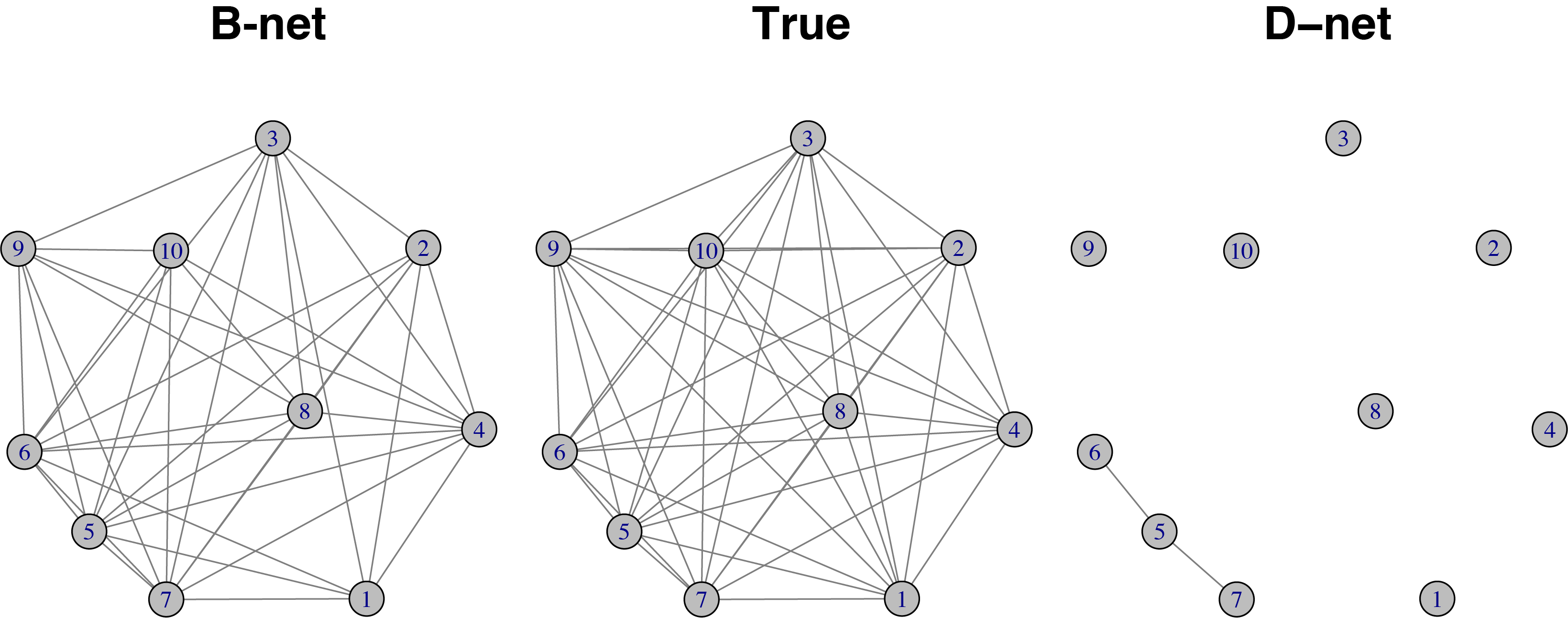}
         \caption{Model 1: AR(1).}
         \label{fig:graph_structure_p10_mod_1}
      \vspace{10mm}         
    \end{subfigure}&
    \begin{subfigure}[c]{0.3\textwidth}
      \includegraphics[width=\textwidth]{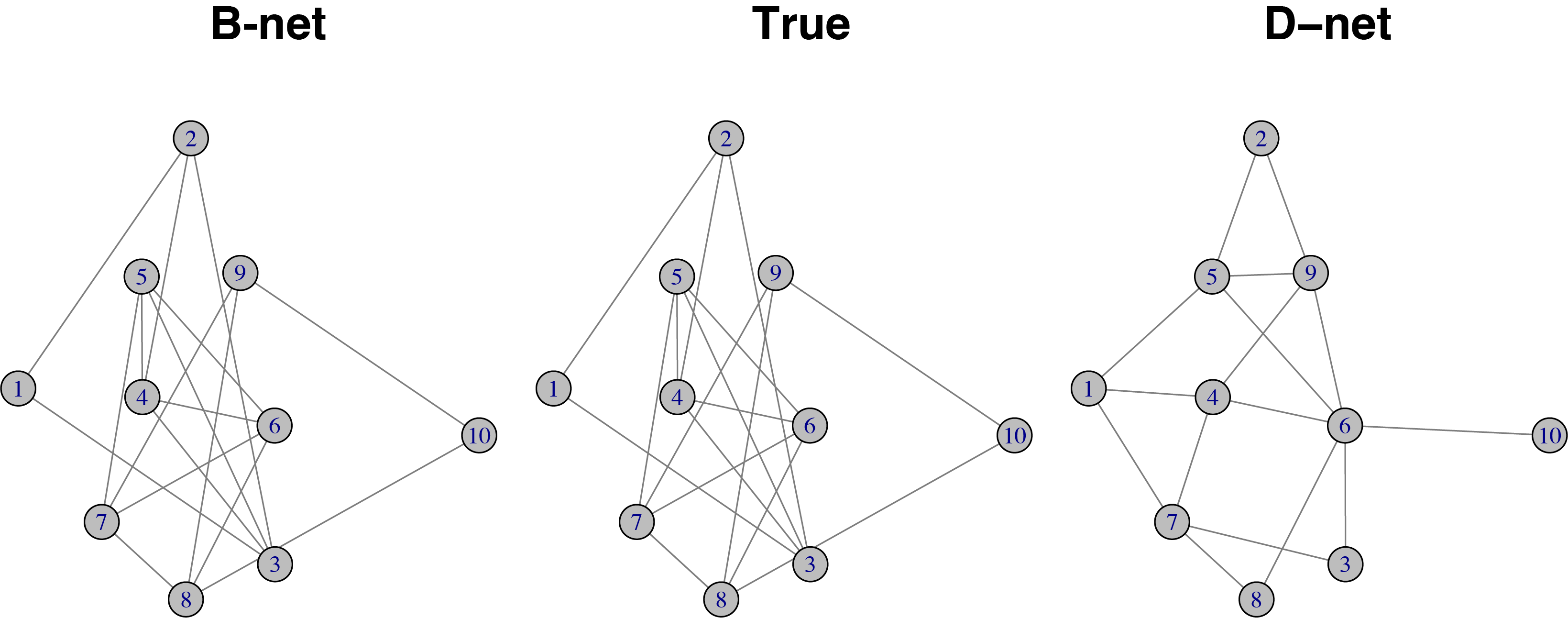}
         \caption{Model 2: AR(2).}
         \label{fig:graph_structure_p10_mod_2}
         \vspace{10mm}
    \end{subfigure}&
    \begin{subfigure}[c]{0.3\textwidth}
      \includegraphics[width=\textwidth]{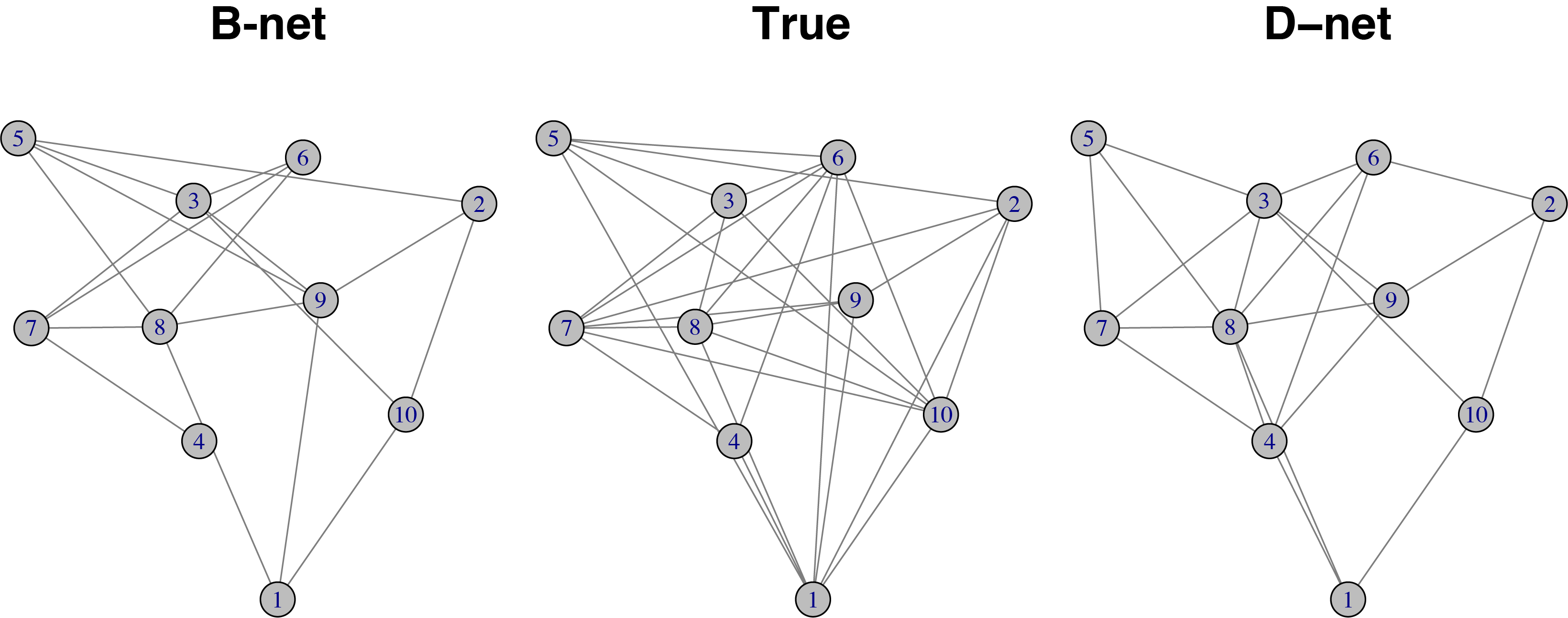}
         \caption{Model 3: at most $80\%$ sparse.}
         \label{fig:graph_structure_p10_mod_3}
         \vspace{10mm}         
    \end{subfigure}\\
    \begin{subfigure}[c]{0.3\textwidth}
      \includegraphics[width=\textwidth]{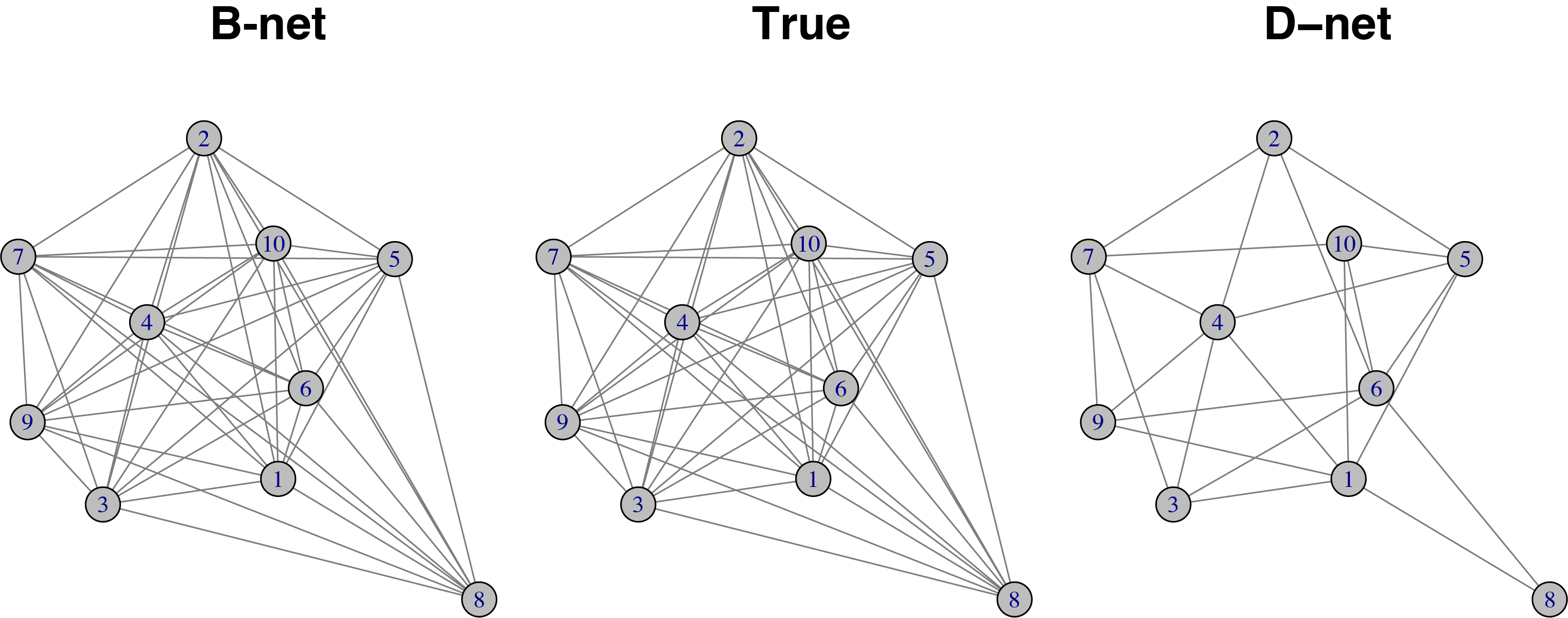}

        \caption{Model 4: at most $40\%$ sparse.}
         \label{fig:graph_structure_p10_mod_4}
         \vspace{10mm}         
    \end{subfigure}&
    \begin{subfigure}[c]{0.3\textwidth}
      \includegraphics[width=\textwidth]{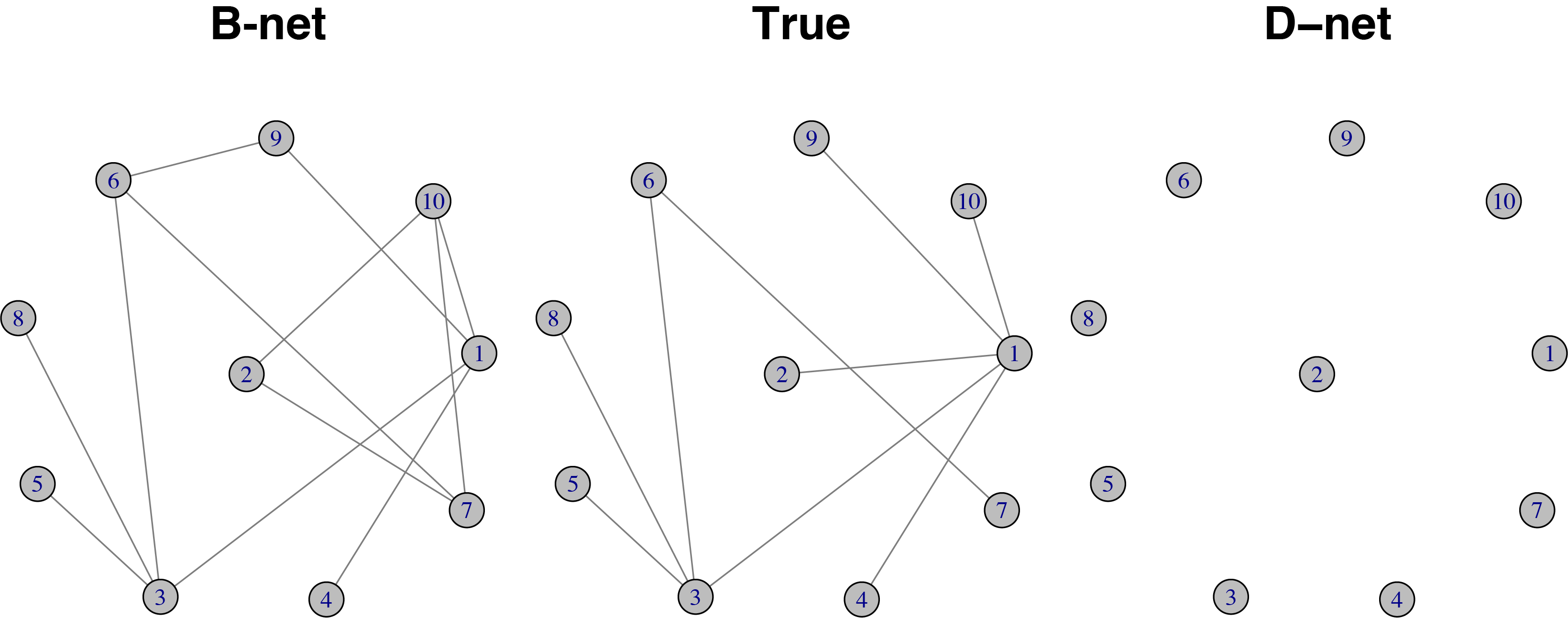}
        \caption{Model 5: scale-free.}
         \vspace{10mm}        
         \label{fig:graph_structure_p10_mod_5}
    \end{subfigure}&
    \begin{subfigure}[c]{0.3\textwidth}
      \includegraphics[width=\textwidth]{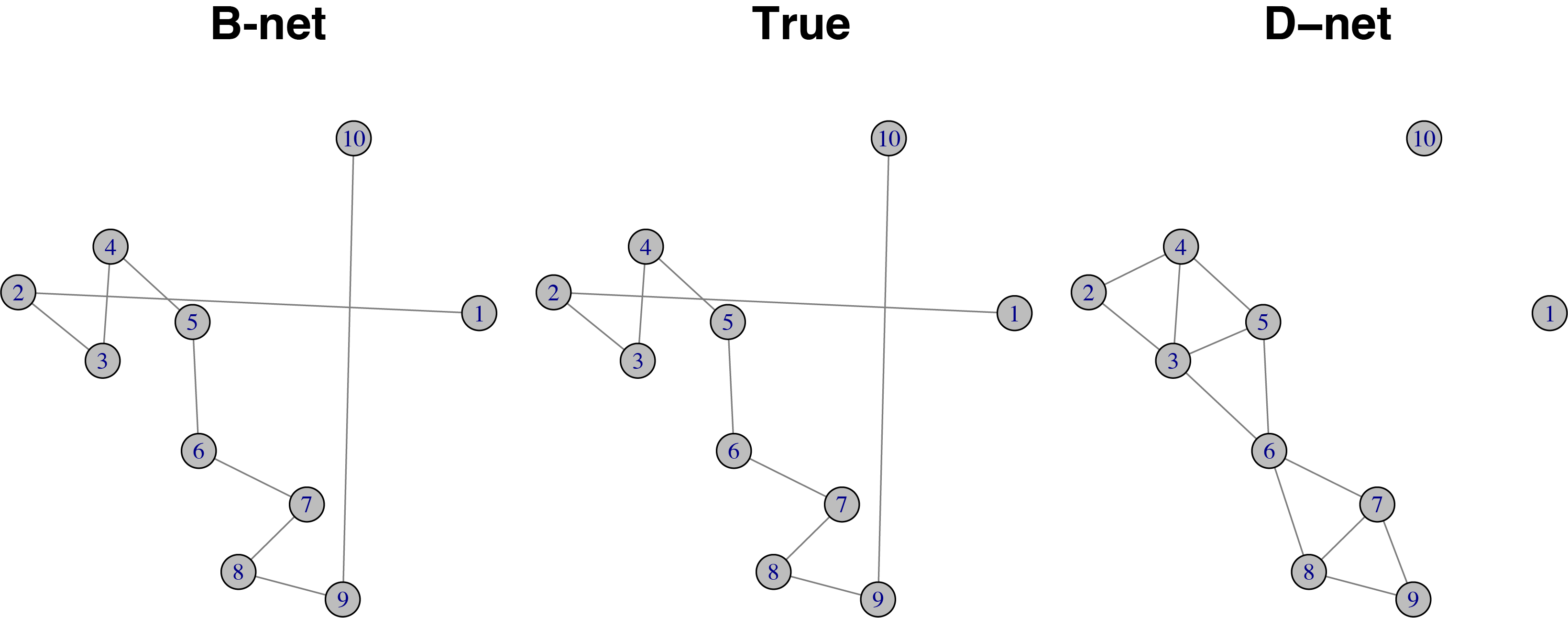}
        \caption{Model 6: band.}
         \vspace{10mm}
         \label{fig:fgraph_structure_p10_mod_6}
    \end{subfigure}\\
    \begin{subfigure}[c]{0.3\textwidth}
      \includegraphics[width=\textwidth]{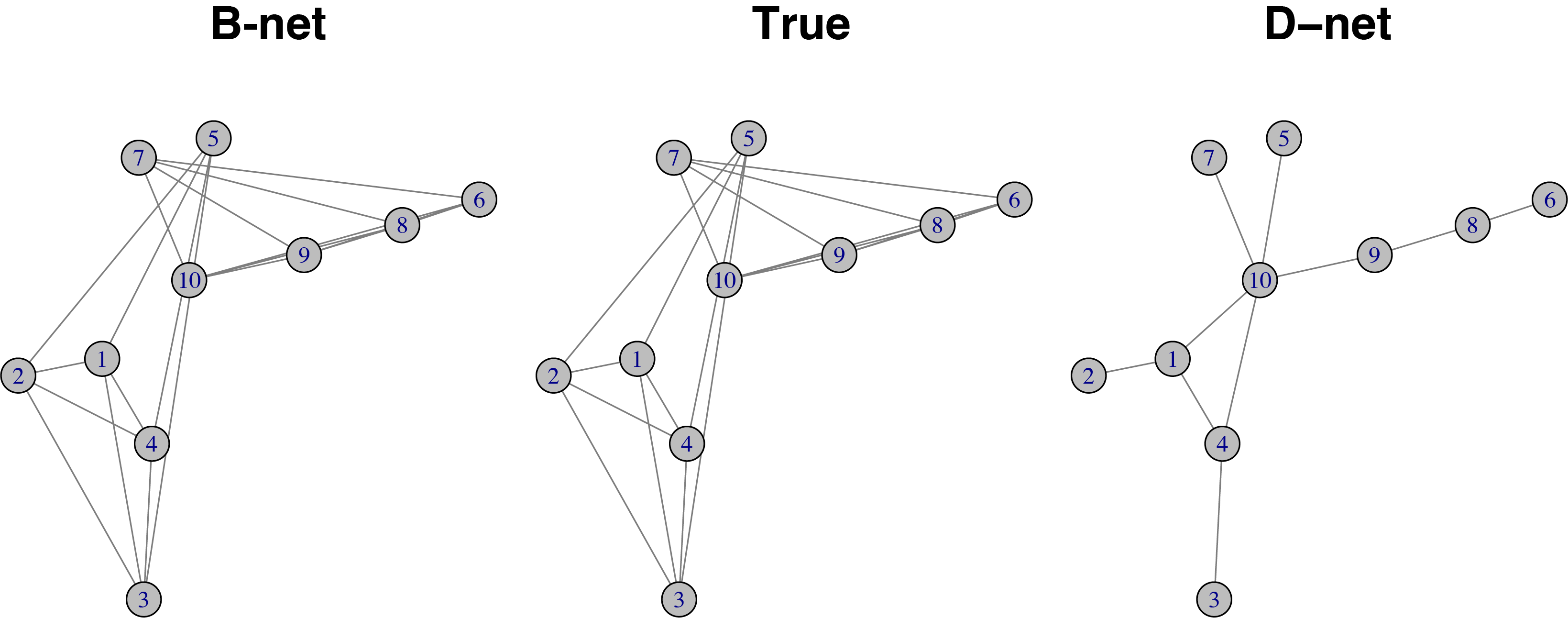}
         \caption{Model 7: cluster.}
         \label{fig:graph_structure_p10_mod_7}
    \end{subfigure}&
    \begin{subfigure}[c]{0.3\textwidth}
      \includegraphics[width=\textwidth]{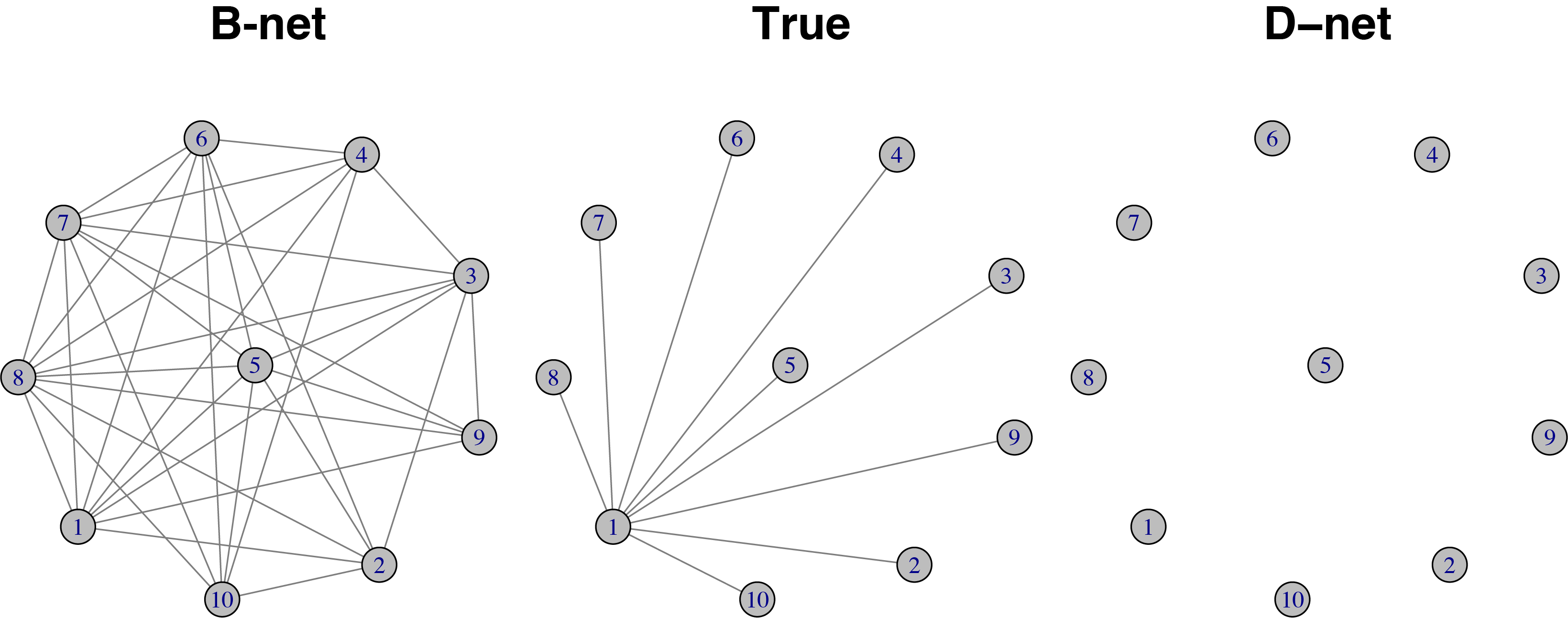}
         \caption{Model 8: star.}
         \label{fig:graph_structure_p10_mod_8}
    \end{subfigure}&
    \begin{subfigure}[c]{0.3\textwidth}
      \includegraphics[width=\textwidth]{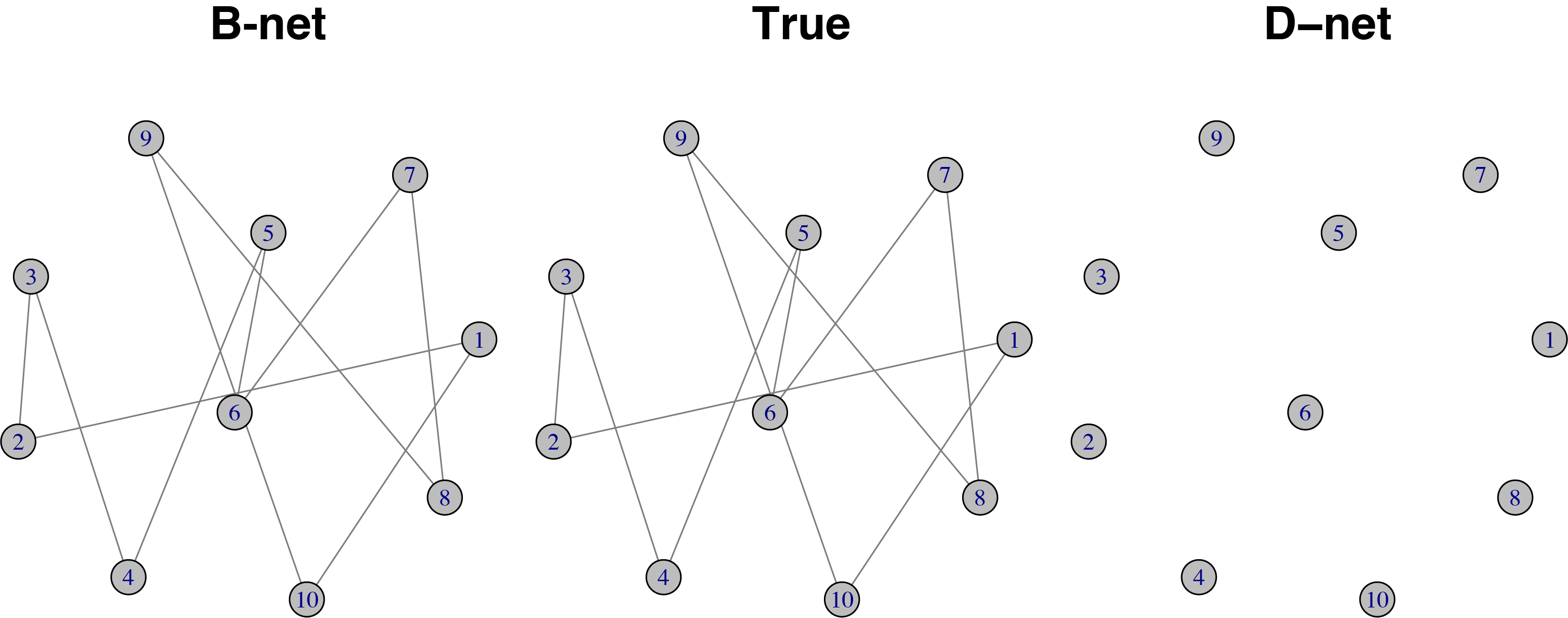}
         \caption{Model 9: circle.}
         \label{fig:graph_structure_p10_mod_9}
    \end{subfigure}\\
  \end{tabular} 
\caption{
\bf Comparison of the true DN, B-net and D-net graphical structure estimates for an AR($1$), AR($2$), sparse random, scale-free, band, cluster, star and circle graphical model and $p=10$.}
\label{fig:bayesian_graphical_structure_p10}
\end{sidewaysfigure}

\begin{table}[ht]
\caption{
\bf Performance measures used to assess classification accuracy of the B-net and D-net graphical models estimates.}
\centering
\begin{tabular}{l|l|l}
\hline                              
\bf Measure & \bf Performance function & \bf Abbreviation  \\
\thickhline
Specificity                     & $\frac{\mathrm{TN}}{\mathrm{TN}+\mathrm{FP}}$              & SP              \\

Sensitivity                        &  $\frac{\mathrm{TP}}{\mathrm{TP}+\mathrm{FN}}$             &  SE             \\

False negative rate                 & $\frac{FP}{FP+TN}$              &   FNR            \\

F1-score                  & $\frac{TP}{TP+\frac{1}{2}(FP+FN)}$              & F1              \\

Matthews Correlation Coefficient                & $\frac{\mathrm{TP}\times\mathrm{TN} - \mathrm{FP}\times\mathrm{FN}}{\sqrt{(\mathrm{TP} + \mathrm{FP})(\mathrm{TP} + \mathrm{FN})(\mathrm{TN} + \mathrm{FP})(\mathrm{TN} + \mathrm{FN})}}$              & MC             \\
   
\hline
\end{tabular}
\label{tab:performance_measures}
\end{table}
The B-net estimator generally outperforms the D-net estimator across all models and all dimensions according to the MCC, f1-score, sensitivity and false negative rate, with the exception of the star case for $p=100$. Figure \ref{fig:bayesian_graphical_structure_p10} (\ref{fig:thresh_study_p10_mod_1}) - (\ref{fig:thresh_study_p10_mod_9}) display the true and inferred undirected DN graphs for both the B-net and D-net estimators for $p=10$; higher dimensions are available in the supplementary material. Lastly, Figure \ref{fig:bayesian_adjacency_matrix_p10} (\ref{fig:adjacency_matrix_p10_mod_1}) - (\ref{fig:adjacency_matrix_p10_mod_9}) display the true and inferred adjacency matrices for $p=10$. Both Figures \ref{fig:bayesian_graphical_structure_p10} and \ref{fig:bayesian_adjacency_matrix_p10} visually demonstrate the superiority of the B-net estimator. \par

\begin{figure}[!h]
\begin{adjustwidth}{-1.5cm}{0in}
  \centering
  \begin{tabular}[c]{ccc}
  \centering
    \begin{subfigure}[c]{0.35\textwidth}
      \includegraphics[width=\textwidth]{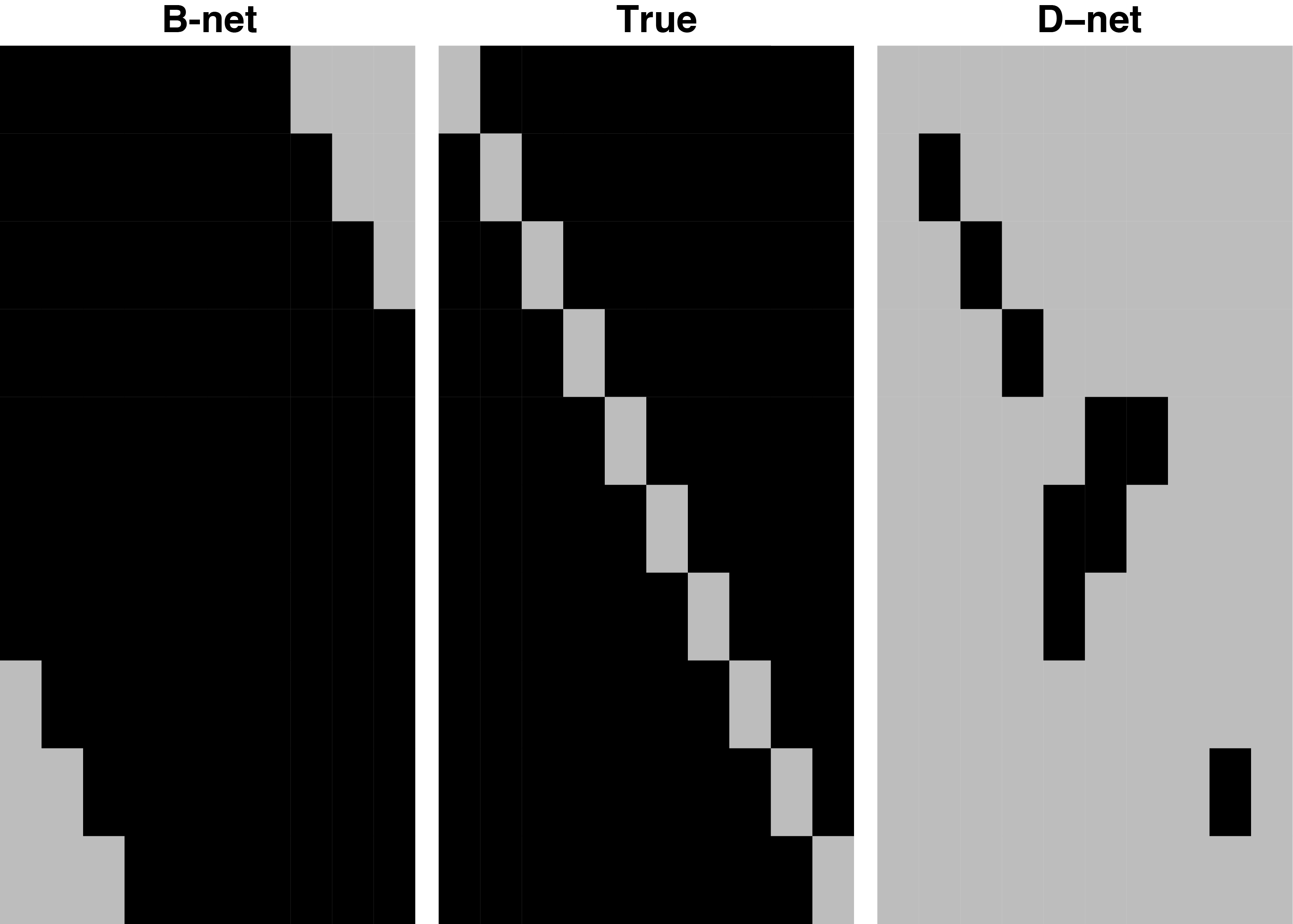}
         \caption{Model 1: AR(1).}
         \label{fig:adjacency_matrix_p10_mod_1}
    \end{subfigure}&
    \begin{subfigure}[c]{0.35\textwidth}
      \includegraphics[width=\textwidth]{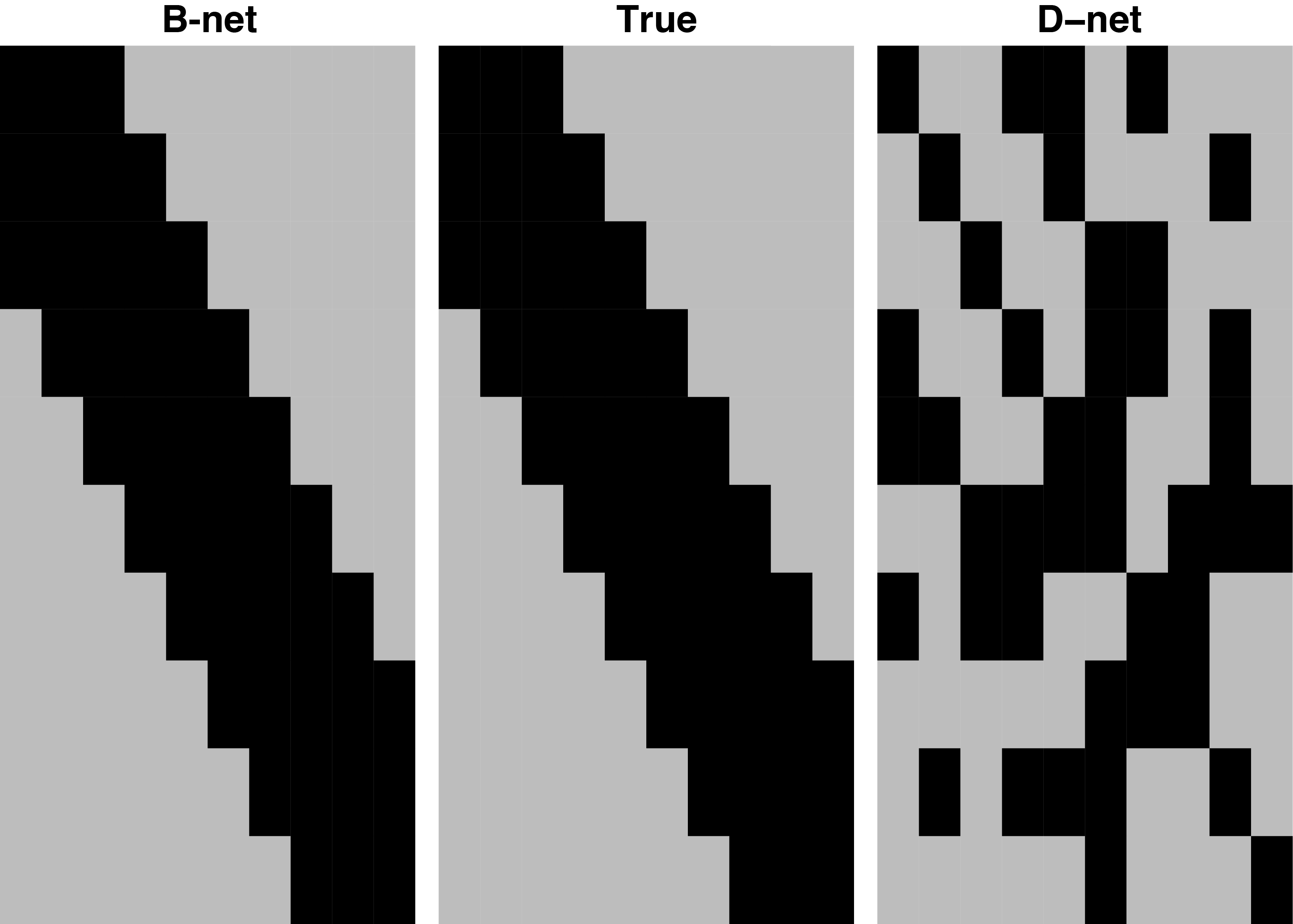}
         \caption{Model 2: AR(2).}
         \label{fig:adjacency_matrix_p10_mod_2}
    \end{subfigure}&
    \begin{subfigure}[c]{0.35\textwidth}
      \includegraphics[width=\textwidth]{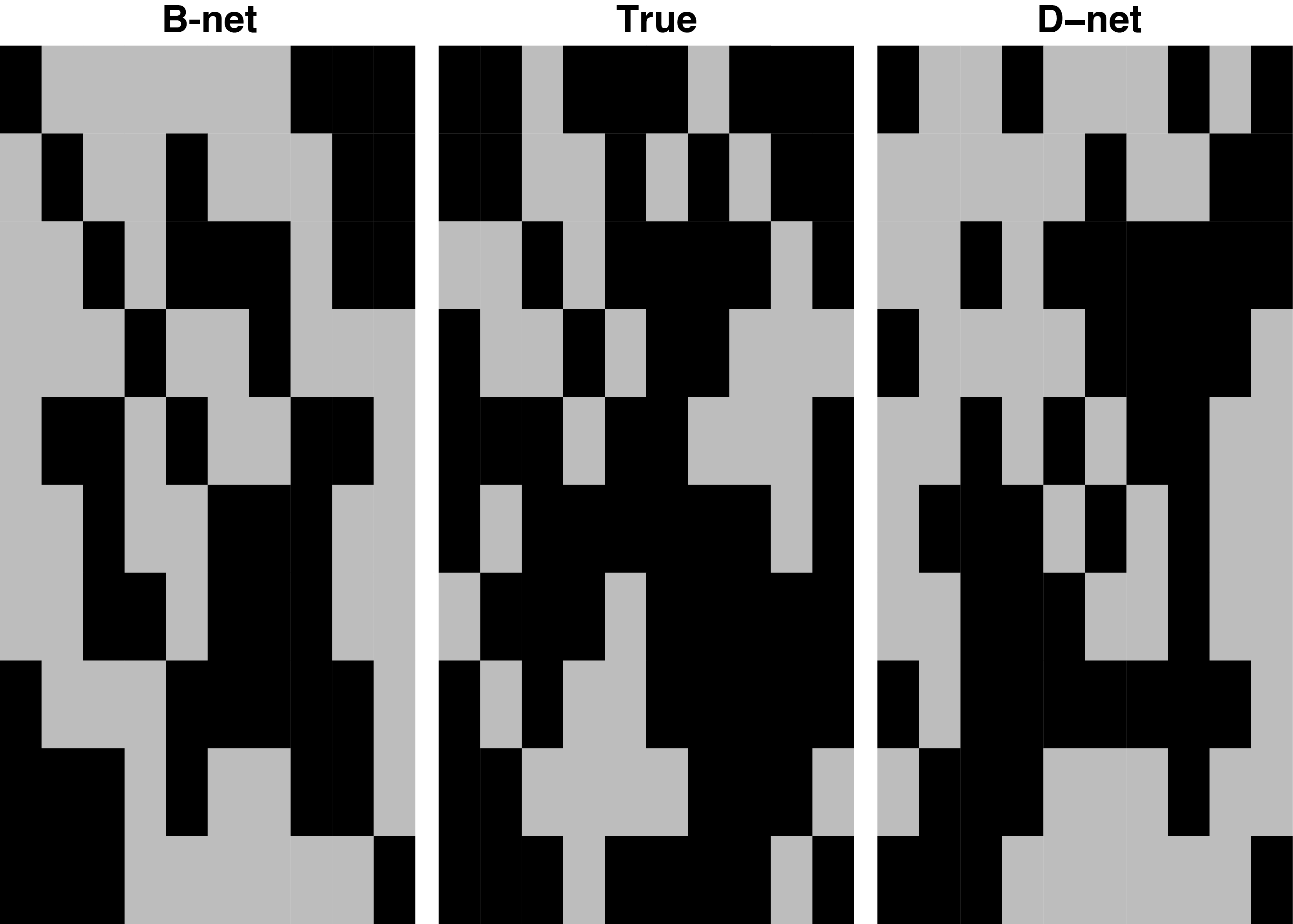}
         \caption{Model 3: at most $80\%$ sparse.}
         \label{fig:adjacency_matrix_p10_mod_3}
    \end{subfigure}\\
    \begin{subfigure}[c]{0.35\textwidth}
      \includegraphics[width=\textwidth]{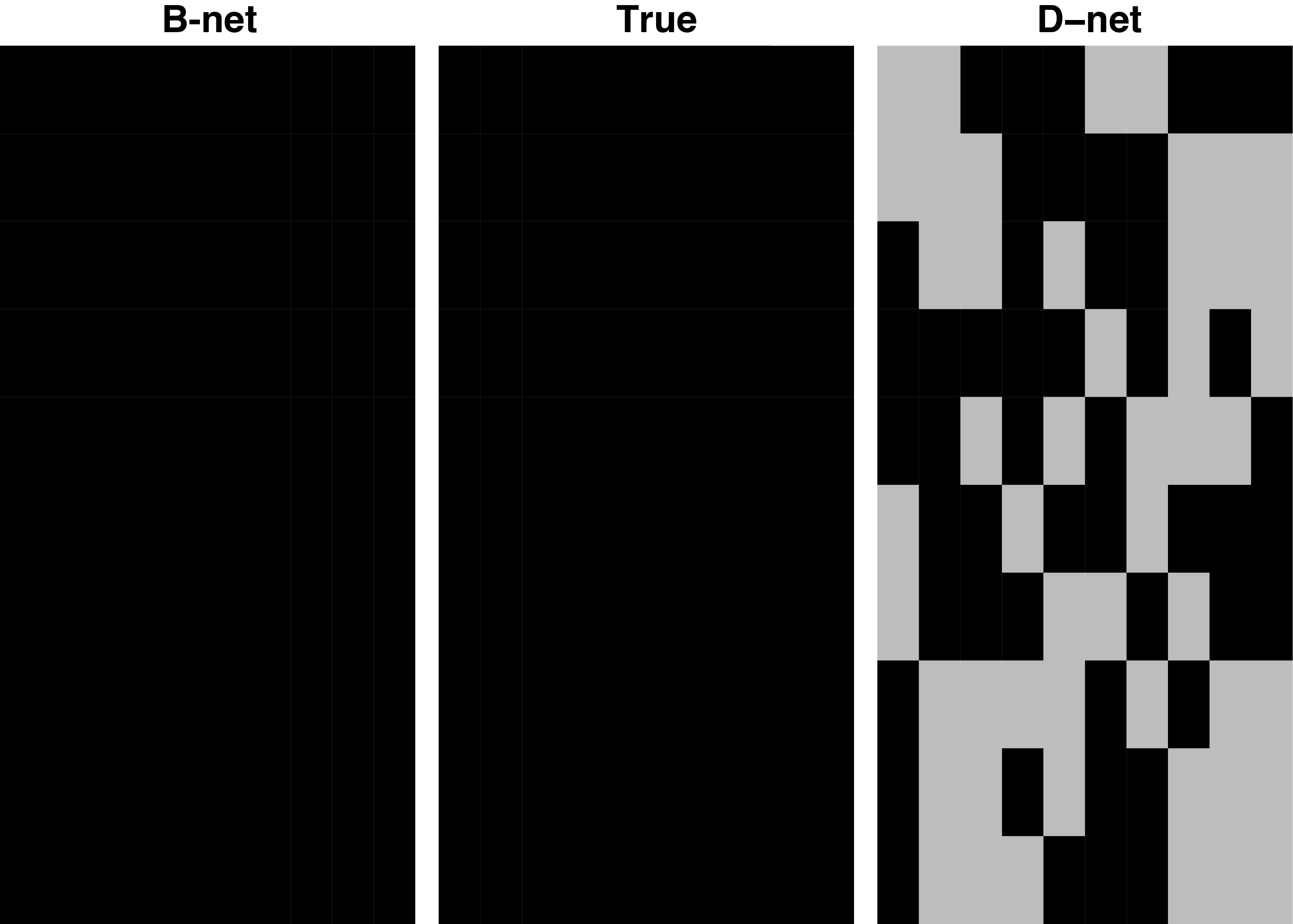}
        \caption{Model 4: at most $40\%$ sparse.}
         \label{fig:adjacency_matrix_p10_mod_4}
    \end{subfigure}&
    \begin{subfigure}[c]{0.35\textwidth}
      \includegraphics[width=\textwidth]{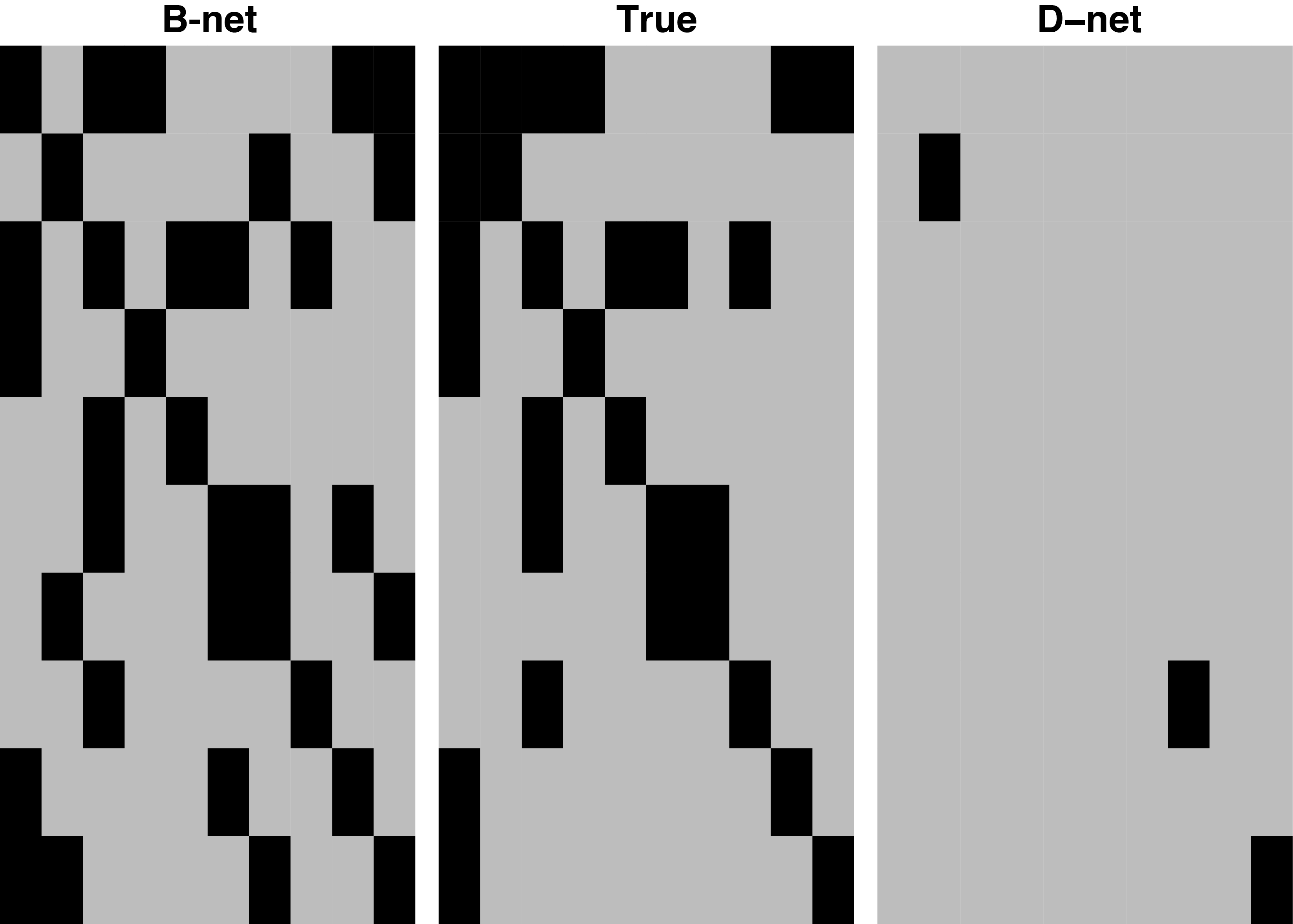}
        \caption{Model 5: scale-free.}
         \label{fig:adjacency_matrix_p10_mod_5}
    \end{subfigure}&
    \begin{subfigure}[c]{0.35\textwidth}
      \includegraphics[width=\textwidth]{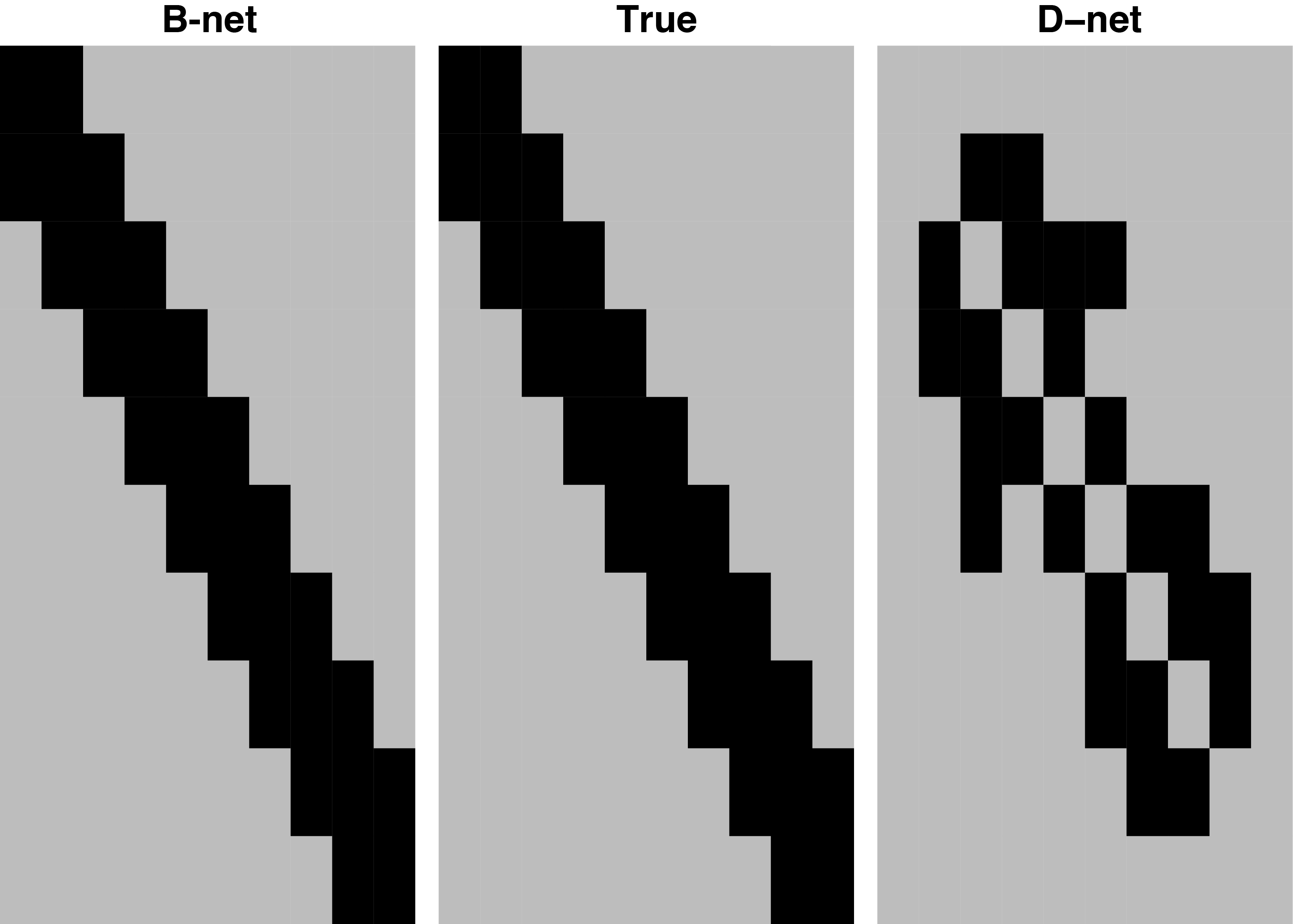}
        \caption{Model 6: band .}
         \label{fig:adjacency_matrix_p10_mod_6}
    \end{subfigure}\\
    \begin{subfigure}[c]{0.35\textwidth}
      \includegraphics[width=\textwidth]{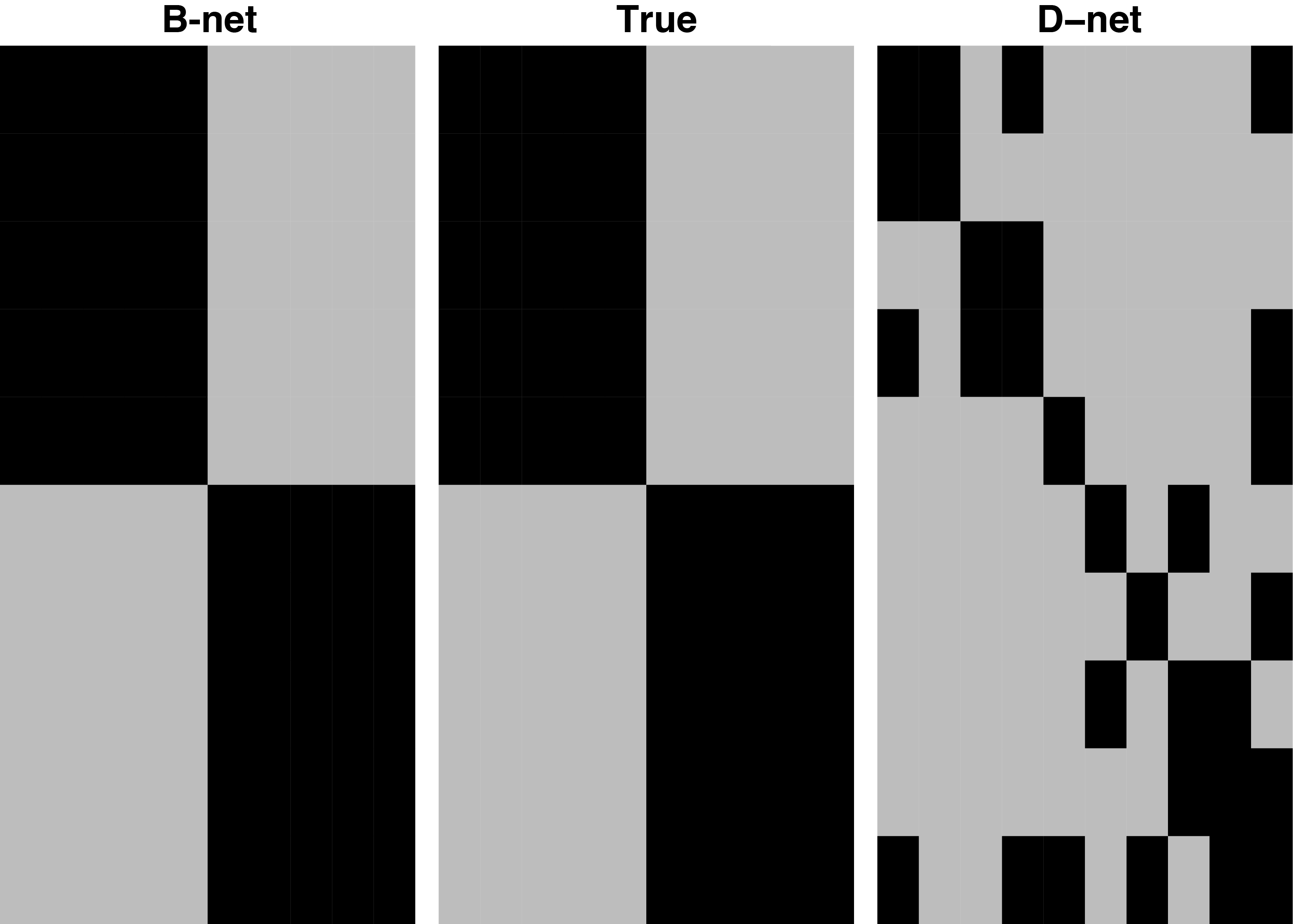}
         \caption{Model 7: cluster.}
         \label{fig:adjacency_matrix_p10_mod_7}
    \end{subfigure}&
    \begin{subfigure}[c]{0.35\textwidth}
      \includegraphics[width=\textwidth]{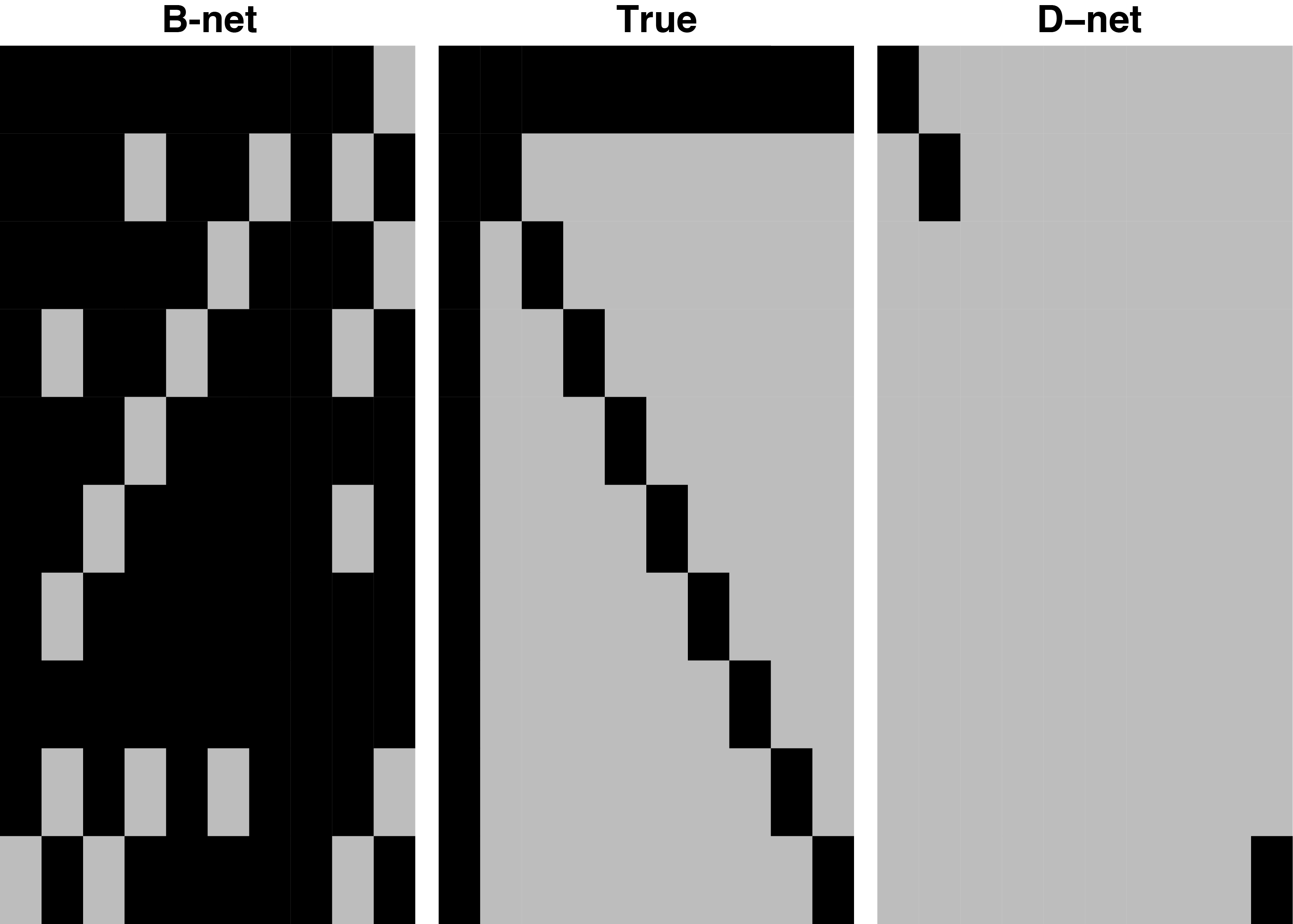}
         \caption{Model 8: star.}
         \label{fig:adjacency_matrix_p10_mod_8}
    \end{subfigure}&
    \begin{subfigure}[c]{0.35\textwidth}
      \includegraphics[width=\textwidth]{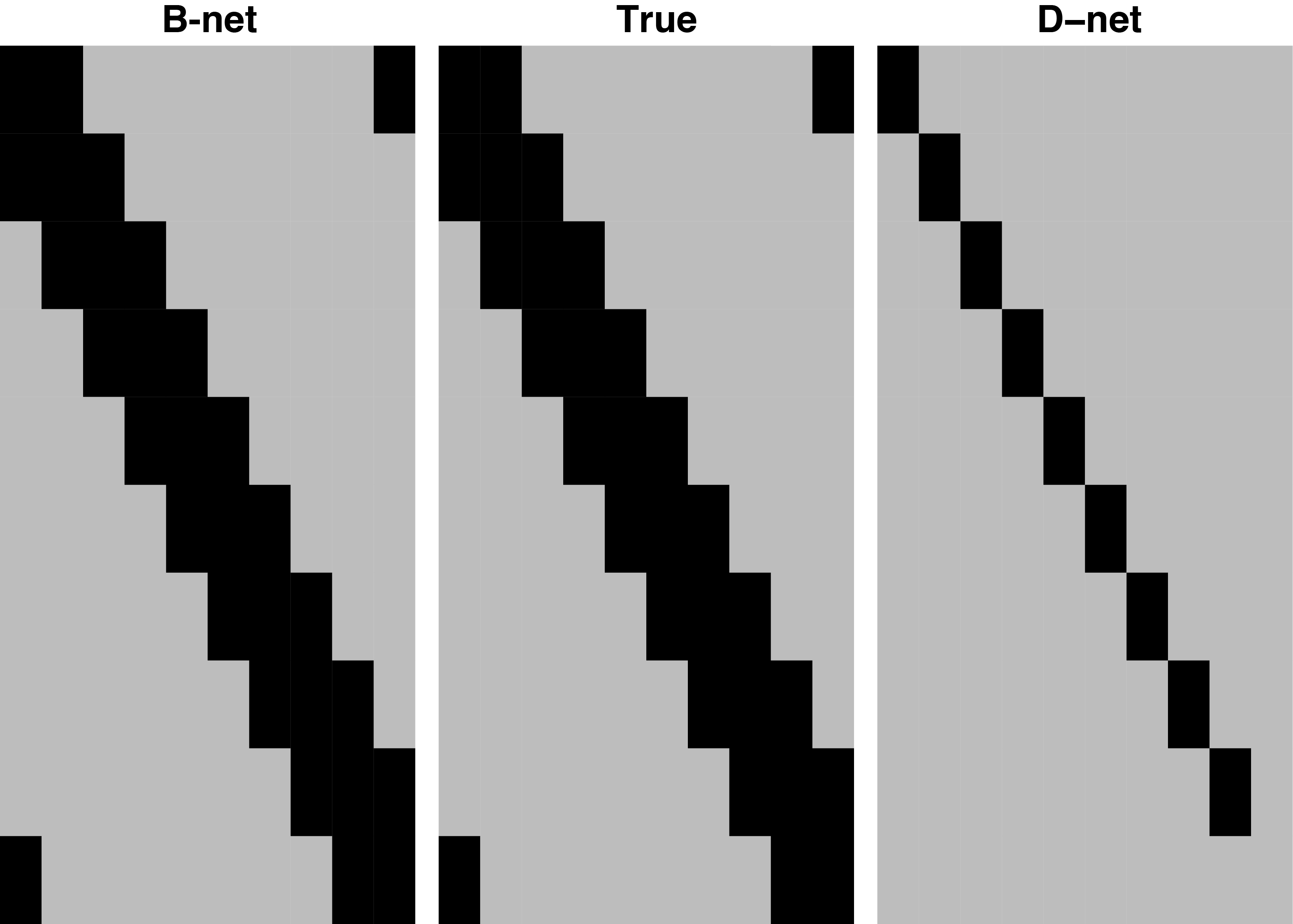}
         \caption{Model 9: circle.}
         \label{fig:adjacency_matrix_p10_mod_9}
    \end{subfigure}\\
  \end{tabular} 
\caption{
\bf Comparison of the true DN, B-net and D-net adjacency matrices for an AR($1$), AR($2$), sparse random, scale-free, band, cluster, star and circle graphical model and $p=10$. }
\label{fig:bayesian_adjacency_matrix_p10}
\end{adjustwidth}       
\end{figure}
\clearpage


\clearpage

\section*{Real data analysis}\label{sec:real_data_analysis}
This section focuses on applying the novel Bayesian DN estimator, B-net, as well as the terative shrinkage-thresholding estimator, D-net, to the spambase dataset, available at \url{ https://archive.ics.uci.edu/ml/datasets/spambase} to investigate changes in DN structure between spam and non-spam data. In addition, the B-net estimator is applied to South African COVID-19 data, obtained from \url{https://www.nicd.ac.za/diseases-a-z-index/covid-19/surveillance-reports} to investigate the change in DN structure between various phases of the pandemic.

\subsection*{Spam data}\label{subsec:spam_data_analysis}
The objective here is to compare the B-net and D-net graphical model estimates of the spam and non-spam emails. The dataset consists of $1813$ spam emails and $2788$ non-spam emails. The attributes include, amongst others, the average length of uninterrupted sequences of capital letters; total number of capital letters in the e-mail; an indicator denoting whether the e-mail was considered spam or not, in this study.

\begin{figure}[ht]
     \centering
     \begin{subfigure}[b]{0.45\textwidth}
         \centering
         \includegraphics[width=\textwidth]{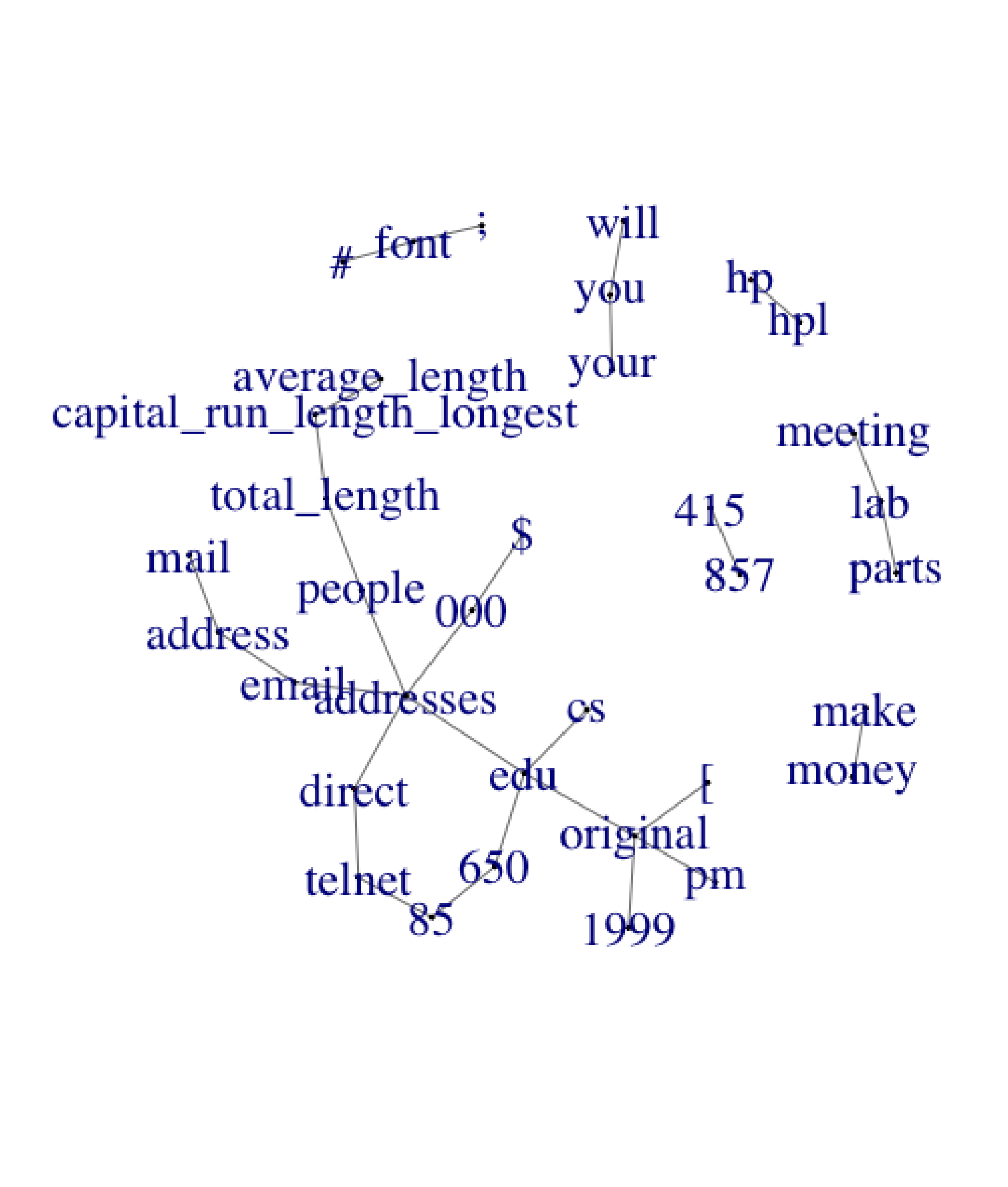}
         \caption{The Bayesian DN for the spam emails dataset.}
         \label{fig:spam_bayes}
     \end{subfigure}
     \hspace{05mm}
     \begin{subfigure}[b]{0.45\textwidth}
         \centering
         \includegraphics[width=\textwidth]{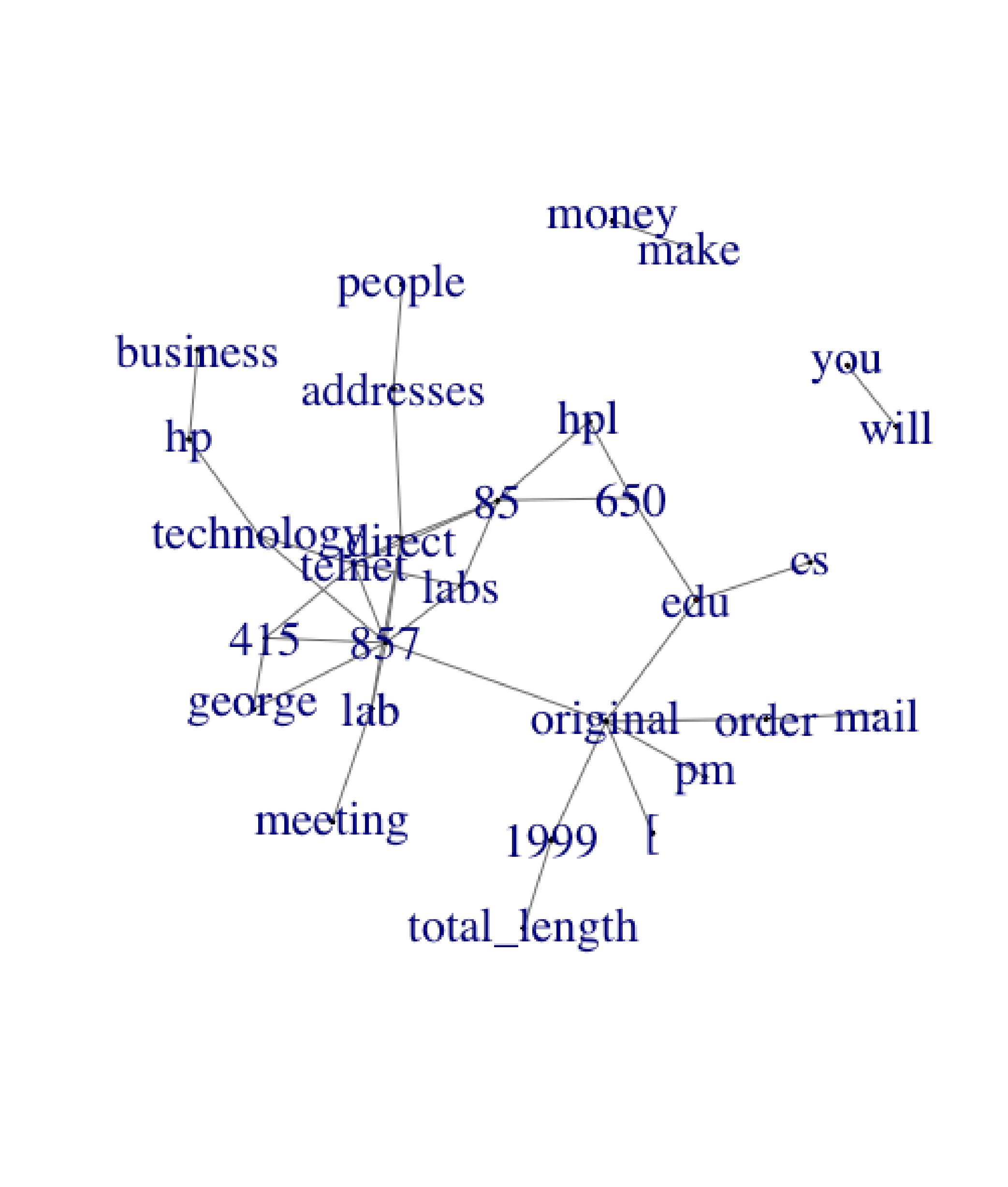}
         \caption{The iterative-shrinkage DN for the spam emails dataset.}
         \label{fig:bayes_diffnet}
     \end{subfigure}

     \caption{
     \bf A comparison of the D-net and B-net DN estimates for the spam emails dataset.}
        \label{fig:spam_data_diff_net}
        

\end{figure}

Following the approach of \cite{tang2020fast}, the data is standardised using a non-paranormal transformation in order to satisfy the Gaussian assumption. The B-net estimates are based on $10000$ iterations of the Monte Carlo sampler after $5000$ burn-in iterations. Figure \ref{fig:spam_data_diff_net} illustrates the difference between the B-net and D-net estimates. Both estimators indicate the presence of several common hub features namely, `edu', `original', `direct', `lab', `telnet' and  `addresses'. It is clear from both panes that the state of the covariance matrix structure between the spam and non-spam emails may very well be different. Furthermore, given that Hewlett-Packard Labs donated the data, words such as `telnet' and `hp' appear more often in the non-spam emails and can be used to distinguish between spam and non-spam emails.

\clearpage
\subsection*{South African COVID-19 data}\label{subsec:sa_covid_analysis}
The 2019
novel coronavirus (COVID-19) has affected more than 180 countries around the world, including South Africa. Understanding the interaction of key metrics and attributes between various phases, cycles or waves of the pandemic may prove to be invaluable in strategic planning and prevention. The goal here, is to use the Bayesian DN, B-net, to illustrate that the interactivity of key daily metrics between suspected homogeneous and heterogeneous phases within the pandemic life cycle is ever changing. In particular, the B-net is used to model the interactivity of daily metrics between the first two peaks or waves; the first wave and the following plateau and finally the difference between the first and second post wave plateaus. The data consists of 446 observations from the  7\textsuperscript{th} of February 2020 to the 27\textsuperscript{th} of April 2021. The daily metrics include, deaths; performed tests; positive test rate; active cases; tests per active case; recoveries; hospital admissions; hospital discharges; ICU admissions and the number of ventilated patients. Due to the irregularities in data capturing and publishing, a seven day moving average is applied across all daily metrics. The data is standardised using a non-paranormal transformation in order to satisfy the Gaussian assumption. The B-net is applied to the data using $10000$ iterations of the Monte Carlo sampler after $5000$ burn-in iterations. \par

\begin{figure}[ht]
    \centering
    \includegraphics[width=0.65\textwidth]{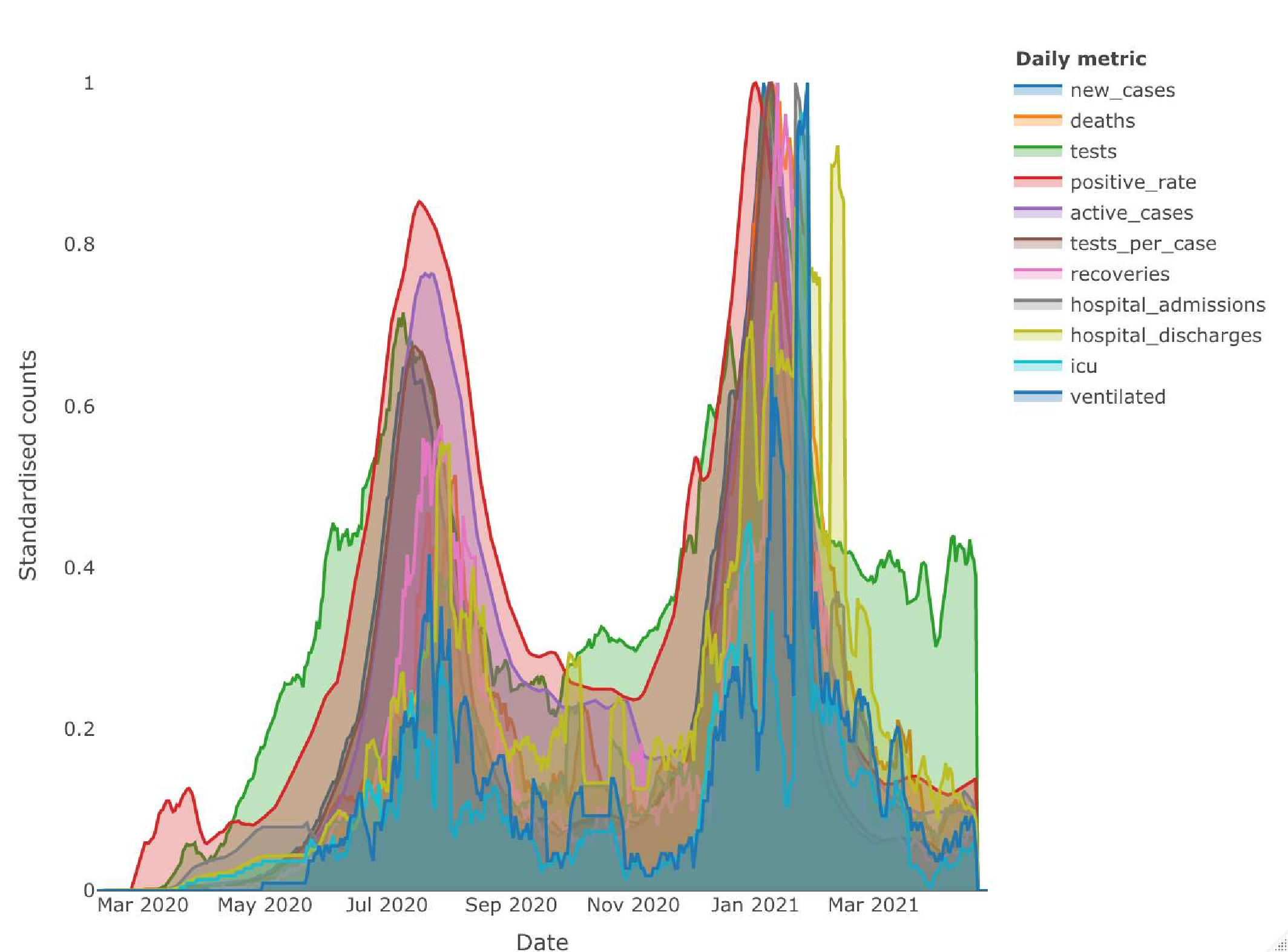}
    \caption{
    \bf 7-day moving average filled area line plots with standardised counts for daily new cases; deaths; tests; positive test rate; active cases; tests per active case; recoveries; hospital admissions; hospital discharges; ICU admissions and ventilated patients.}
    \label{fig:layered_covid_plots}
\end{figure}

Figure \ref{fig:layered_covid_plots} highlights the temporal nature of the pandemic between suspected homogeneous and heterogeneous phases. In other words, comparing the cyclical behaviour of individual daily metrics may seem clearly distinctive over time; a peak or wave is always followed by a plateau. Furthermore, extrapolation of the temporal behaviour of individual daily metrics may incorrectly allude to distinct multi modality of multiple daily metrics. Upon observing multiple metrics simultaneously, the crisp group-wise multi modality diminishes rather rapidly. The figures in Table \ref{tab:covid_networks} illustrate the higher proportions of hub features present in the DNs. Interestingly, the Bayesian DN provides insight to the change in interaction between daily metrics between perceived homogeneous pandemic phases, that is comparisons between the two peaks and two post-peak plateaus. This change in behaviour could be as a result of the change in population adherence to public sanitation awareness; weather conditions; virus mutations or complacency of over time.        
\newpage

\begin{table}[!ht]
\begin{adjustwidth}{-1cm}{0in}
\centering
\begin{tabular}{l+l l|l}
\hline
\bf Bayesian DN & \bf First precision & \bf Second precision & $\mathbf{p-value}$  \\ 
\thickhline
\begin{subfigure}[c]{0.28\textwidth}
      \includegraphics[width=\textwidth]{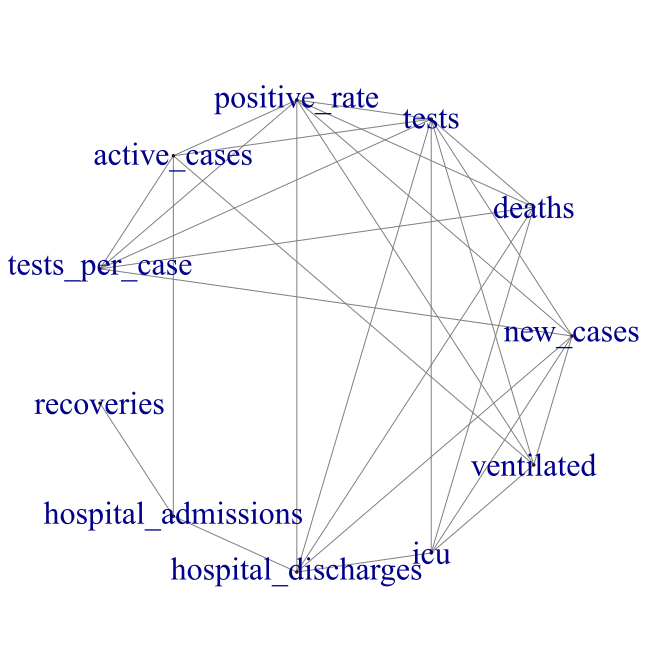}
         \caption{Bayesian DN for first wave and first post wave plateau.}
         \label{fig:bayes_dn_first_wave_first_plat}
    \end{subfigure}&
\begin{subfigure}[c]{0.28\textwidth}
      \includegraphics[width=\textwidth]{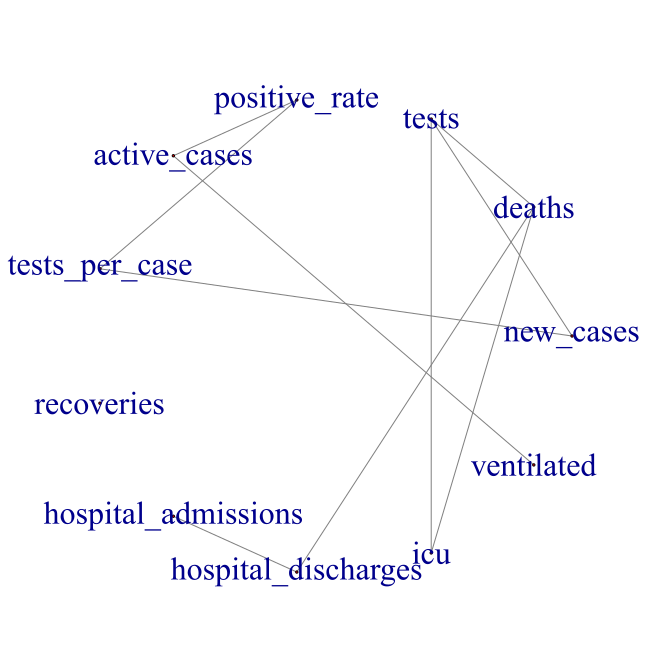}
         \caption{BAGLASSO graphical model of the first wave.}
         \label{fig:first_wave_prec}
    \end{subfigure}&    \begin{subfigure}[c]{0.28\textwidth}
      \includegraphics[width=\textwidth]{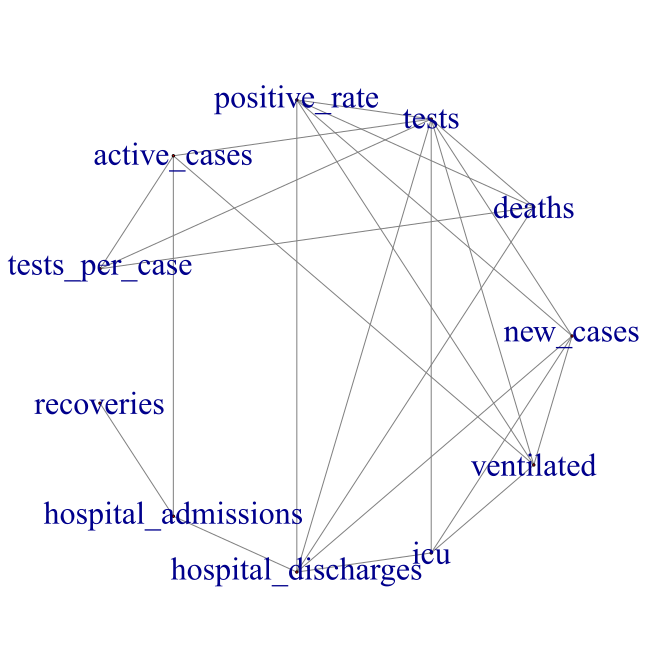}
         \caption{BAGLASSO graphical model of the first post wave plateau.}
         \label{fig:first_plat_prec}
    \end{subfigure}&     $<0.001$\\
\hline
\begin{subfigure}[c]{0.28\textwidth}
      \includegraphics[width=\textwidth]{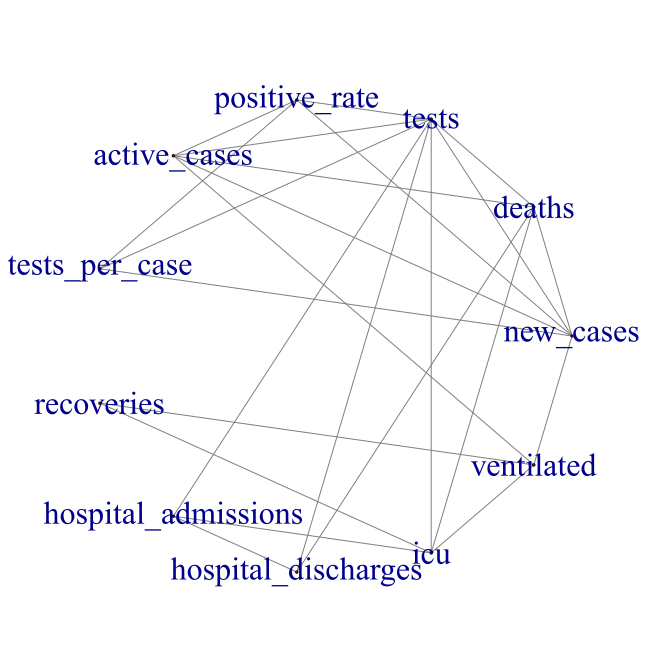}
         \caption{Bayesian DN for first and second wave.}
         \label{fig:bayes_dn_first_wave_second_wave}
    \end{subfigure}&    \begin{subfigure}[c]{0.28\textwidth}
      \includegraphics[width=\textwidth]{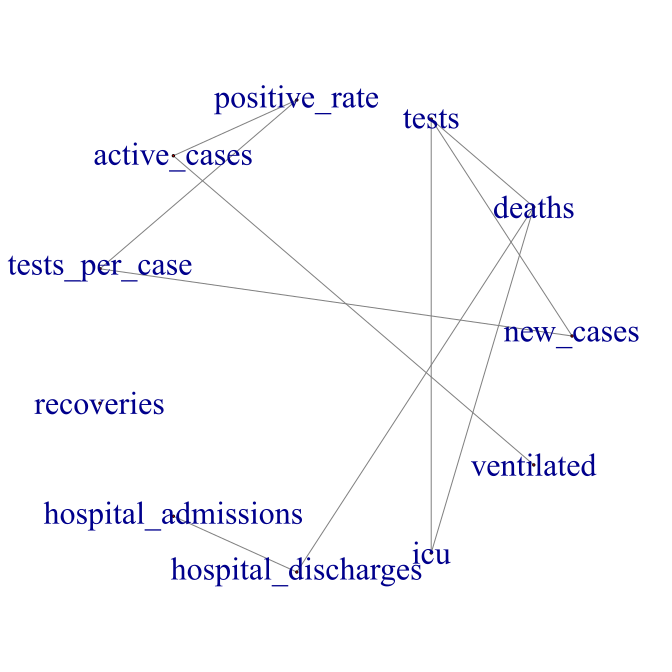}
         \caption{BAGLASSO graphical model of the first wave.}
         \label{fig:first_wave_prec_1}
    \end{subfigure}&    \begin{subfigure}[c]{0.28\textwidth}
      \includegraphics[width=\textwidth]{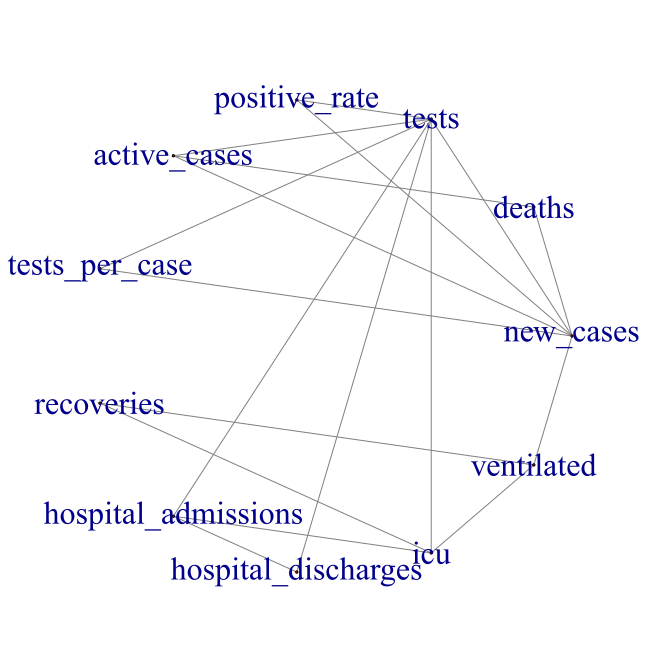}
         \caption{BAGLASSO graphical model of the second wave.}
         \label{fig:second_wave_prec}
    \end{subfigure}&    $<0.001$\\
\hline
\begin{subfigure}[c]{0.28\textwidth}
      \includegraphics[width=\textwidth]{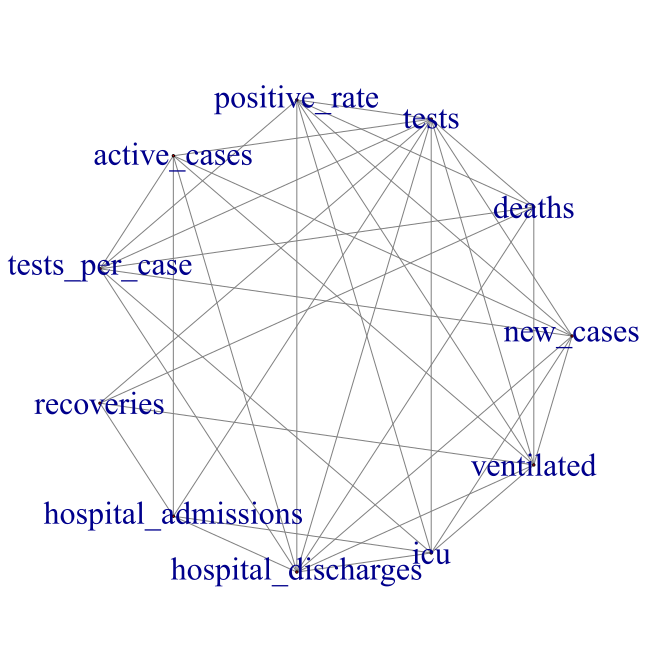}
         \caption{Bayesian DN for first and second post wave plateaus.}
         \label{fig:bayes_dn_first_plat_second_plat}
    \end{subfigure}&    \begin{subfigure}[c]{0.28\textwidth}
      \includegraphics[width=\textwidth]{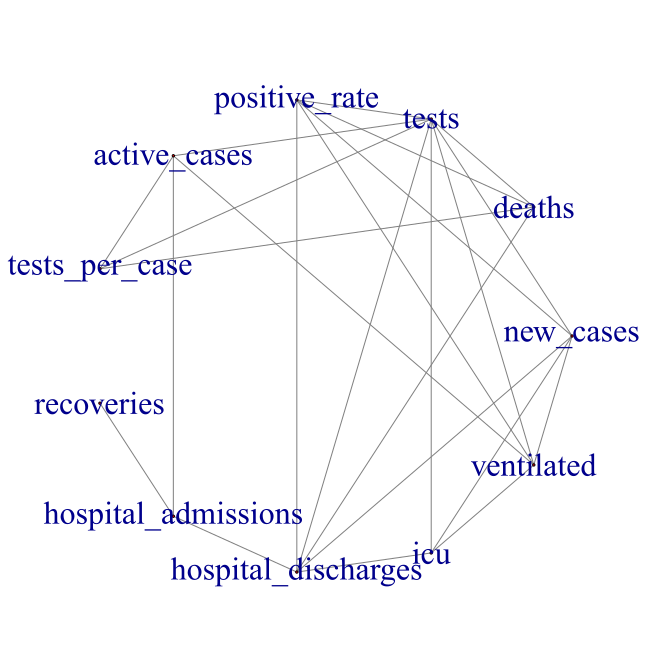}
         \caption{BAGLASSO graphical model of the first post wave plateau.}
         \label{fig:first_plat_prec_2}
    \end{subfigure}&    \begin{subfigure}[c]{0.28\textwidth}
      \includegraphics[width=\textwidth]{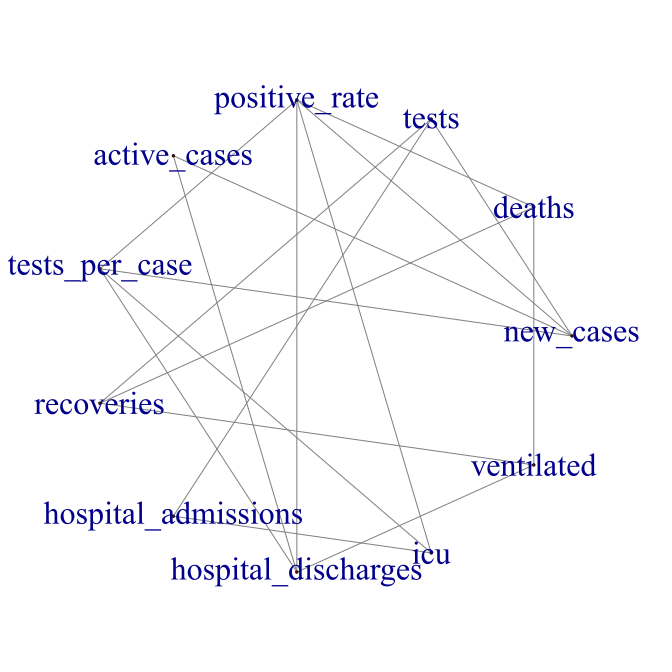}
         \caption{BAGLASSO graphical model of the second post wave plateau.}
         \label{fig:second_plat_prec}
    \end{subfigure}&    $<0.001$          
\end{tabular}
\vspace{5mm}
\caption{
\bf The Bayesian DN and corresponding BAGLASSO graphical models between the first two waves; the first wave and the following plateau and finally the difference between the first and second post wave plateaus. This figures are accompanied by the associated $p-values$ from the Box's M-test for homogeneity of covariance matrices between the contributing precision matrices \cite{box1949general}.}

\label{tab:covid_networks}
\end{adjustwidth}
\end{table}
\clearpage

\section*{Discussion}\label{sec:discussion}

The Bayesian differential network estimator is the first of its kind which utilises the excellent graphical structure determination and matrix estimation of the Bayesian graphical lasso \cite{wang_2012_bayes}. In comparison with the state of the art iterative shrinkage-thresholding approach, the Bayesian differential network offers MCMC outputs that allow the user to gain deeper insight and inference in the estimation procedure. The numerical accuracy and graphical structure determination of the Bayesian differential network are, in general, superior to that of the iterative shrinkage-thresholding estimator. The graphical structure learning is a crucial component of the Bayesian differential network estimator. The ad hoc approach provided in Eq \eqref{threshold_rule_smith} suggests a suitable sparsity threshold under the varying graph structures. The Bayesian differential network also provides key insights to changes in the interactive behaviour of real data metrics ranging from filtering spam emails to COVID-19 life cycles. For high-dimensional data, the block Gibbs sampler may be adjusted to incorporate the singular normal distribution presented in \cite{bland1966note} in the hierarchical representation Eq \eqref{unique_lambda_hier_glasso}. Furthermore, research on simultaneous Bayesian estimation and optimisation of both $\mathbf{\Sigma_1}^{-1}$ and $\mathbf{\Sigma_2}^{-1}$ in the construction of the differential network is underway.

\section*{Supporting information}

\paragraph*{S1 Appendix.}
\label{s1_appendix}
{\bf Supplementary file containing a block Gibbs sampler, as well as, additional optimal threshold; adjacency heatmaps and graphical network figures for dimensions $p=30$ and $p=100$.}\\
\noindent (PDF)

\section*{Acknowledgments}
This work was based upon research supported in part by the National Research Foundation (NRF) of South Africa, SARChI Research Chair UID: 71199; Ref.: IFR170227223754 grant No. 109214; Ref.: SRUG190308422768 grant No. 120839. The opinions expressed and conclusions arrived at are those of the authors and are not necessarily to be attributed to the CoE-MaSS or the NRF. The research of the corresponding author is supported by a grant from Ferdowsi University of Mashhad (N.$2$/$54034$).

\section*{Competing interests}
The authors have declared that no competing interests exist.
\section*{Author Contributions}

\paragraph*{Formal analysis:}
J. Smith, M.Arashi, A. Bekker.
\paragraph*{Methodology:}
J. Smith, M.Arashi, A. Bekker.
\paragraph*{Project administration:}
J. Smith, M.Arashi, A. Bekker.
\paragraph*{Funding acquisition:}
M.Arashi.
\paragraph*{Software:}
J. Smith.
\paragraph*{Supervision:}
M.Arashi, A. Bekker.
\paragraph*{Validation:}
J. Smith, M.Arashi, A. Bekker.
\paragraph*{Visualization:}
J. Smith.
\paragraph*{Writing - original draft:}
J. Smith.
\paragraph*{Writing - review and editing:}
J. Smith, M.Arashi, A. Bekker.


\bibliography{bibliography}

%
%
%





\end{document}


\vspace*{0.2in}

\begin{flushleft}
{\Large
\textbf\newline{Supplementary material: Empowering differential networks using Bayesian analysis} 
}
\newline
\\

\newcommand{\orcidauthorA}{0000-0003-4235-6147}
\newcommand{\orcidauthorB}{0000-0003-4793-5674}
\newcommand{\orcidauthorC}{0000-0002-5881-9241}

Jarod Smith \textsuperscript{\orcidA{} 1\Yinyang},
Andri{\"e}tte Bekker\textsuperscript{\orcidB{}1\Yinyang},
Mohammad Arashi\textsuperscript{\orcidC{}1,2\Yinyang*}
\\
\bigskip
\textbf{1} Department of Statistics, University of Pretoria,
Pretoria, 0002, South Africa
\\
\textbf{2} Department of Statistics, Faculty of Mathematical Sciences, Ferdowsi University of Mashhad, Mashhad, Iran
\\
\bigskip

%
%
\Yinyang These authors contributed equally to this work.\\
* arashi@um.ac.ir
\end{flushleft}

\section*{Appendix A: Block Gibbs sampler}\label{subsubsec_3_technicalties}

The block Gibbs sampler in \cite{wang_2012_bayes} for the hierarchical representation Eq ($7$) of Eq ($3$), allowing for different $\lambda_{ij}$'s in the target distribution Eq ($5$) is re-presented here for completeness.\par

To update a single column and row of $\boldsymbol{\Theta}$ per iteration - focusing on the last column and row - let $\boldsymbol{\Upsilon}$ be a $p\times p$ symmetric matrix with a zero main diagonal and $\boldsymbol{\tau}$ in the upper off diagonal entries. Secondly, partition the matrices $\boldsymbol{\Theta}$, $\boldsymbol{S}$ and $\boldsymbol{\Upsilon}$ as follows 

\begin{equation} \label{partition_gibbs}
    \boldsymbol{\Theta}=
        \begin{pmatrix}
         \boldsymbol{\Theta_{11}} & \boldsymbol{\theta_{12}}\\
         \boldsymbol{\theta_{21}} & \theta_{22}
        \end{pmatrix},
    \:\:\:\:
    \boldsymbol{S} = 
        \begin{pmatrix}
         \boldsymbol{S_{11}} & \boldsymbol{s_{12}}\\
         \boldsymbol{s_{21}} & s_{22}
        \end{pmatrix},
    \:\:\:\:
    \boldsymbol{\Upsilon} = 
        \begin{pmatrix}
         \boldsymbol{\Upsilon_{11}} & \boldsymbol{\tau_{12}}\\
         \boldsymbol{\tau_{21}} & 0
        \end{pmatrix}.
\end{equation}

\noindent Notice that the conditional distribution of the last column in Eq ($5$) can be presented as 

\begin{equation*}
    \begin{aligned}
        p(\boldsymbol{\theta_{12}},\theta_{22}\:|\: \boldsymbol{\Theta}_{11},\boldsymbol{\Upsilon},\boldsymbol{S},\lambda)&\propto
        (\theta_{22}-\boldsymbol{\theta}_{21}\boldsymbol{\Theta}^{-1}_{11}\boldsymbol{\theta}_{12})^{\frac{n}{2}}\\
        &\times exp\left[-\frac{1}{2}\{\boldsymbol{\theta_{21}}\boldsymbol{D}_{\boldsymbol{\tau}}^{-1}\boldsymbol{\theta_{12}}  +2\boldsymbol{s}_{21}\boldsymbol{\theta_{12}} + (\lambda+s_{22})\theta_{22}\}\right],
    \end{aligned}
\end{equation*}

\noindent here $\boldsymbol{D}_{\boldsymbol{\tau}}$ is the diagonalised matrix of the vector $\boldsymbol{\tau}_{12}$. The following change of variables 

\begin{equation*}
    \begin{aligned}
        \boldsymbol{\beta} &=\boldsymbol{\theta}_{12}\\
        \gamma &= \theta_{22}-\boldsymbol{\theta}_{21}\boldsymbol{\Theta}^{-1}_{11}\boldsymbol{\theta}_{12}
    \end{aligned}
\end{equation*}

\noindent with Jacobian independent of $(\boldsymbol{\beta},\gamma)$, yields the following conditional distribution

\begin{equation*}
    \begin{aligned}
        p(\boldsymbol{\beta},\gamma\:|\: \boldsymbol{\Theta}_{11},\boldsymbol{\Upsilon},\boldsymbol{S},\lambda)
    \propto
    &(\gamma)^{\frac{n}{2}}exp(-\frac{s_{22}+\lambda}{2}\gamma)\\
    &\times
    exp\left[-\frac{1}{2}(\boldsymbol{\beta}^{\top}\{\boldsymbol{\tau}^{-1} + (s_{22 +\lambda})\boldsymbol{\Theta}_{11}^{-1} \}\boldsymbol{\beta+2\boldsymbol{s}_{21}\boldsymbol{\beta}})\right].
    \end{aligned}
\end{equation*}

\noindent It follows that
\begin{equation*}
    (\boldsymbol{\beta},\gamma\:|\: \boldsymbol{\Theta}_{11},\boldsymbol{\Upsilon},\boldsymbol{S},\lambda) 
    \sim \mathrm{GA}(\frac{n}{2}+1,\frac{s_{22}+\lambda}{2})
    \mathrm{N}(-\boldsymbol{C}\boldsymbol{s}_{21},\boldsymbol{C}),
\end{equation*}

\noindent where $\boldsymbol{C}=\{(s_{22}+\lambda)\boldsymbol{\Theta}_{11}+\boldsymbol{D}_{\boldsymbol{\tau}}^{-1}\}^{-1}$. See \cite{wang_2012_bayes} for a discussion on how the positive definite constraint on $\boldsymbol{\Theta}$ is adhered to in the formulation above. \par

\noindent Moreover, $\boldsymbol{\tau}$ can be updated by  observing that the conditional posterior distribution of the $1/\tau_{ij}$'s in Eq ($5$) are independently inverse Gaussian ($\mathrm{INV}-\mathrm{GAU}$) with $\lambda_{ij}^{'}=\lambda_{ij}^2$ and $\mu^{'}=\sqrt{(\lambda_{ij}^2/\theta_{ij}^2)}$

\begin{equation*}
    p(x)=(\frac{\lambda_{ij}^{'}}{2x\pi^3})^\frac{1}{2}
    exp\bigg\{\frac{-\lambda_{ij}^{'}(x-\mu^{'})^2}{2(\mu^{'})^2x}\bigg\}, \:\:\:x\:>\:0.
\end{equation*}

\noindent The block Gibbs sampler algorithm is summarised by

\begin{algorithm}
	\caption{Block Gibbs sampler} 
	\begin{algorithmic}[1]
	    \State Set the hyperparameters $(s,r)$ for the gamma distribution in Eq ($7$) and initialise $\{\lambda_{ii}\}_{i=1}^p$ and  $\boldsymbol{\tau}$.
		\For {$i=1,2,\ldots, p $}
            \State Partition  $\boldsymbol{\Theta}$, $\boldsymbol{S}$ and $\boldsymbol{\Upsilon}$ as in Eq \eqref{partition_gibbs}.
            \State Sample $\gamma \sim \mathrm{GA}(n/2+1,(s_{22}+\lambda)/2)$ and 
            $\boldsymbol{\beta} \sim \mathrm{N}(-\boldsymbol{C}\boldsymbol{s}_{21},\boldsymbol{C})$.
            \State Update $\boldsymbol{\theta}_{12}=\boldsymbol{\beta}, \boldsymbol{\theta}_{21}=\boldsymbol{\beta}^{\top}, 
            \theta_{22}=\gamma+\boldsymbol{\beta}^{\top}\boldsymbol{\Theta}_{11}^{-1}\boldsymbol{\beta}$ 
		\EndFor
		\For {$i \neq j$}
		    \State Sample $\lambda_{ij}\sim \mathrm{GA}(1+r,|\theta_{ij}|+s)$.
		    \State Sample $\delta_{ij} \sim \mathrm{INV}-\mathrm{GAU}(\mu^{'},\lambda^{'})$ and update $\tau_{ij}=1/\delta_{ij}$
		\EndFor
	\end{algorithmic} 
\end{algorithm}

\noindent For the choice of the hyperparameters $(r,s)$, \cite{wang_2012_bayes} suggest $r=10^{-2}$ and $s=10^{-6}$ for excellent performance in the adaptability of $\lambda_{ij}$ to each $\theta_{ij}$ and $\lambda_{ii}=1$. \par

\section*{Appendix B: Additional synthetic data analysis figures}
\begin{figure}[hbt!]
  \centering
   \begin{tabular}[c]{cc}
  \vspace*{1cm}
    \begin{subfigure}[c]{0.9\textwidth}
      \includegraphics[width=\textwidth]{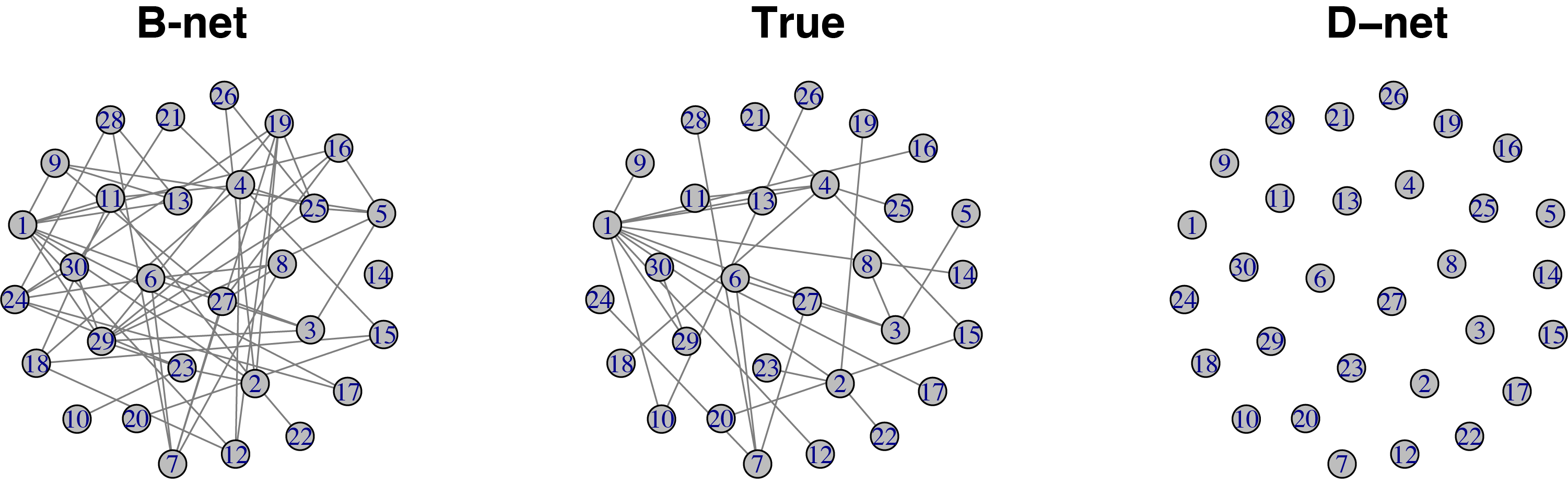}
         \caption{Model 5: scale-free.}
         \label{fig:adjacency_matrix_p10_mod_3}
    \end{subfigure} &\\
  \vspace*{1cm}
    \begin{subfigure}[c]{0.9\textwidth}
      \includegraphics[width=\textwidth]{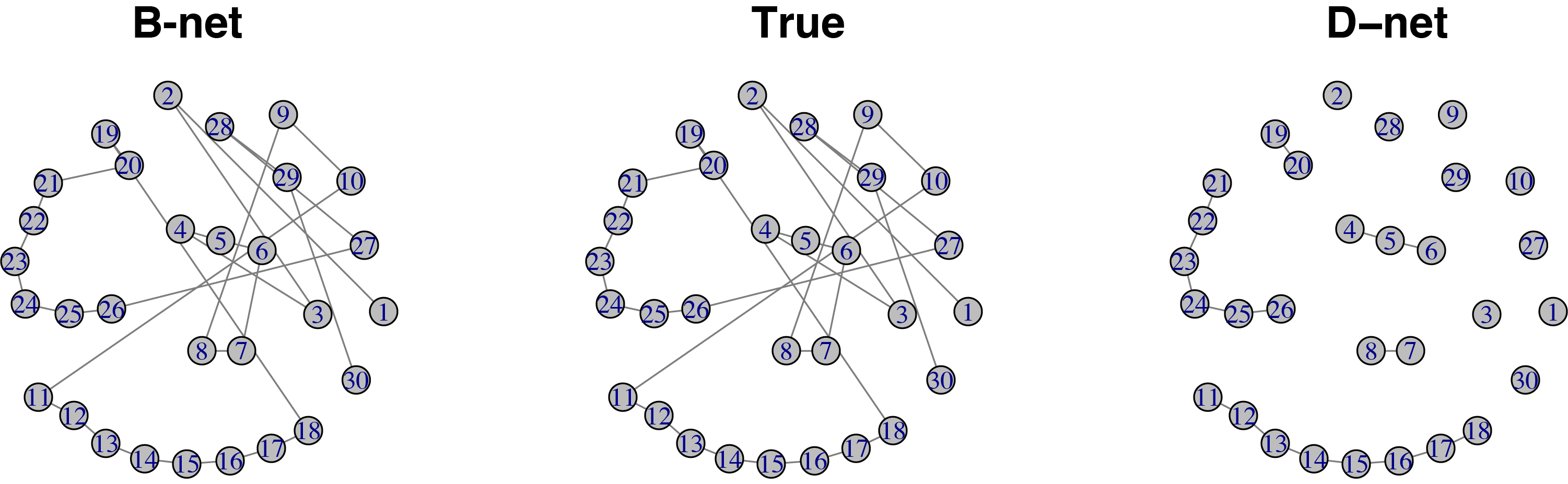}
        \caption{Model 6: band.}
         \label{fig:adjacency_matrix_p10_mod_4}
    \end{subfigure} & \\
    \begin{subfigure}[c]{0.9\textwidth}
      \includegraphics[width=\textwidth]{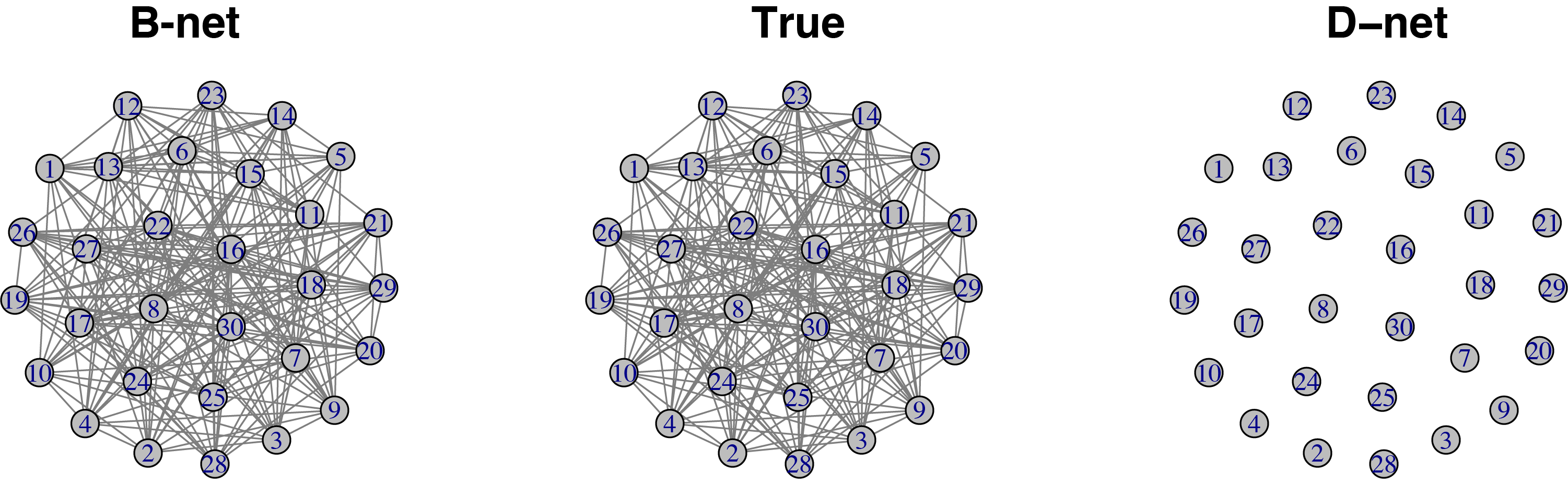}
        \caption{Model 7: cluster.}
         \label{fig:adjacency_matrix_p10_mod_4}
    \end{subfigure} & \\    
  \end{tabular} 
\caption{Comparison of the true DN, B-net and D-net graphical structure estimates for the scale-free, band and cluster structure for $p=30$. }
\label{fig:bayesian_bn_gm_p30}
        
\end{figure}

\begin{figure}[hbt!]
  \centering
   \begin{tabular}[c]{cc}
    \begin{subfigure}[c]{0.6\textwidth}
      \includegraphics[width=\textwidth]{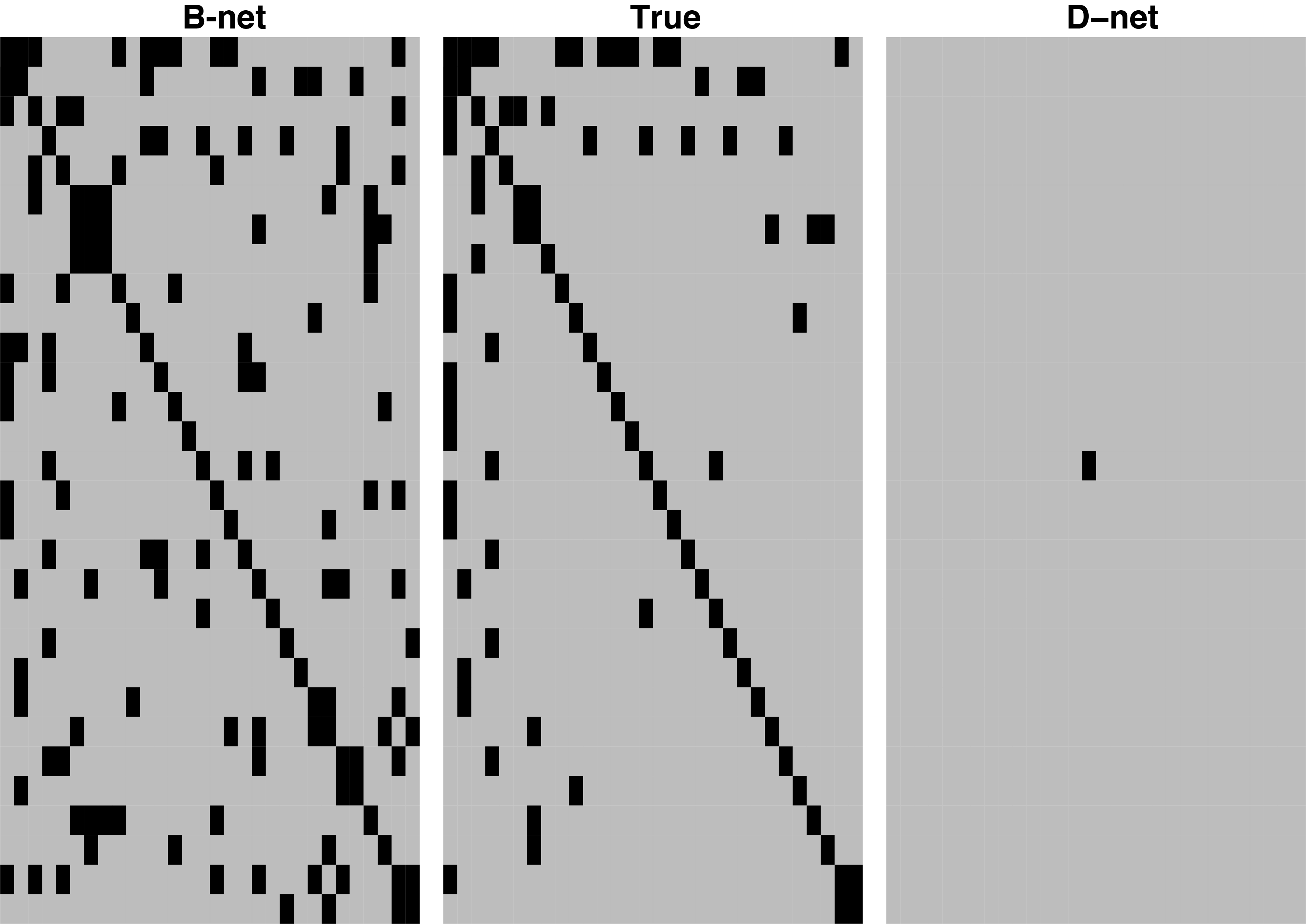}
         \caption{Model 5: scale-free.}
         \label{fig:adjacency_matrix_p10_mod_3}
    \end{subfigure} &\\
    \begin{subfigure}[c]{0.6\textwidth}
      \includegraphics[width=\textwidth]{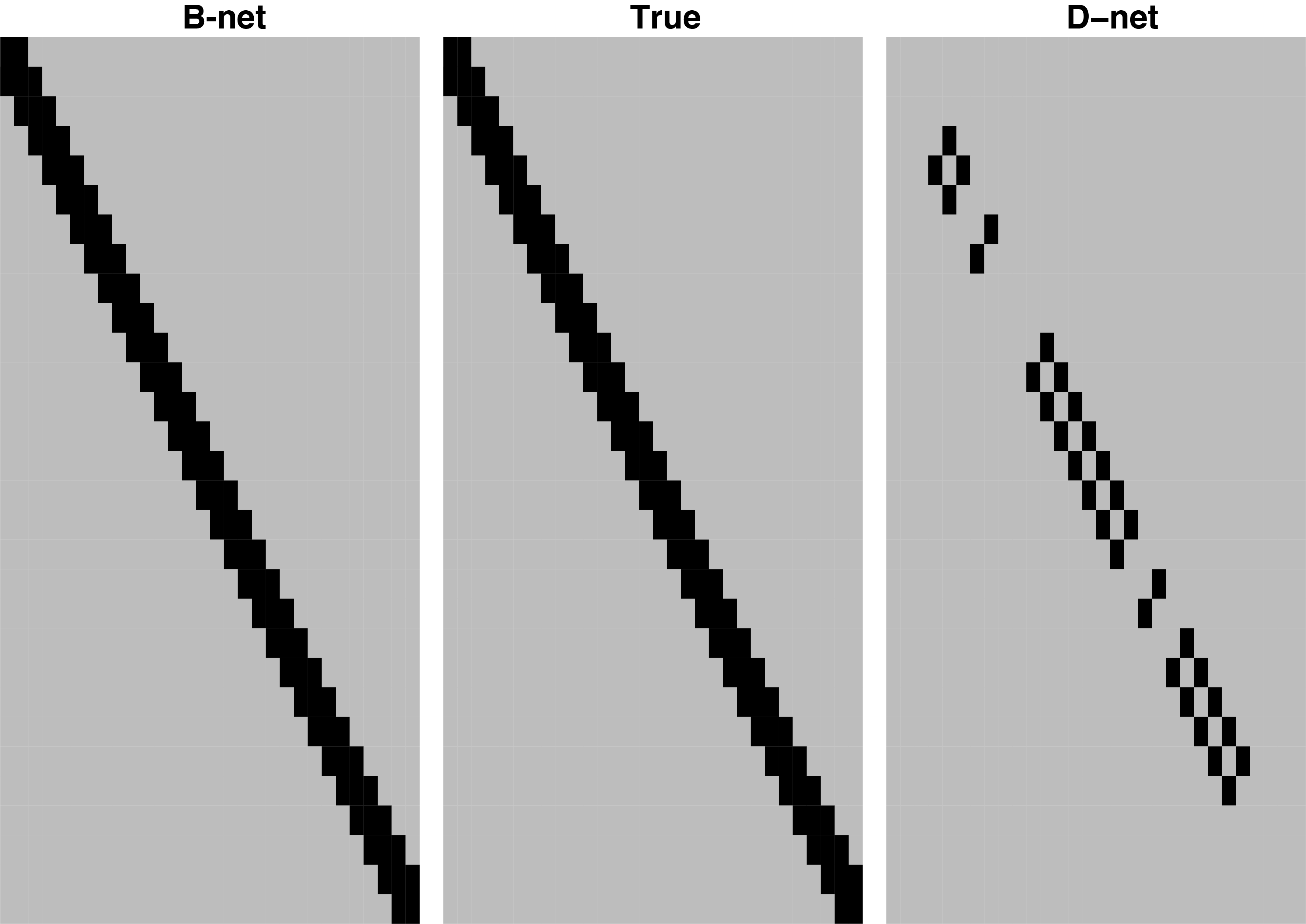}
        \caption{Model 6: band.}
         \label{fig:adjacency_matrix_p10_mod_4}
    \end{subfigure} & \\
    \begin{subfigure}[c]{0.6\textwidth}
      \includegraphics[width=\textwidth]{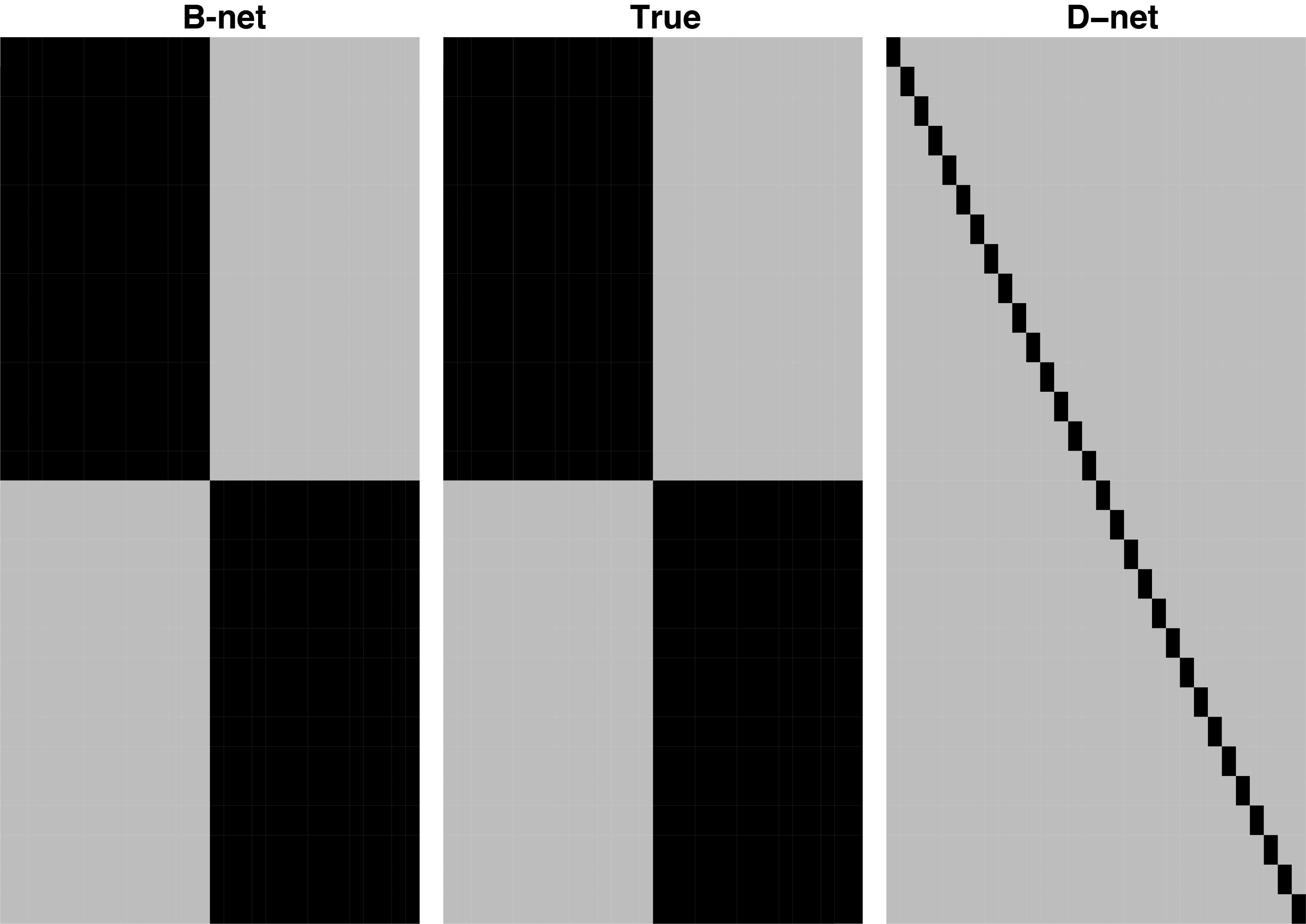}
        \caption{Model 7: cluster.}
         \label{fig:adjacency_matrix_p10_mod_4}
    \end{subfigure} & \\    
  \end{tabular} 
\caption{
\bf Comparison of the true DN, B-net and D-net adjacency matrix estimates for the scale-free, band and cluster structure for $p=30$. }
\label{fig:bayesian_adjacency_matrix_p30}
        
\end{figure}

\begin{figure}[hbt!]
  \centering
   \begin{tabular}[c]{cc}
  \vspace*{1cm}
    \begin{subfigure}[c]{0.9\textwidth}
      \includegraphics[width=\textwidth]{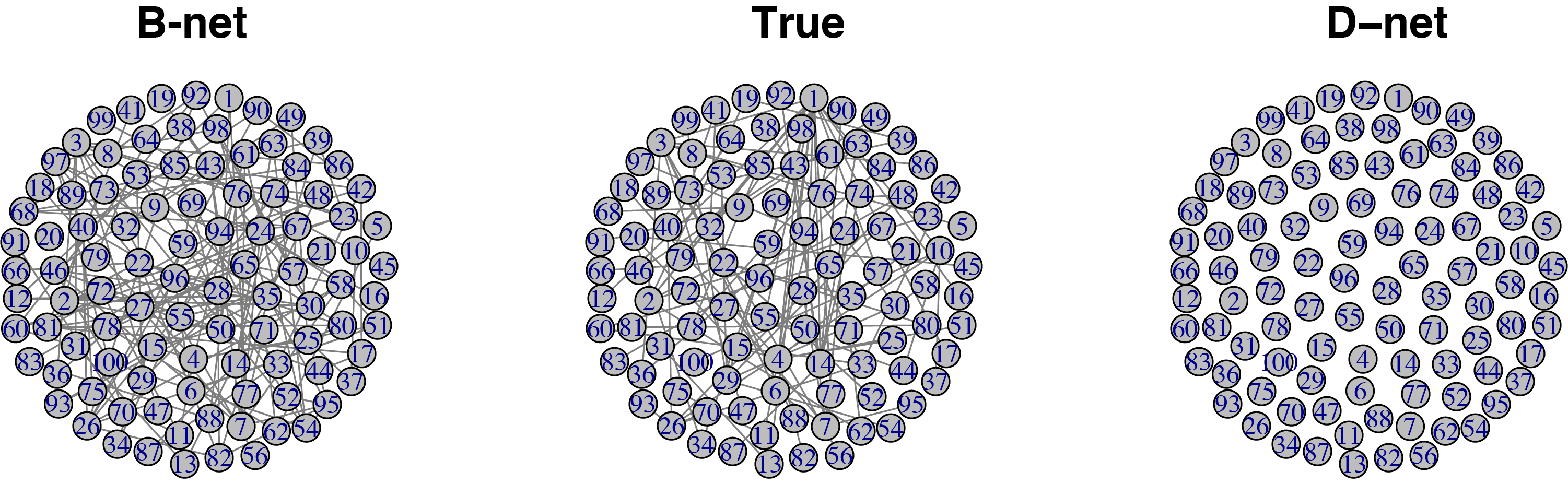}
         \caption{Model 5: scale-free.}
         \label{fig:adjacency_matrix_p10_mod_3}
    \end{subfigure} &\\
  \vspace*{1cm}
    \begin{subfigure}[c]{0.9\textwidth}
      \includegraphics[width=\textwidth]{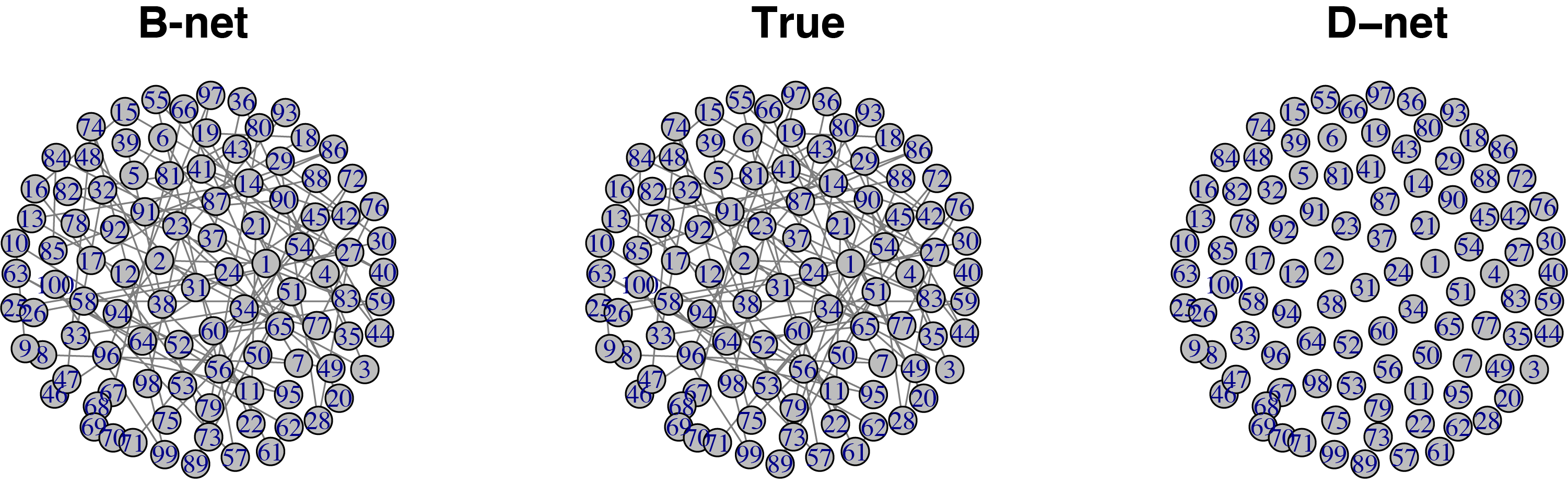}
        \caption{Model 6: band.}
         \label{fig:adjacency_matrix_p10_mod_4}
    \end{subfigure} & \\
    \begin{subfigure}[c]{0.9\textwidth}
      \includegraphics[width=\textwidth]{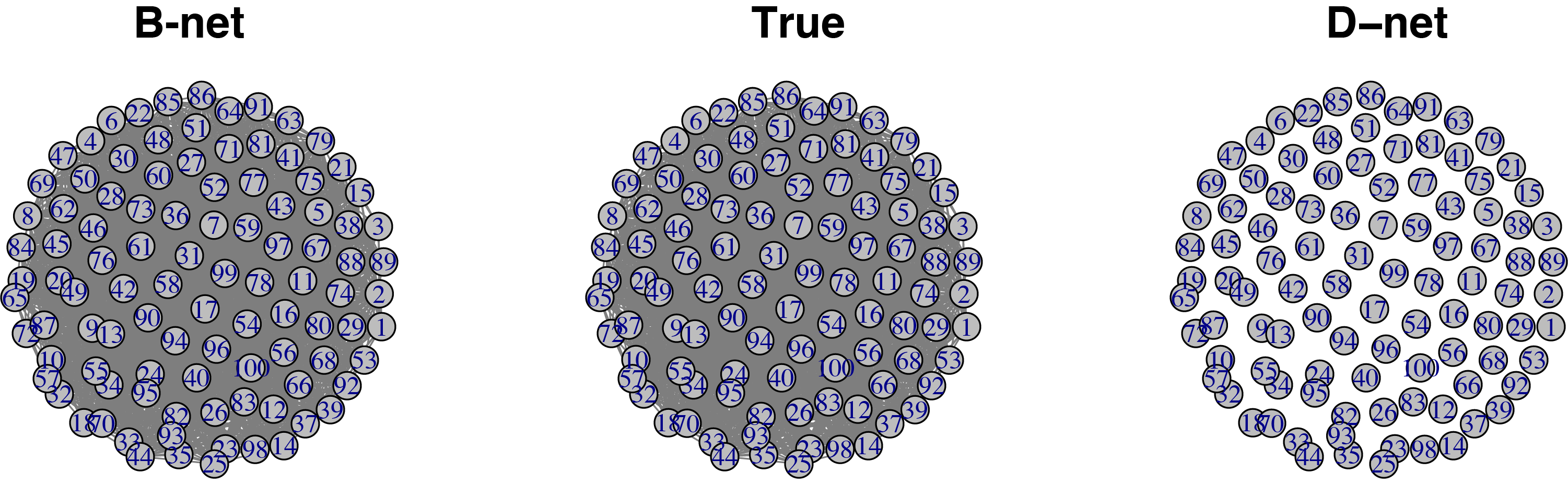}
        \caption{Model 7: cluster.}
         \label{fig:adjacency_matrix_p10_mod_4}
    \end{subfigure} & \\    
  \end{tabular} 
\caption{
\bf Comparison of the true DN, B-net and D-net graphical structure estimates for the scale-free, band and cluster structure for $p=100$. }
\label{fig:bayesian_bn_gm_p100}
        
\end{figure}

\begin{figure}[hbt!]
  \centering
   \begin{tabular}[c]{cc}
    \begin{subfigure}[c]{0.6\textwidth}
      \includegraphics[width=\textwidth]{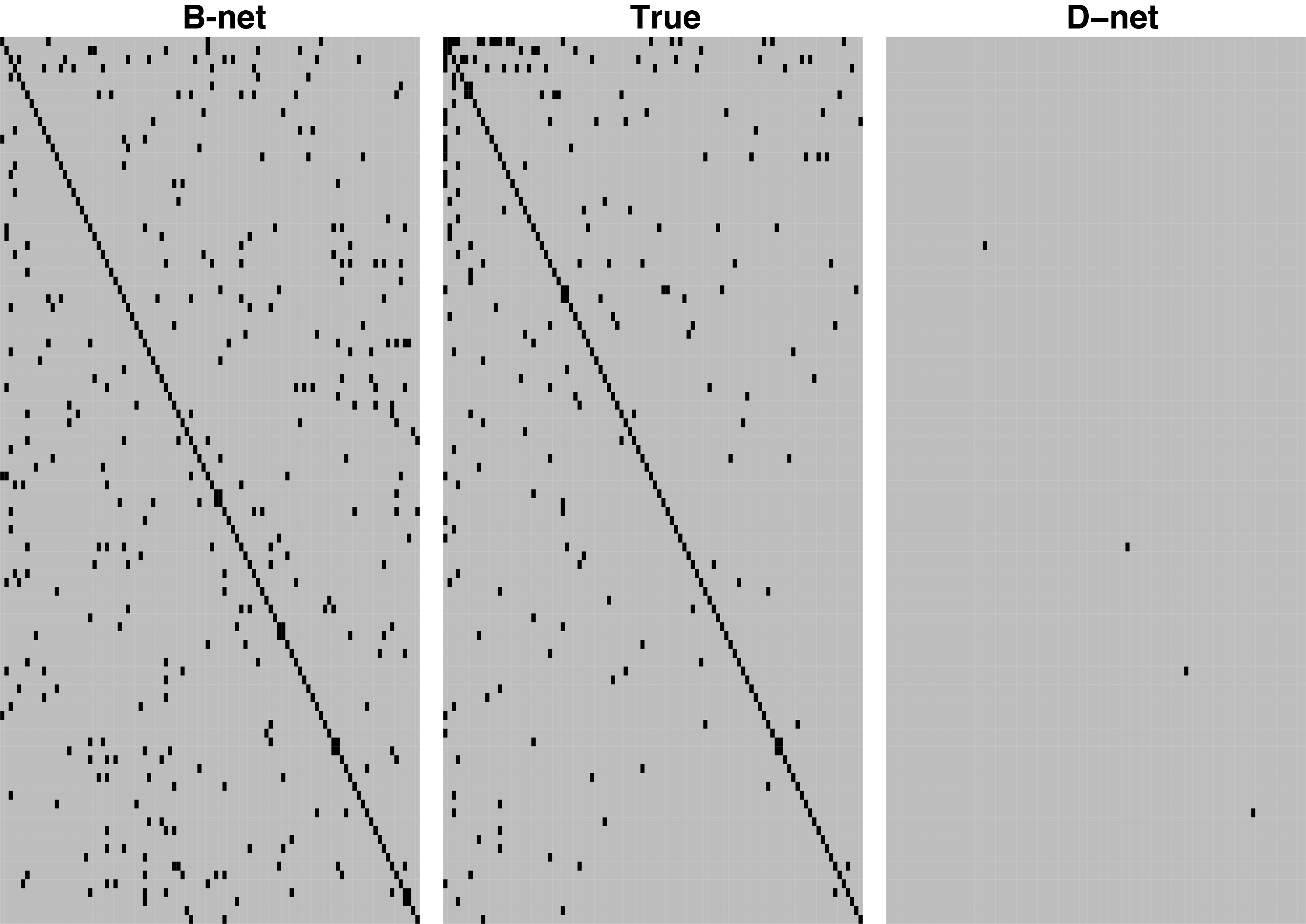}
         \caption{Model 5: scale-free.}
         \label{fig:adjacency_matrix_p10_mod_3}
    \end{subfigure} &\\
    \begin{subfigure}[c]{0.6\textwidth}
      \includegraphics[width=\textwidth]{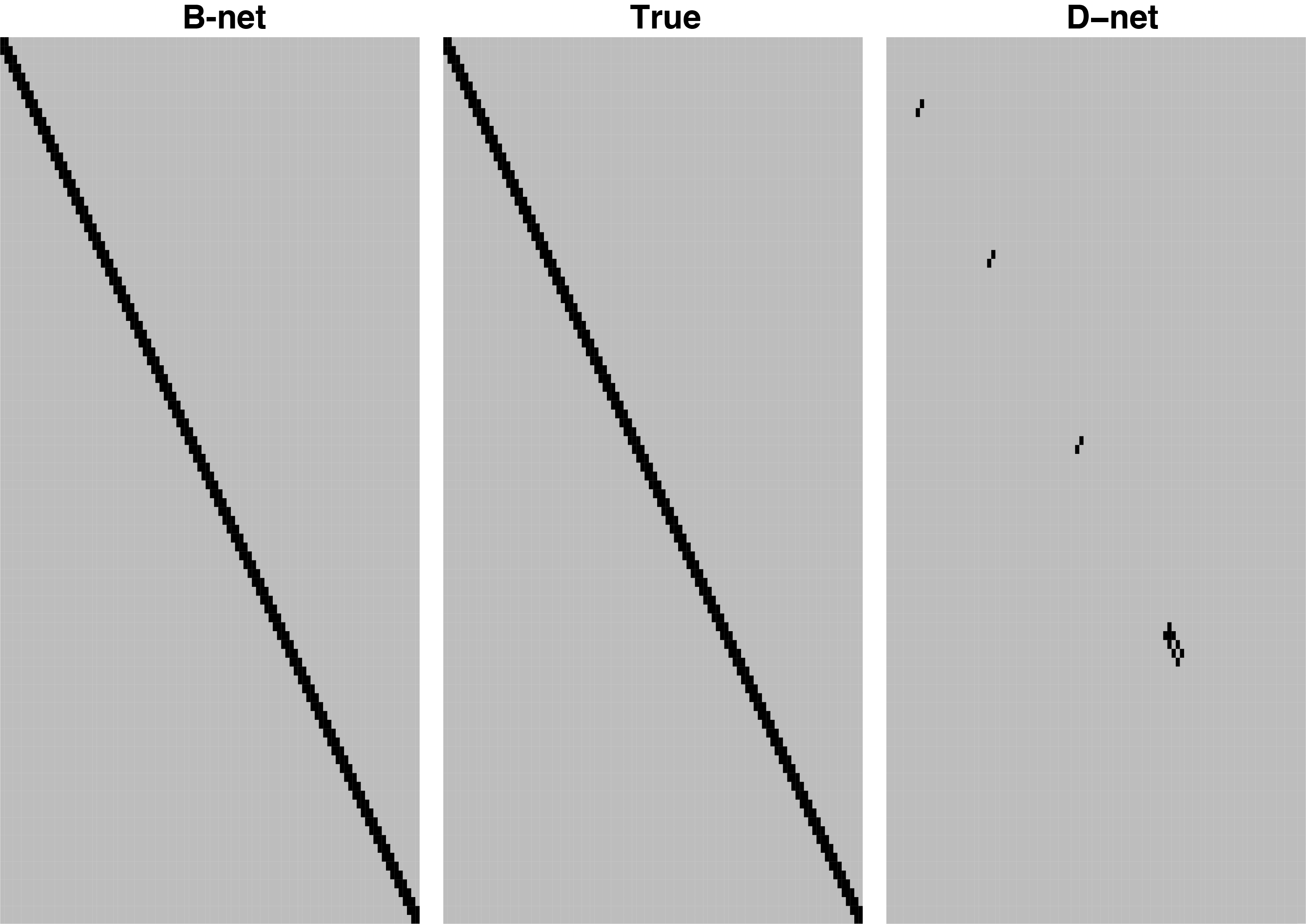}
        \caption{Model 6: band.}
         \label{fig:adjacency_matrix_p10_mod_4}
    \end{subfigure} & \\
    \begin{subfigure}[c]{0.6\textwidth}
      \includegraphics[width=\textwidth]{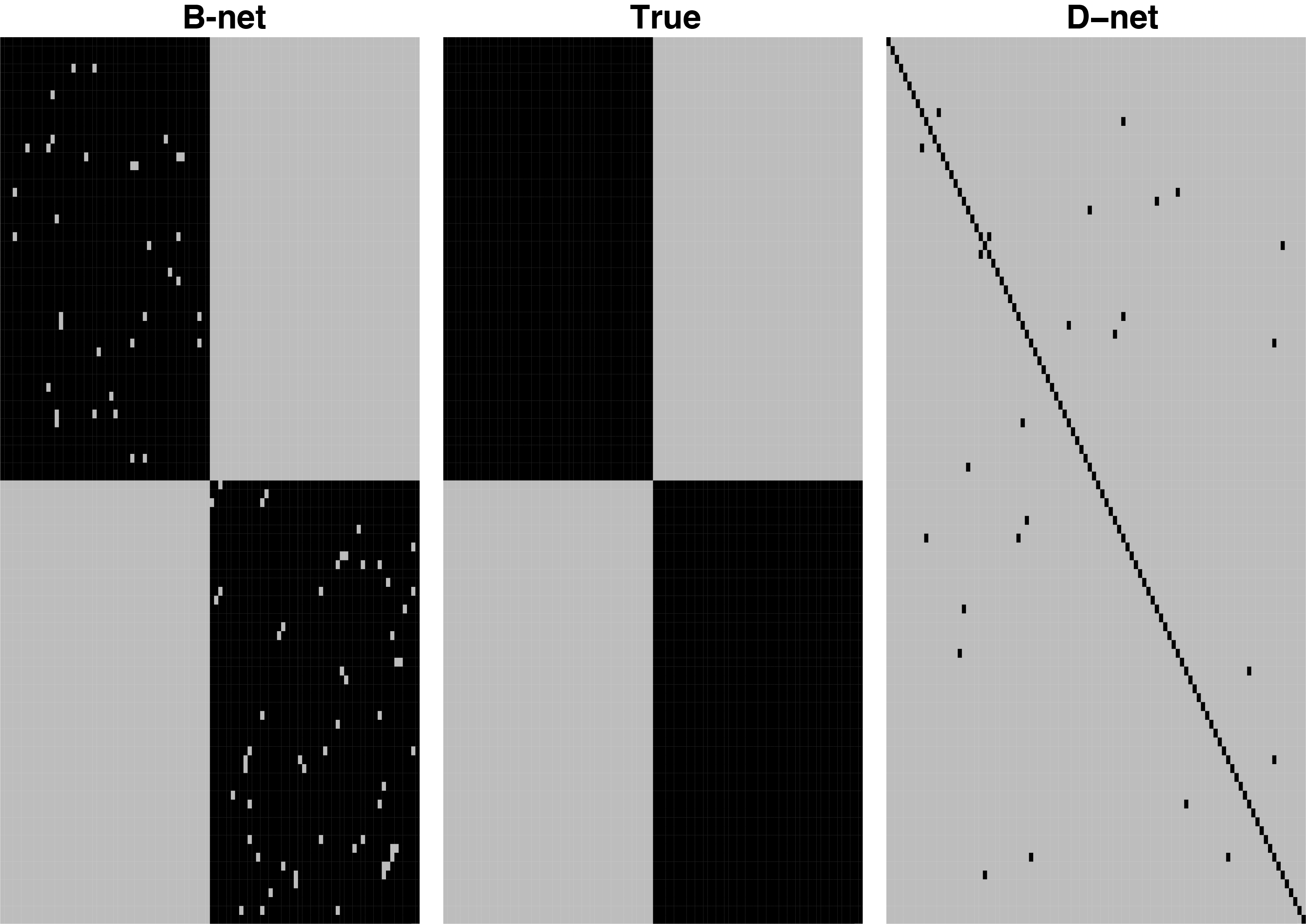}
        \caption{Model 7: cluster.}
         \label{fig:adjacency_matrix_p10_mod_4}
    \end{subfigure} & \\    
  \end{tabular} 
\caption{
\bf Comparison of the true DN, B-net and D-net adjacency matrix estimates for the scale-free, band and cluster structure for $p=100$. }
\label{fig:bayesian_adjacency_matrix_p100}
        
\end{figure}

\begin{figure}[hbt!]
  \centering
  \begin{tabular}[c]{ccc}
  \centering
    \begin{subfigure}[c]{0.3\textwidth}
      \includegraphics[width=\textwidth,height=3.3cm]{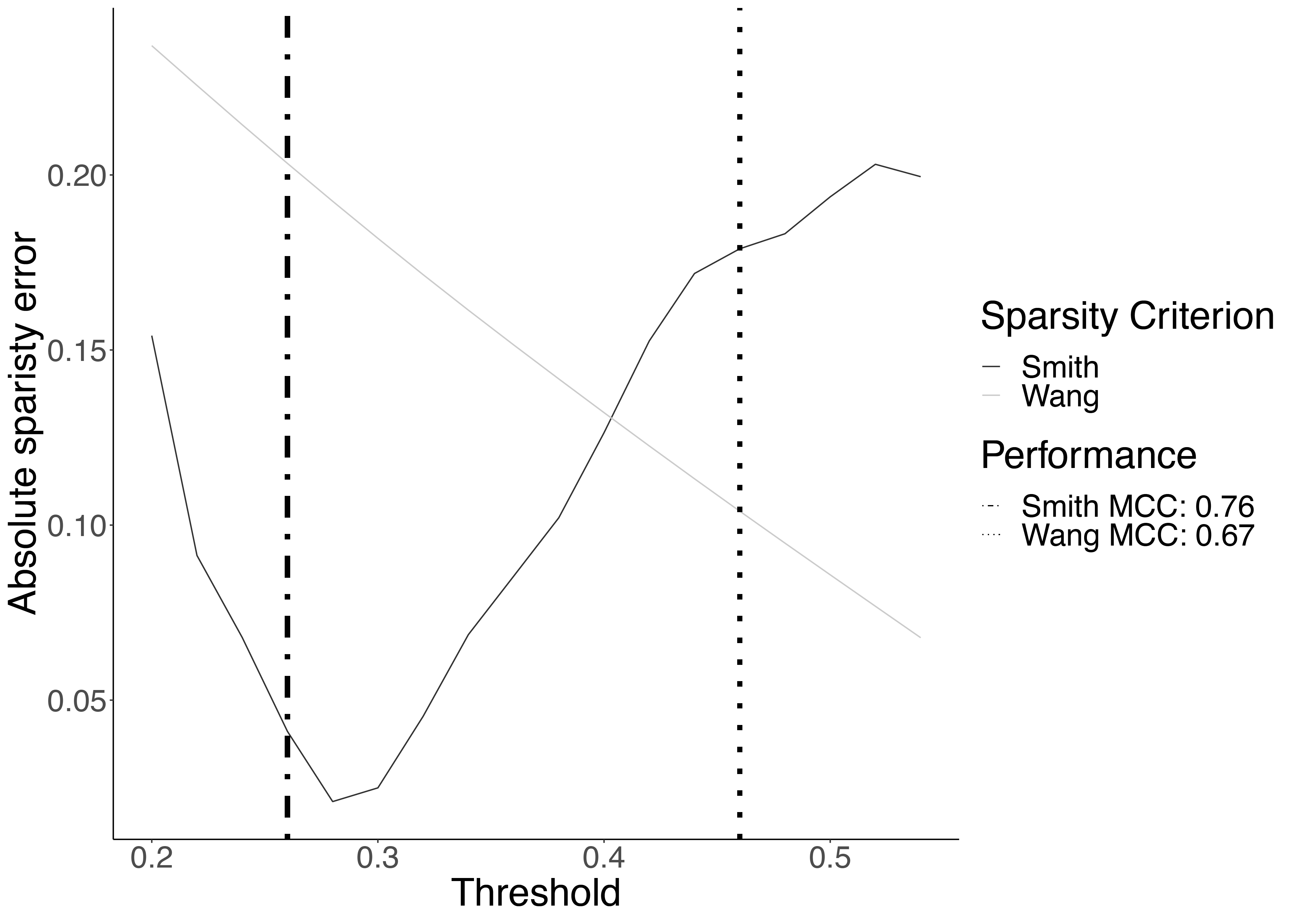}
         \caption{Model 5: scale-free.}
         \label{fig:thresh_study_p10_mod_1}
    \end{subfigure}&
    \begin{subfigure}[c]{0.3\textwidth}
      \includegraphics[width=\textwidth,height=3.3cm]{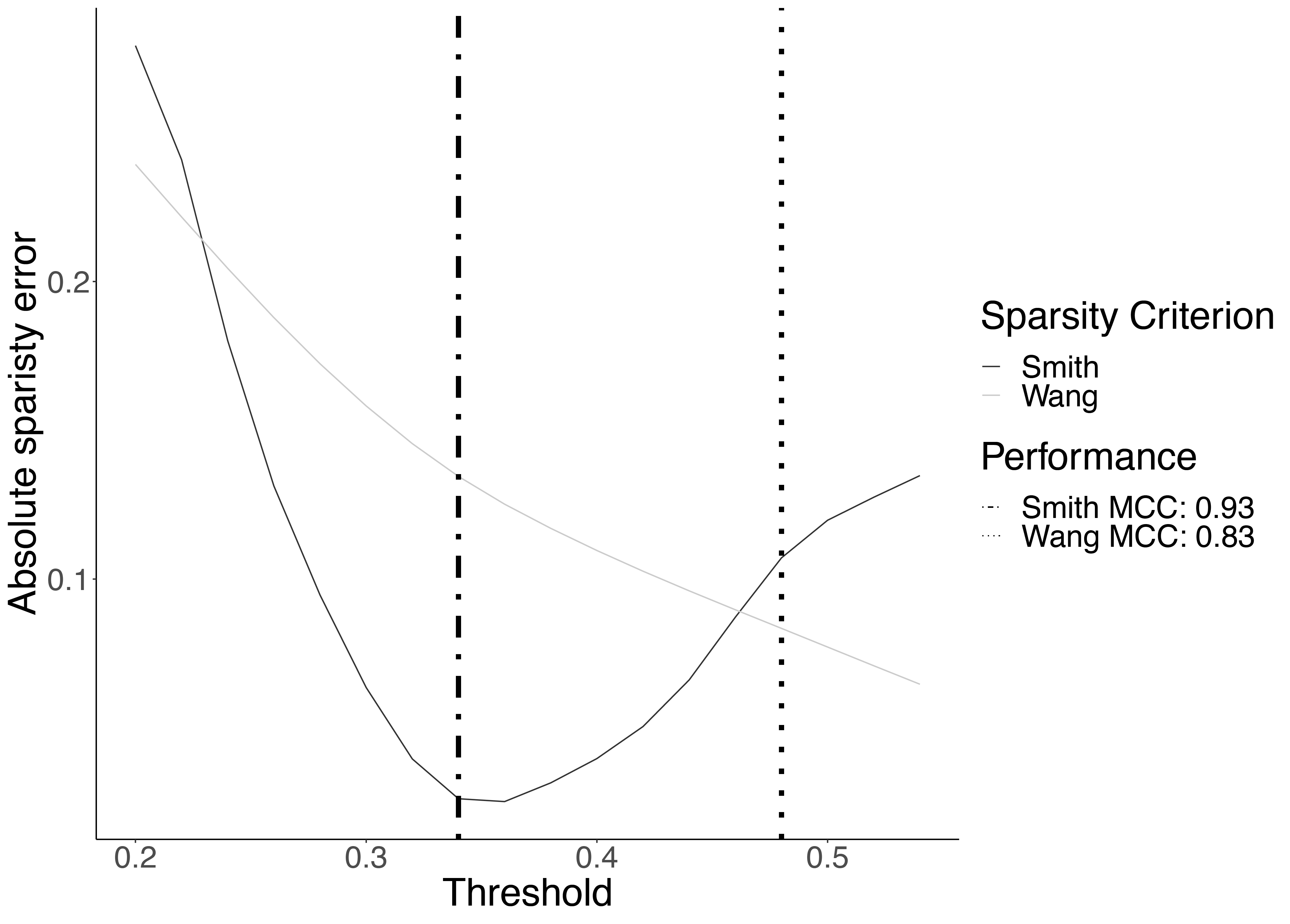}
         \caption{Model 6: band.}
         \label{fig:thresh_study_p10_mod_2}
    \end{subfigure}&
    \begin{subfigure}[c]{0.3\textwidth}
      \includegraphics[width=\textwidth,height=3.3cm]{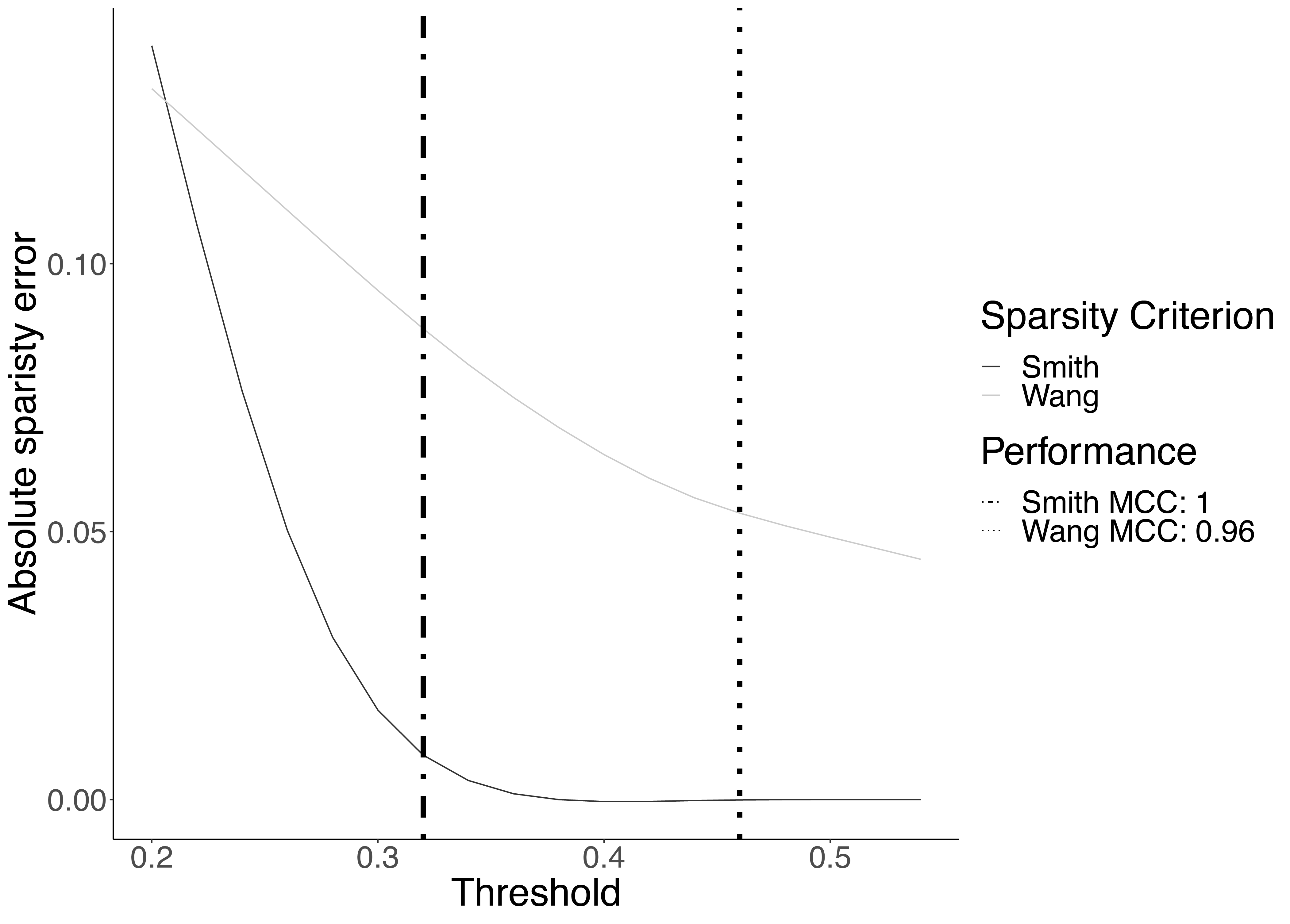}
         \caption{Model 7: cluster.}
         \label{fig:thresh_study_p10_mod_3}
    \end{subfigure}\\
  \end{tabular} 
  \hspace*{\fill}
    \caption{
    \bf The median of the absolute sparsity error and best performing MCC for the scale-free, band and cluster structure under varying thresholds for each Bayesian sparsity criterion in Eq ($9$) (dotted) and Eq ($8$) (dot-dash) for dimension $p=30$. The best performing threshold is indicated by a vertical line with the accompanying MCC value displayed in the legend.}
    \label{fig:smooth_wang_thresh_comparison_p30}
        
\end{figure}

\begin{figure}[hbt!]
  \centering
  \begin{tabular}[c]{ccc}
  \centering
    \begin{subfigure}[c]{0.3\textwidth}
      \includegraphics[width=\textwidth,height=3.3cm]{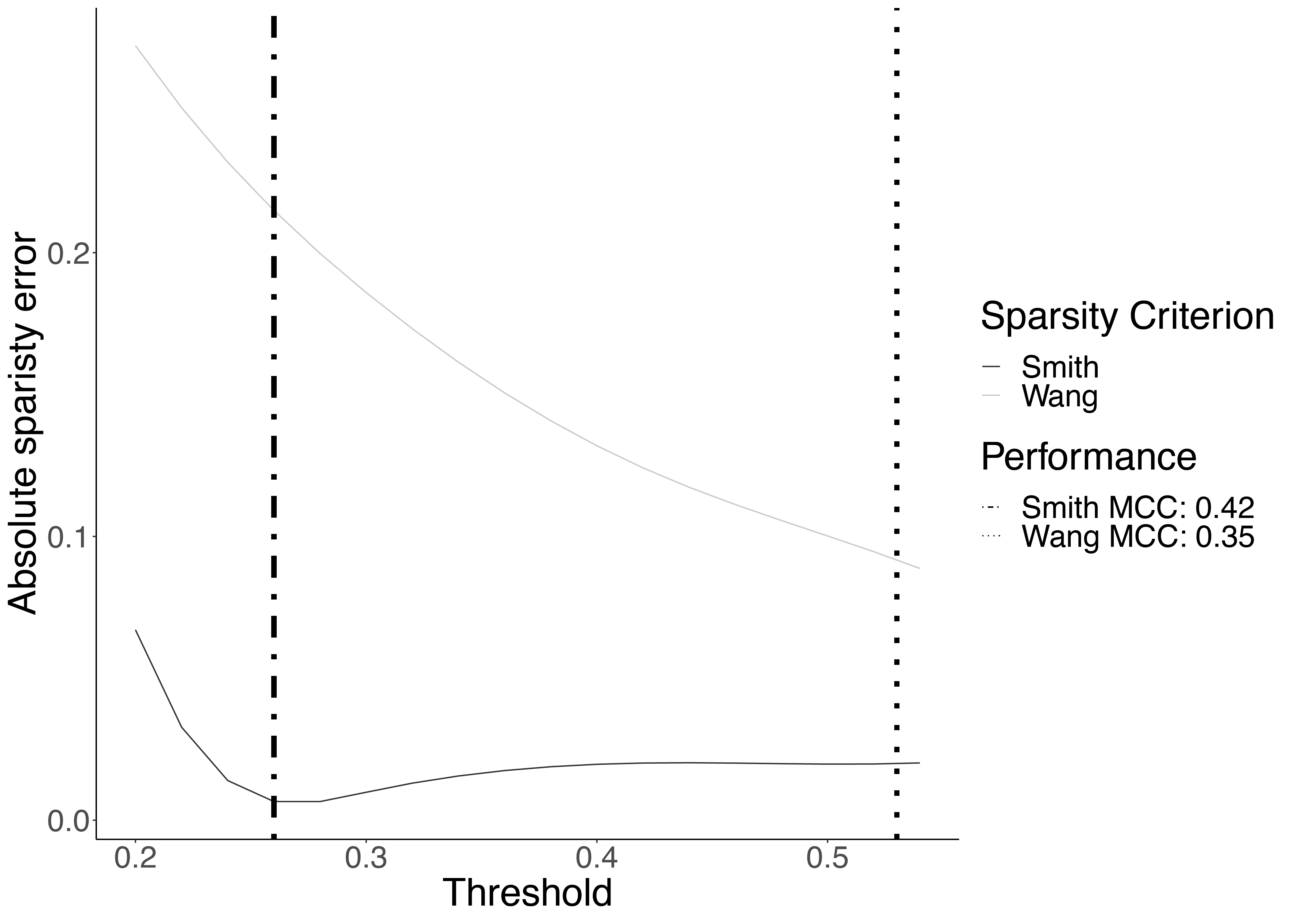}
         \caption{Model 5: scale-free.}
         \label{fig:thresh_study_p10_mod_1}
    \end{subfigure}&
    \begin{subfigure}[c]{0.3\textwidth}
      \includegraphics[width=\textwidth,height=3.3cm]{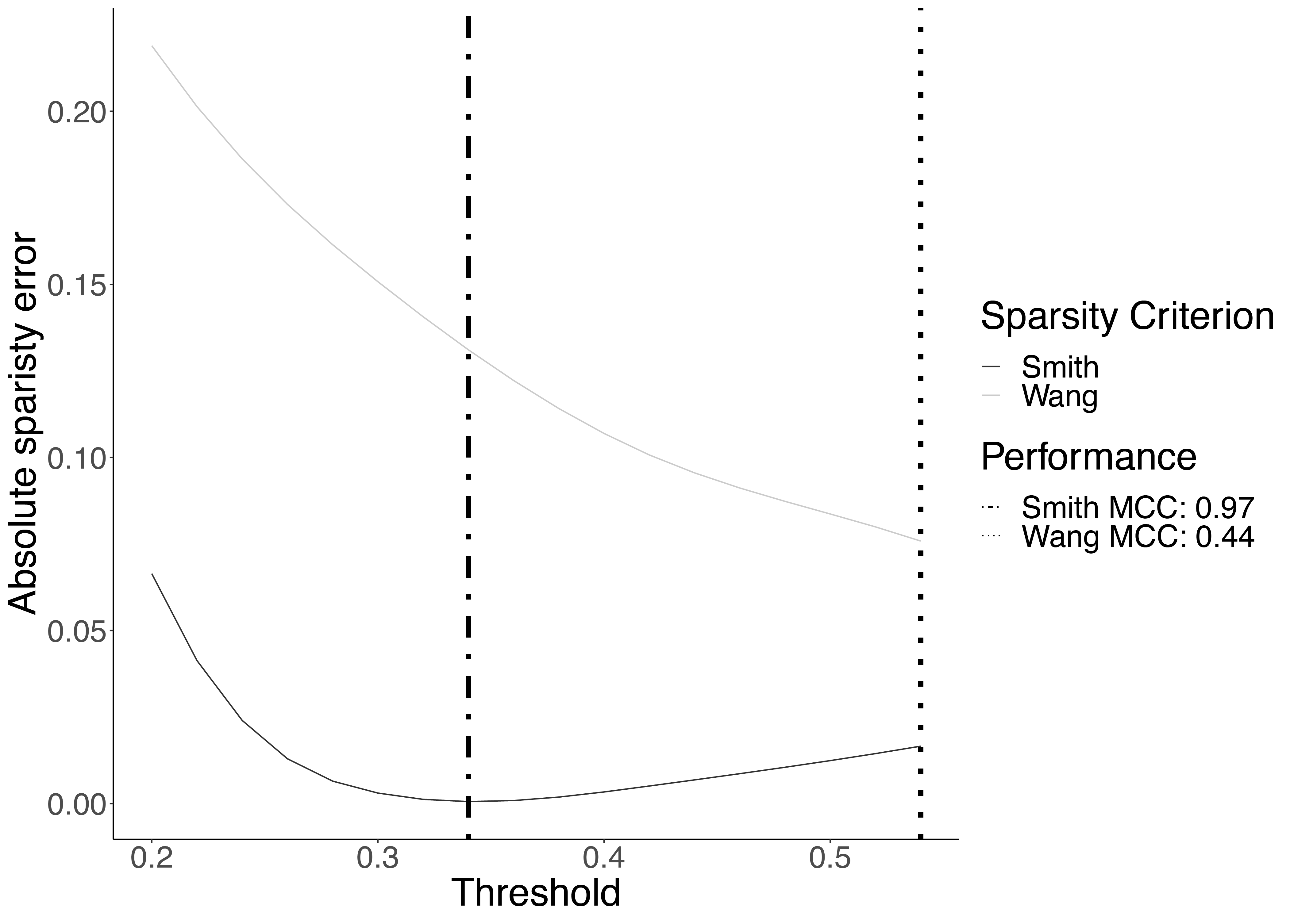}
         \caption{Model 6: band.}
         \label{fig:thresh_study_p10_mod_2}
    \end{subfigure}&
    \begin{subfigure}[c]{0.3\textwidth}
      \includegraphics[width=\textwidth,height=3.3cm]{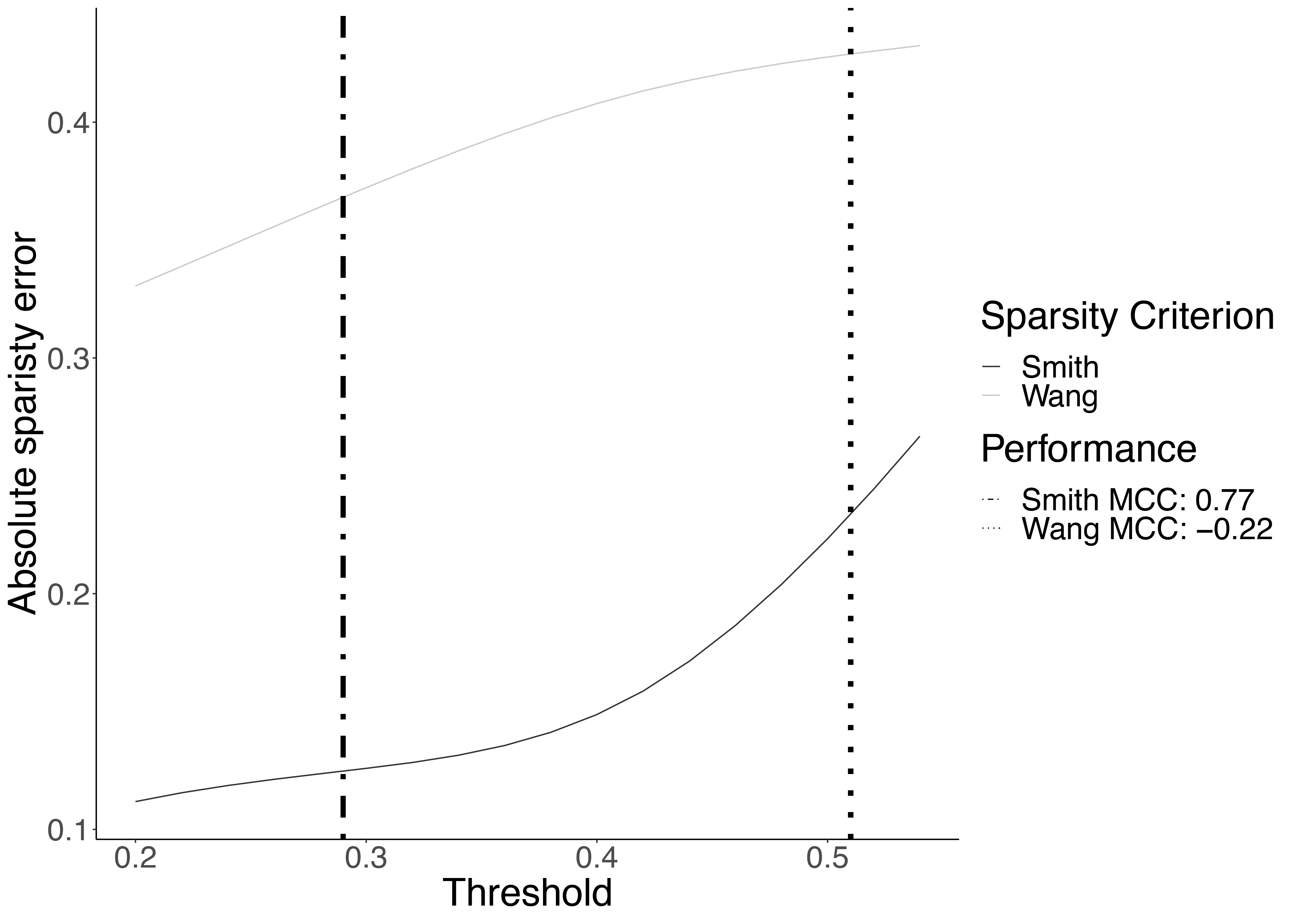}
         \caption{Model 7: cluster.}
         \label{fig:thresh_study_p10_mod_3}
    \end{subfigure}\\
  \end{tabular} 
  \hspace*{\fill}
    \caption{
    \bf The median of the absolute sparsity error and best performing MCC for the scale-free, band and cluster structure  under varying thresholds for each Bayesian sparsity criterion in Eq ($9$) (dotted) and Eq ($8$) (dot-dash) for dimension $p=100$. The best performing threshold is indicated by a vertical line with the accompanying MCC value displayed in the legend.}
    \label{fig:smooth_wang_thresh_comparison_p100}
        
\end{figure}

Fig \ref{fig:bayesian_bn_gm_p30} and \ref{fig:bayesian_adjacency_matrix_p30} provide additional insight into the comparative performance of the of the B-net estimator with regards to the estimated scale-free, band and cluster network structures. The best suited sparsity threshold for $p=30$ for said structures is presented in Fig \ref{fig:smooth_wang_thresh_comparison_p30}. Similarly, Fig \ref{fig:bayesian_bn_gm_p100}, \ref{fig:bayesian_adjacency_matrix_p100} and \ref{fig:smooth_wang_thresh_comparison_p100} depict the aforementioned insights for $p=100$.     


\clearpage

\bibliographystyle{ba}
\bibliography{bibliography}